%% file: main.tex
\documentclass[prd,amsmath,aps,floats,amssymb, floatfix,
  superscriptaddress,nofootinbib,twocolumn,preprintnumbers]{revtex4-2}
\usepackage{tabularx}
\usepackage{amsmath,amssymb,natbib,latexsym,times}

\usepackage{booktabs}
\usepackage{multirow}

\usepackage{graphicx}
\usepackage{dcolumn}
\usepackage{bm}

\usepackage{url}
\usepackage{color}
\usepackage{comment}
\usepackage{mathrsfs}
\usepackage{hyperref}
\usepackage[dvipsnames]{xcolor}
\usepackage{orcidlink}

\newcommand{\dnnz}{{\sc DNNz}~}
\newcommand{\mizuki}{{\sc Mizuki}~}
\newcommand{\dempz}{{\sc DEmPz}~}

\newcommand{\mnras}{Mon.~Not.~Roy.~Astron.~Soc.}
\newcommand{\physrep}{Phys.~Rep.}
\newcommand{\jcap}{JCAP}
\newcommand{\pasj}{Publ.~Astron.~Soc.~Japan}
\newcommand{\apjl}{\apj~Lett.}
\newcommand{\apjs}{\apj~Suppl.}
\newcommand{\aap}{Astronomy~\&~Astrophysics}
\newcommand{\aj}{The~Astronomical~J.}

\newcommand{\mpch}{h^{-1}{\rm Mpc}}

\newcommand{\dSigma}{\Delta\!\Sigma}
\newcommand{\wproj}{w_{\rm p}}
\newcommand{\avrg}[1]{\left\langle#1\right\rangle}
\newcommand{\simgt}{\lower.5ex\hbox{$\; \buildrel > \over \sim \;$}}
\newcommand{\sqdeg}{deg$^2$}
\newcommand{\deltapz}{\Delta z_{\rm ph}}

\begin{document}


\title{Hyper Suprime-Cam Year 3 Results: 
Cosmology from  Galaxy Clustering and Weak Lensing with HSC and SDSS using the Emulator Based Halo Model}

\author{Hironao~Miyatake\orcidlink{0000-0001-7964-9766}}
\email{miyatake@kmi.nagoya-u.ac.jp}
\affiliation{Kobayashi-Maskawa Institute for the Origin of Particles and the Universe (KMI),
Nagoya University, Nagoya, 464-8602, Japan}
\affiliation{Institute for Advanced Research, Nagoya University, Nagoya 464-8601, Japan}
\affiliation{Kavli Institute for the Physics and Mathematics of the Universe (WPI), The University of Tokyo Institutes for Advanced Study (UTIAS), The University of Tokyo, Chiba 277-8583, Japan}

\author{Sunao~Sugiyama\orcidlink{0000-0003-1153-6735}}
\email{sunao.sugiyama@ipmu.jp}
\affiliation{Kavli Institute for the Physics and Mathematics of the Universe (WPI), The University of Tokyo Institutes for Advanced Study (UTIAS), The University of Tokyo, Chiba 277-8583, Japan}
\affiliation{Department of Physics, The University of Tokyo, Bunkyo, Tokyo 113-0031, Japan}

\author{Masahiro~Takada\orcidlink{0000-0002-5578-6472}}
\email{masahiro.takada@ipmu.jp}
\affiliation{Kavli Institute for the Physics and Mathematics of the Universe (WPI), The University of Tokyo Institutes for Advanced Study (UTIAS), The University of Tokyo, Chiba 277-8583, Japan}

\author{Takahiro~Nishimichi\orcidlink{0000-0002-9664-0760}}
\affiliation{Center for Gravitational Physics and Quantum Information, Yukawa Institute for Theoretical Physics, Kyoto University, Kyoto 606-8502, Japan}
\affiliation{Kavli Institute for the Physics and Mathematics of the Universe
(WPI), The University of Tokyo Institutes for Advanced Study (UTIAS),
The University of Tokyo, Chiba 277-8583, Japan}
\affiliation{Department of Astrophysics and Atmospheric Sciences, Faculty of Science, Kyoto Sangyo University, Motoyama, Kamigamo, Kita-ku, Kyoto 603-8555, Japan}

\author{Xiangchong~Li\orcidlink{0000-0003-2880-5102}}
\affiliation{McWilliams Center for Cosmology, Department of Physics, Carnegie Mellon University, Pittsburgh, PA 15213, USA}
\affiliation{Kavli Institute for the Physics and Mathematics of the Universe
(WPI), The University of Tokyo Institutes for Advanced Study (UTIAS),
The University of Tokyo, Chiba 277-8583, Japan}

\author{Masato~Shirasaki\orcidlink{0000-0002-1706-5797}}
\affiliation{National Astronomical Observatory of Japan, National Institutes of Natural Sciences, Mitaka, Tokyo 181-8588, Japan}
\affiliation{The Institute of Statistical Mathematics,
Tachikawa, Tokyo 190-8562, Japan}

\author{Surhud~More\orcidlink{0000-0002-2986-2371}}
\affiliation{The Inter-University Centre for Astronomy and Astrophysics, Post bag 4, Ganeshkhind, Pune 411007, India}
\affiliation{Kavli Institute for the Physics and Mathematics of the Universe
(WPI), The University of Tokyo Institutes for Advanced Study (UTIAS),
The University of Tokyo, Chiba 277-8583, Japan}

\author{Yosuke~Kobayashi\orcidlink{0000-0002-6633-5036}}
\affiliation{Department of Astronomy and Steward Observatory, University of Arizona, 933 N Cherry Ave, Tucson, AZ 85719, USA}
\affiliation{Kavli Institute for the Physics and Mathematics of the Universe
(WPI), The University of Tokyo Institutes for Advanced Study (UTIAS),
The University of Tokyo, Chiba 277-8583, Japan}

\author{Atsushi~J.~Nishizawa\orcidlink{0000-0002-6109-2397}}
\affiliation{Gifu Shotoku Gakuen University, Gifu 501-6194, Japan}
\affiliation{Institute for Advanced Research, Nagoya University, Nagoya 464-8601, Japan}
\affiliation{Kobayashi-Maskawa Institute for the Origin of Particles and the Universe (KMI),
Nagoya University, Nagoya, 464-8602, Japan}

\author{Markus~M.~Rau\orcidlink{0000-0003-3709-1324}}
\affiliation{McWilliams Center for Cosmology, Department of Physics, Carnegie Mellon University, Pittsburgh, PA 15213, USA}
\affiliation{High Energy Physics Division, Argonne National Laboratory, Lemont, IL 60439, USA}

\author{Tianqing~Zhang\orcidlink{0000-0002-5596-198X}}
\affiliation{McWilliams Center for Cosmology, Department of Physics, Carnegie Mellon University, Pittsburgh, PA 15213, USA}

\author{Ryuichi~Takahashi}
\affiliation{Faculty of Science and Technology, Hirosaki University, 3 Bunkyo-cho, Hirosaki, Aomori 036-8561, Japan}

\author{Roohi~Dalal\orcidlink{0000-0002-7998-9899}}
\affiliation{Department of Astrophysical Sciences, Princeton University, Princeton, NJ 08544, USA}

\author{Rachel~Mandelbaum\orcidlink{0000-0003-2271-1527}}
\affiliation{McWilliams Center for Cosmology, Department of Physics, Carnegie Mellon University, Pittsburgh, PA 15213, USA}

\author{Michael~A.~Strauss\orcidlink{0000-0002-0106-7755}}
\affiliation{Department of Astrophysical Sciences, Princeton University, Princeton, NJ 08544, USA}

\author{Takashi~Hamana}
\affiliation{National Astronomical Observatory of Japan, National Institutes of Natural Sciences, Mitaka, Tokyo 181-8588, Japan}

\author{Masamune~Oguri\orcidlink{0000-0003-3484-399X}}
\affiliation{Center for Frontier Science, Chiba University, Chiba 263-8522, Japan}
\affiliation{Research Center for the Early Universe, The University of Tokyo, Bunkyo, Tokyo 113-0031, Japan}
\affiliation{Department of Physics, The University of Tokyo, Bunkyo, Tokyo 113-0031, Japan}
\affiliation{Kavli Institute for the Physics and Mathematics of the Universe (WPI), The University of Tokyo Institutes for Advanced Study (UTIAS), The University of Tokyo, Chiba 277-8583, Japan}

\author{Ken~Osato\orcidlink{0000-0002-7934-2569}}
\affiliation{Center for Frontier Science, Chiba University, Chiba 263-8522, Japan}
\affiliation{Department of Physics, Graduate School of Science, Chiba University, Chiba 263-8522, Japan}

\author{Wentao~Luo}
\affiliation{School of Physical Sciences, University of Science and Technology of China, Hefei, Anhui 230026, China}
\affiliation{CAS Key Laboratory for Researches in Galaxies and Cosmology/Department of Astronomy, School of Astronomy and Space Science, University of Science and Technology of China, Hefei, Anhui 230026, China}

\author{Arun~Kannawadi\orcidlink{0000-0001-8783-6529}}
\affiliation{Department of Astrophysical Sciences, Princeton University, Princeton, NJ 08544, USA}

\author{Bau-Ching~Hsieh\orcidlink{0000-0001-5615-4904}}
\affiliation{Academia Sinica Institute of Astronomy and Astrophysics, No. 1, Sec. 4, Roosevelt Rd., Taipei 10617, Taiwan}

\author{Robert~Armstrong}
\affiliation{Lawrence Livermore National Laboratory, Livermore, CA 94551, USA}

\author{Yutaka~Komiyama\orcidlink{0000-0002-3852-6329}}
\affiliation{Department of Advanced Sciences, Faculty of Science and Engineering, Hosei University, 3-7-2 Kajino-cho, Koganei-shi, Tokyo 184-8584, Japan}

\author{Robert~H.~Lupton\orcidlink{0000-0003-1666-0962}}
\affiliation{Department of Astrophysical Sciences, Princeton University, Princeton, NJ 08544, USA}

\author{Nate~B.~Lust\orcidlink{0000-0002-4122-9384}}
\affiliation{Department of Astrophysical Sciences, Princeton University, Princeton, NJ 08544, USA}

\author{Lauren~A.~MacArthur}
\affiliation{Department of Astrophysical Sciences, Princeton University, Princeton, NJ 08544, USA}

\author{Satoshi Miyazaki\orcidlink{0000-0002-1962-904X}}
\affiliation{Subaru Telescope,  National Astronomical Observatory of Japan, 650 N Aohoku Place Hilo HI 96720 USA}

\author{Hitoshi~Murayama\orcidlink{0000-0001-5769-9471}}
\affiliation{Berkeley Center for Theoretical Physics, University of California, Berkeley, CA 94720, USA}
\affiliation{Theory Group, Lawrence Berkeley National Laboratory, Berkeley, CA 94720, USA}
\affiliation{Kavli Institute for the Physics and Mathematics of the Universe (WPI), The University of Tokyo Institutes for Advanced Study (UTIAS), The University of Tokyo, Chiba 277-8583, Japan}

\author{Yuki~Okura\orcidlink{0000-0001-6623-4190}}
\affiliation{National Astronomical Observatory of Japan, National Institutes of Natural Sciences, Mitaka, Tokyo 181-8588, Japan}

\author{Paul~A.~Price\orcidlink{0000-0003-0511-0228}}
\affiliation{Department of Astrophysical Sciences, Princeton University, Princeton, NJ 08544, USA}

\author{Tomomi~Sunayama\orcidlink{0009-0004-6387-5784}}
\affiliation{Department of Astronomy and Steward Observatory, University of Arizona, 933 N Cherry Ave, Tucson, AZ 85719, USA}
\affiliation{Kobayashi-Maskawa Institute for the Origin of Particles and the Universe (KMI), Nagoya University, Nagoya, 464-8602, Japan}

\author{Philip~J.~Tait}
\affiliation{Subaru Telescope,  National Astronomical Observatory of Japan, 650 N Aohoku Place Hilo HI 96720 USA}

\author{Masayuki~Tanaka}
\affiliation{National Astronomical Observatory of Japan, National Institutes of Natural Sciences, Mitaka, Tokyo 181-8588, Japan}

\author{Shiang-Yu~Wang}
\affiliation{Academia Sinica Institute of Astronomy and Astrophysics, No. 1, Sec. 4, Roosevelt Rd., Taipei 10617, Taiwan}

\date{\today}

\begin{abstract}
We present cosmology results from a blinded joint analysis of cosmic shear, $\xi_{\pm}(\vartheta)$, galaxy-galaxy weak lensing, $\dSigma(R)$, and projected galaxy clustering, $\wproj(R)$, measured from the Hyper Suprime-Cam three-year (HSC-Y3) shape catalog and the Sloan Digital Sky Survey (SDSS) DR11 spectroscopic galaxy catalog -- a 3$\times$2pt cosmology analysis.  We define luminosity-cut, and therefore nearly volume-limited, samples of SDSS galaxies to serve as the tracers of $\wproj$ and as the lens samples for $\dSigma$ in three spectroscopic redshift bins spanning the range $0.15<z<0.7$.  For the $\xi_{\pm}$ and $\dSigma$ measurements, we use a single sample of about seven million source galaxies over 416\,\sqdeg, selected from HSC-Y3 based on having photometric redshifts (photo-$z$) greater than 0.75. The deep, high-quality HSC-Y3 data enable significant detections of the $\dSigma$ signals, with integrated signal-to-noise ratio $S/N\sim 24$ in the range $3\le R/[h^{-1}{\rm Mpc}]\le 30$ over the three lens samples. $\xi_{\pm}$ has $S/N\sim 19$ in the range $8'\le \vartheta\le 50'$ and $30'\le \vartheta\le 150'$ for $\xi_+$ and $\xi_-$, respectively. For cosmological parameter inference, we use the {\tt Dark Emulator} package, combined with a halo occupation distribution prescription for the relation between galaxies and halos, to model $\wproj$ and $\dSigma$ down to quasi-nonlinear scales, and we estimate cosmological parameters after marginalizing over nuisance parameters. In our baseline analysis we employ an {\it uninformative} flat prior of the residual photo-$z$ error, given by $\Pi(\deltapz)={\cal U}(-1,1)$, to model a residual bias in the mean redshift of HSC source galaxies. Comparing the relative lensing amplitudes for $\dSigma$ in the three redshift bins and for $\xi_{\pm}$ with the single HSC source galaxy sample allows us to calibrate the photo-$z$ parameter $\deltapz$ to the precision of $\sigma(\deltapz)\simeq 0.09$.  With these methods, we obtain a robust constraint on the cosmological parameters for the flat $\Lambda$CDM model: $S_8=\sigma_8(\Omega_{\rm m}/0.3)^{0.5}=0.763^{+0.040}_{-0.036}$ (68\% C.I.), or the best-constrained parameter given by $S'_8=\sigma_8(\Omega_{\rm m}/0.3)^{0.22}=0.721\pm 0.028$, determined with about 4\% fractional precision.  Our HSC-Y3 data exhibits about 2.5$\sigma$ tension with the {\it Planck} inferred $S_8$ value for the $\Lambda$CDM model, and hints at a non-zero residual photo-$z$ bias implying that the true mean redshift of the HSC galaxies at $z\gtrsim 0.75$ is higher than that implied by the original photo-$z$ estimates.
\end{abstract}

\maketitle

\section{Introduction}
\label{sec:introduction}
The cosmological standard model assuming the initial conditions predicted by an inflationary scenario, $\Lambda$ Cold Dark Matter ($\Lambda$CDM) model, has been successful in explaining a variety of observations \citep[e.g.][]{2020moco.book.....D}. 
Wide-area galaxy imaging surveys in optical and near-infrared wavelengths enables us to investigate fundamental problems in cosmology, such as the nature of dark matter and the origin of cosmic acceleration \citep[e.g.][]{Weinbergetal:13}. Precise measurements of weak gravitational lensing by the ongoing Stave-III surveys, such as the Subaru Hyper Suprime-Cam \footnote{\url{https://hsc.mtk.nao.ac.jp/ssp/}} \citep[HSC;][]{HSCoverview:17,2019PASJ...71...43H,2020PASJ...72...16H,  Miyatake:2022b, Sugiyama:2022}, the Dark Energy Survey \footnote{\url{https://www.darkenergysurvey.org}} \citep[DES;][]{DES-Y3}, and the Kilo-Degree Survey \footnote{\url{http://kids.strw.leidenuniv.nl}} \citep[KiDS;][]{2020arXiv200701844J}, have provided tight constraints on cosmological parameters. It is intriguing that, under the $\Lambda$CDM assumption, the weak lensing measurements infers a lower value of $\sigma_8$ or $S_8$, which characterized the clustering amplitude of large-scale structure in the present day universe \citep[Ref.][for a recent review]{2022JHEAp..34...49A}, than the {\it Planck} cosmic microwave background (CMB) measurements \citep{2020A&A...641A...6P} does. This discrepancy might hint at the possibility of new physics beyond the standard $\Lambda$CDM model.

Challenges of large-scale structure probes lie in systematic effects/errors inherent both in observations and theory. One of the important observational systematic effects, relevant to weak lensing cosmology, arises from imperfect photometric redshift estimates (hereafter referred to as photo-$z$).  Due to the limited information carried by broad-band photometry and/or difficulties in uniform and accurate characterization of individual galaxy photometry, photometric redshift estimates are not perfect. Hence, photo-$z$ estimates need to be calibrated using a representative calibration sample of galaxies that have accurate redshift estimates; ideally we need a representative spectroscopic sample but the COSMOS catalog which provides  30-band photo-$z$'s is currently a main calibration sample for photo-$z$'s of faint galaxies \citep{2020arXiv200301511N}.

The main systematic effects on the theory side lie in the difficulties in accurately modeling nonlinear structure formation, and the unknown relation between the distributions of matter and galaxies, where the latter is referred to as the galaxy bias uncertainty. The physical processes inherent in the formation and evolution of galaxies cannot yet be accurately and fully modeled from first principles. Nevertheless, on large scales, i.e., beyond a few 10~Mpc where gravity is a driving force of structure formation, the linear theory of structure formation is quite accurate, and predicts that the galaxy distribution for any type of galaxy is related to the underlying matter distribution by a scale-independent factor, i.e the linear bias parameter \cite{Kaiser:1984}. On smaller scales, the bias function is scale-dependent, due to the mode coupling in nonlinear structure formation. 

Combining multiple cosmological probes provides a promising way to mitigate the aforementioned systematic effects in cosmological inference. In this paper, we combine the projected correlation function of galaxies ($\wproj$), galaxy-galaxy weak lensing ($\dSigma$), and cosmic shear correlations ($\xi_{\pm}$) -- so-called 3$\times$2pt cosmology analysis, measured from the photometric HSC three-year (hereafter HSC-Y3) data covering about 416~deg$^2$ of the sky and the spectroscopic SDSS galaxy catalogs. By cross-correlating the positions of SDSS galaxies with shapes of the background HSC galaxies, we can measure the $\dSigma$ signal,  which in turn allows us to infer the average matter distribution around the SDSS galaxies. We then combine the $\dSigma(R)$ measurement with the auto-correlation function of galaxies in the same sample, $\wproj(R)$, as a function of projected separation $R$ to observationally infer the galaxy bias function of the SDSS galaxies, including its scale-dependence. We will use a {\it single} sample of the HSC source galaxies to perform  weak lensing measurements for $\dSigma$ for each of the SDSS galaxy subsamples that are subdivided into three spectroscopic redshift bins spanning the range $z=[0.15,0.7]$. Comparing the relative $\dSigma$ amplitudes in the three redshift bins and the cosmic shear signal $\xi_\pm$, for a given sample of HSC source galaxies, enables us to {\it calibrate} any residual error in the mean photometric redshifts of HSC source galaxies, as proposed in \citet{OguriTakada:11} \citep[also see][]{Miyatake:2022a,Miyatake:2022b}. 

To resolve the modeling difficulties of  clustering observables on small scales, we use the halo model approach \citep{Seljak:00,MaFry:00,PeacockSmith:00,2001ApJ...546...20S}. Dark matter halos are self-gravitating systems where galaxies form. Clustering statistics of halos such as the halo mass function and the halo-matter and halo-halo correlation functions can be accurately modeled down to small scales using $N$-body simulations for a given cosmological model. Based on this motivation, \citet{2018arXiv181109504N} used an ensemble of $N$-body simulations for different cosmologies to build  an emulation package, dubbed as {\tt Dark Emulator}, that enables fast and accurate computations of the halo clustering quantities as a function of halo masses, redshift and separations for an input cosmological model. As shown in \citet{2018arXiv181109504N} \citep[also see][]{Miyatake:2022a,Miyatake:2022b,Kobayashi:2021}, the ``scale-dependent'' halo bias in the halo-matter and halo-halo correlation functions, relative to the matter correlation function, carries useful cosmological information beyond the linear theory.

The purpose of this paper is to use the combined 3$\times$2pt measurements from the photometric HSC-Y3 galaxies and the spectroscopic SDSS galaxies to estimate cosmological parameters while mitigating the impact of the systematic photo-$z$ error and the galaxy bias uncertainty. We carried out a similar analysis with the HSC-Y1 data \citep{Miyatake:2022b}, but there are some important differences. First, while the HSC-Y1 analysis used only measurements of $\dSigma$ and $\wproj$, in this paper we include the cosmic shear measurements, $\xi_{\pm}$, to improve the precision of cosmological parameter inference and the calibration of the residual photo-$z$ error parameter (hereafter $\deltapz$). Second, we employ a completely uninformative flat prior of $\deltapz$, ${\cal U}(-1,1)$, in our baseline analysis method. Much narrower priors have been used in other weak lensing analyses; e.g., the HSC-Y1 $2\times$2pt analysis \citep{Miyatake:2022b} used a Gaussian prior with width $\sigma(\deltapz)=0.1$ and many other weak lensing cosmology analyses use a prior with width ${\cal O}(10^{-2})$ \citep[e.g.][]{DES-Y3,Heymansetal:2021} for the mean redshift of source galaxies. We will show that the statistical power of the HSC-Y3 data enables us to calibrate the $\deltapz$ parameter to a precision of $\sigma(\deltapz)\sim 0.1$. For theoretical templates, we combine {\tt Dark Emulator} and the halo occupation distribution, which gives a phenomenological description of the galaxy-halo connection, to model the $\dSigma$ and $\wproj$ observables down to quasi-nonlinear scales. We will validate our model and method using a synthetic data vector of the clustering observables, taking into account the covariance matrix for the HSC-Y3 and SDSS observables. In this paper, we will pay particular attention to a stringent test of the flat $\Lambda$CDM model, especially whether the HSC-Y3 data exhibits a tension in the $S_8$ constraint with the {\it Planck} result. 

We perform  a {\it blinded} cosmology analysis at the catalog and analysis levels to avoid confirmation bias. We carry out various tests for systematic errors in the measurements and do extensive validation tests of the method and model. During the blinded analysis stage, we determine the analysis setup, including the uninformative uniform prior of $\deltapz$, without access to the values of cosmological parameters, and we agree not to make any changes in our analysis methodology after we unblind. We will explicitly mention any results that were found ``post-unblinding''. This paper is one of a series of the HSC-Y3 cosmology papers: More, Sugiyama et al. \citep{more2023} give detailed descriptions of the measurements used in the 3$\times$2pt analysis, \citet{sugiyama2023} use exactly the same  3$\times$2pt observables as those in this paper to perform a cosmology analysis using a perturbation theory based model, \citet{li2023} show cosmology results using the real-space cosmic shear tomography, and \citet{dalal2023} show cosmology results using the Fourier-space cosmic shear tomography. The two 3$\times$2pt papers (this paper and \citet{sugiyama2023}) use the same blinded shape catalog of the HSC data. \citet{li2023} and \citet{dalal2023} use different blinded catalogs. Thus we use three different blinded catalogs for our cosmology analyses. We compared the cosmological parameters from the 3$\times$2pt analyses and the real- and Fourier-space cosmic shear analyses only after unblinding. We believe that our analysis strategy and method allow us to obtain a robust, convincing result for both the cosmological parameters and the residual photo-$z$ error, without being subject to confirmation bias.

This paper is organized as follows. In Section~\ref{sec:data} we describe the HSC three-year shape catalog and the spectroscopic SDSS galaxy catalog that are used in this paper. In Section~\ref{sec:analysis} we describe our analysis method: the theoretical templates based on the halo model and the likelihood analysis. In Section~\ref{sec:blinding} we describe our blinding strategy for the cosmology analysis. In Section~\ref{sec:results} we show the main results of this paper: our cosmological constraints, the robustness to different systematics, and the degree of tension of our results with the {\it Planck} inferred cosmology. In Section~\ref{sec:discussion} we give a detailed discussion of our cosmology results: the impact of the residual photo-$z$ error and the assembly bias, and the cosmological results when combined with external constraints on $\Omega_{\rm m}$. Finally we give our conclusions in Section~\ref{sec:conclusion}. We give technical details of our method and tests of systematic effects in several Appendices. 

Throughout this paper we use the natural unit $c=1$ for the speed of light. Unless stated otherwise, we quote the central value of a parameter from the mode value of the posterior parameter that has the highest probability in the marginalized 1D posterior distribution in the chain: ${\cal P}(p_{\rm mode})={\rm maximum}$. The justification of the use of mode as the central value is described in \citet{dalal2023}. We quote the 68\% credible interval for the parameter(s) from the highest density interval of parameter(s) satisfying
\begin{align}
\int_{{\bf p}\in {\cal P}>{\cal P}_{68}}\mathrm{d}{\bf p}~ {\cal P}({\bf p})=0.68,
\end{align}
where ${\cal P}({\bf p})$ is the 1D or 2D marginalized posterior distribution. The 95\% credible interval is similarly defined. 

\section{Data}
\label{sec:data}
\subsection{HSC-Y3 Data: Source Galaxies for Galaxy-galaxy Weak Lensin{}g}
\label{sec:HSC-Y3}
HSC is a wide-field imaging camera on the prime focus of the 8.2m Subaru Telescope \citep{2018PASJ...70S...1M,2018PASJ...70S...2K,2018PASJ...70S...3F,2018PASJ...70...66K}. The HSC Subaru Strategic Program (HSC SSP) survey conducted a five-band ($grizy$) wide-area imaging survey \citep{HSCoverview:17} from 2014 to 2021, spending 330 nights. HSC is one of the most powerful instruments for a weak lensing survey because of the combination of its wide field-of-view (1.77~\sqdeg), superb image quality (typically $0.6^{\prime\prime}$ seeing FWHM in $i$ band), and large photon-collecting power. The HSC SSP survey consists of three layers; Wide, Deep, and Ultradeep. Among them the Wide layer is designed for weak lensing cosmology, covering about 1,100~\sqdeg~ of the sky with a $5\sigma$ depth of $i\sim26$ ($2^{\prime\prime}$ aperture for a point source). The $i$-band images are are taken under good seeing conditions, since they are used for galaxy shape measurements in weak lensing analyses.

In this paper, we use the HSC three-year (hereafter HSC-Y3) galaxy shape \cite{Li2021} and photo-$z$ catalogs \citep{Nishizawa_inprep}, constructed from the S19A internal data release (released in September 2019) of data acquired from March 2014 to April 2019. In the following subsections, we describe details of the shape and photo-$z$ catalogs.

\subsubsection{HSC-Y3 galaxy shape catalog}
\label{sec:HSC-Y3_shape}
In this paper, we use the HSC-Y3 shape catalog \citep{Li2021} from the S19A images that were processed with {\tt hscPipe}~v7 \cite{2018PASJ...70S...5B}. In {\tt hscPipe}~v7, there were a number of improvements to the PSF modelling, image warping kernel, background subtraction and bright star masks, which have improved the quality of the shape catalog in HSC-Y3 compared to the  HSC Year~1 shape catalog  \citep{HSCDR1_shear:17,2018MNRAS.481.3170M}. The detailed selection of galaxies that form the shape catalog is presented in \citet{Li2021}. Briefly, the shape catalog consists of galaxies selected from the ``full-depth full-color region'' in all five filters. Apart from some basic quality cuts related to pixel level information, we select extended objects with an extinction corrected cmodel magnitude $i<24.5$, $i$-band SNR$\ge 10$, resolution $>0.3$, $>5\sigma$ detection in at least two bands other than $i$, a 1 arcsec diameter aperture magnitude cut of $i<25.5$, and a blendedness cut in the $i$-band of $10^{-3.8}$.

The shape catalog consists of 35.7 million galaxies spanning an area of about 430~\sqdeg, with an effective number density of 19.9~arcmin$^{-2}$. It is divided into six disjoint regions: XMM, VVDS, GAMA09H, WIDE12H, GAMA15H and HECTOMAP \citep[see Fig.~2 in Ref.][]{Li2021}. The shape measurements in the catalog were calibrated using detailed image simulations, such that the galaxy-property-dependent multiplicative shear bias uncertainty is less than $\sim 10^{-2}$. \citet{Li2021} also presented a number of systematics tests and null tests, and quantify the level of residual systematics in the shape catalog that could affect the cosmological science analyses carried out using the data. \citet{Li2021} flag residual additive biases due to PSF model shape residual correlations and star galaxy shape correlations as systematics requiring special attention and marginalization, so we will also investigate the effect of these systematics on the cosmic shear measurements.

As described in detail in companion papers, \citet{more2023}, \citet{li2023} and \citet{dalal2023}, we find a significant source of $B$-mode systematics in the cosmic shear correlation functions for a $\sim 20$~\sqdeg~patch in the GAMA09H region, and we remove this problematic region from the following analysis. The resultant total area of the HSC data is about 416~\sqdeg.

\subsubsection{Source galaxy catalog for galaxy-galaxy weak lensing}
\label{sec:source_galaxies}

Given the depth of the HSC-Y3 data, we can define a secure sample of source galaxies behind lens galaxies. In this paper we use three samples of lens galaxies as a function of redshift, selected from the Data Release 11 (DR11) of spectroscopic SDSS galaxies up to $z=0.7$, as described below. To select background galaxies behind the SDSS galaxies, we use photo-$z$ estimates of each HSC source galaxy. The HSC-Y3 shape catalog is accompanied by a photo-$z$ catalog of galaxies based on three different methods \citep{2020arXiv200301511N}. \mizuki \cite{Tanaka:2015} is a template fitting based photo-$z$ estimation code. \dempz \cite{Hsieh:2014} and \dnnz \cite{Nishizawa_inprep} on the other hand provide machine-learning-based estimates of the galaxy photo-$z$'s. Each of these methods provides an estimate of the posterior distribution of the redshift for individual galaxies, denoted as $P(z_{\rm s})$. In this paper we employ the \dempz photo-$z$ catalog as our fiducial choice to define a sample of background galaxies by requiring that the posterior that the galaxy has redshift less than 0.75 be less than 1\% \citep{2014MNRAS.444..147O,2018PASJ...70...30M,2019ApJ...875...63M}:
\begin{equation}
\int_{z_{\rm l, max}+0.05}^{7}\!\mathrm{d}z_{\rm s} ~ P_i(z_{\rm s}) 
\ge 0.99 \,,
\label{eq:source-selection}
\end{equation} 
where $z_{\rm l, max}$=0.70 is the maximum redshift of the lens samples. Such cuts significantly reduce the contamination of source galaxies which are physically associated with the lens galaxies and which would dilute the weak lensing signal. The total number of galaxies in our source sample is $\sim 24\%$ of the original HSC-Y3 shape catalog, with an effective number density of $4.9$~galaxies per square arcmin. The mean redshift of the sample, estimated from the stacked photo-$z$ posterior, is $\langle z_{\rm s}\rangle\simeq 1.3$.

Photo-$z$ uncertainties are one of the most important systematic effects in weak lensing cosmology, and could cause significant biases in the cosmological parameters if unknown residual systematic errors in photo-$z$ exist. To minimize the impact of possible systematic photo-$z$ error, we will employ the method in \citet{OguriTakada:11} that enables a self-calibration of such residual photo-$z$ errors, using a {\it single} sample of photometric source galaxies for the weak lensing measurements as we will later describe in detail. 

\subsection{Lens Galaxy Sample}
\label{sec:lens_sample}
We use the large-scale structure sample compiled as part of DR11\footnote{\url{https://www.sdss.org/dr11/}} \cite{Alam:2015} of the SDSS-III Baryon Oscillation Spectroscopic Survey (BOSS) project \citep{2013AJ....145...10D} for measurements of the clustering of galaxies and as lens galaxies for the weak lensing signal measurements. The lens galaxy sample used in this paper is the same as that used in the first year analysis of HSC data (\citet{Miyatake:2022a,Sugiyama:2022}) \citep[also see][]{Miyatakeetal:15}. We use a luminosity-limited catalog of SDSS galaxies in order for it to be approximately volume limited
(see \citet{more2023} for the details). We describe
the resultant catalog here briefly.

The BOSS is a spectroscopic survey of galaxies and quasars selected from the imaging data obtained by the SDSS-I/II and covers an area of approximately 11,000 deg$^2$ \cite{2009ApJS..182..543A} using the dedicated 2.5m SDSS Telescope \cite{2006AJ....131.2332G}. Imaging data obtained in five photometric bands ($ugriz$) as part of the SDSS I/II surveys \cite{1996AJ....111.1748F,2002AJ....123.2121S,2010AJ....139.1628D} were augmented with an additional 3,000 deg$^2$ in SDSS DR9 to cover a larger portion of the sky in the southern region \citep{2011AJ....142...72E,2012ApJS..203...21A,2013AJ....145...10D,2011ApJS..193...29A}. These data were processed by photometric processing pipelines \citep{2001ASPC..238..269L,2003AJ....125.1559P,2008ApJ...674.1217P}, and corrected for Galactic extinction  \citep{1998ApJ...500..525S} to obtain a reliable photometric catalog which is used as an input to select targets for spectroscopy \citep{2013AJ....145...10D}. The BOSS spectra were processed by an automated pipeline to perform redshift determination and spectral classification \cite{2012AJ....144..144B}. The BOSS large-scale structure (LSS) catalog consists of two samples: LOWZ at $0.15<z<0.35$ and CMASS at $0.43<z<0.7$. In addition to the BOSS galaxies we also use galaxies which pass the target selection but had already been observed in the SDSS-I/II project. These galaxies are subsampled in each sector so that they follow the same completeness as that of the LOWZ and CMASS samples in their redshift ranges \citep{2014MNRAS.441...24A}.

We define three redshift subsamples ``LOWZ'' galaxies in the redshift range $z=[0.15,0.35]$ and  the CMASS galaxies divided into redshift bins, $z=[0.43,0.55]$ and $z=[0.55,0.70]$, hereafter called ``CMASS1'' and ``CMASS2'', respectively. As shown in Fig.~1 of \citet{Miyatake:2022b}, we define the subsamples by selecting galaxies with absolute magnitudes $M_i-5\log {h}<-21.5$, $-21.9$ and $-22.2$ for the LOWZ, CMASS1 and CMASS2 subsamples, respectively, to construct nealy volume-limites samples. The comoving number densities for the {\it Planck} cosmology are $\bar{n}_{\rm g}/[10^{-4}\,(h^{-1}{\rm Mpc})^{-3}] \simeq 1.8, 0.74$ and 0.45, respectively, which are a few times smaller than those of the parent LOWZ and CMASS samples. 

\section{Modeling and Analysis method}
\label{sec:analysis}
In this paper, we use three clustering observables to perform the cosmological parameter inference -- the so-called 3$\times$2pt analysis. To be more precise, we use (i) the average excess surface mass density profile, denoted as $\dSigma(R)$, that is measured from the galaxy-galaxy weak lensing combining the photometric HSC source galaxy sample and each of the three spectroscopic SDSS lens subsamples over the overlapping 416~\sqdeg~area of HSC-Y3 and BOSS, (ii) the projected correlation function, denoted as $\wproj(R)$, for each of the spectroscopic SDSS subsamples used as lens samples in the $\dSigma$ analysis measured from the entire BOSS regions of about 8,300~\sqdeg~area, and (iii) the cosmic shear correlation functions, denoted as $\xi_{\pm}(\vartheta)$, for the HSC source sample measured from the HSC-Y3 416~\sqdeg~area. The details of the measurements, null and systematics tests and covariance matrix are described in the  companion paper, \citet{more2023}. In this section we describe our model of these clustering observables within the $\Lambda$CDM framework and our method of Bayesian based parameter inference. 

\subsection{Model}
\label{sec:model}
\subsubsection{Dark Emulator}
\label{sec:dark_emulator}
\begin{table}
\begin{center}
\caption{The set of six cosmological parameters used in our analysis, which specify a model within the flat-geometry $\Lambda$CDM framework. For an input $\Lambda$CDM model, {\tt Dark Emulator} outputs the halo clustering quantities (see text for details). The column labeled  ``parameters'' lists the six cosmological parameters. The column labeled ``supported range'' denotes the range of parameters that is supported by {\tt Dark Emulator}.
\label{tab:cosmological_parameters_supportingrange}}
\begin{tabular}{l|l} \hline\hline
parameters & supported range [min,max]  
\\ \hline
$\Omega_{\rm de}$ &  $[0.54752,0.82128]$\\ 
$\ln (10^{10}A_{\rm s})$ &  $[2.4752,3.7128]$\\ \hline
$\omega_{\rm b}\equiv\Omega_{\rm b}h^2$&  $[0.0211375,0.0233625]$\\
$\omega_{\rm c}\equiv\Omega_{\rm c}h^2$& $[0.10782,0.13178]$\\
$n_{\rm s}$ &  $[0.916275,1.012725]$\\
\hline\hline
\end{tabular}
\end{center}
\end{table}
To model $\dSigma$ and $\wproj$, we use the publicly-available code, {\tt Dark Emulator}\footnote{\url{https://github.com/DarkQuestCosmology/dark_emulator_public}}, developed in \citet{2018arXiv181109504N}. {\tt Dark Emulator} is a software package enabling fast, accurate computations of halo clustering quantities for an input flat $w$CDM cosmological model. {\tt Dark Emulator}  is based on an ensemble set of cosmological $N$-body simulations, each of which was performed with $2048^3$ particles for a box with length $1$ or $2~h^{-1}{\rm Gpc}$ on a side, for 101 flat $w$CDM cosmological models. The $w$CDM cosmology is parametrized by six parameters, ${\bf p}=\{\omega_{\rm b},\omega_{\rm c}, \Omega_{\rm de},\ln(10^{10}A_{\rm s}),n_{\rm s}, w_{\rm de}\}$, where $\omega_{\rm b}(\equiv \Omega_{\rm b}h^2)$ and $\omega_{\rm c}(\equiv \Omega_{\rm c}h^2)$ are the physical density parameters of baryons and CDM, respectively, $h$ is the Hubble parameter, $\Omega_{\rm de}\equiv 1-(\omega_{\rm b}+\omega_{\rm c}+\omega_\nu)/h^2$ is the density parameter of dark energy for a flat-geometry universe, $A_{\rm s}$ and $n_{\rm s}$ are the amplitude and tilt parameters of the primordial curvature power spectrum normalized at $k_{\rm pivot}=0.05~{\rm Mpc}^{-1}$, and $w_{\rm de}$ is the equation of state parameter for dark energy. In the following we focus on flat $\Lambda$CDM cosmological models with $w_{\rm de}=-1$.

For the $N$-body simulations, the effect of finite neutrino mass was included by fixing the neutrino density parameter $\omega_\nu\equiv \Omega_\nu h^2$ to 0.00064. This value corresponds to a total mass of three neutrino species of 0.06~eV, the lower bound of the normal mass hierarchy \citep{2019JHEP...01..106E}. The presence of massive neutrinos affects the linear transfer function, where the total matter fluctuation was computed including massive neutrinos by CAMB \cite{Lewis:2000} and was scaled back to the initial redshift of the simulations using the linear growth factor with the neutrino density included in the matter content. The subsequent nonlinear growth was followed consistently in an $N$-body simulation, including the neutrino density as a part of matter density \citep[see][for details]{2018arXiv181109504N}. Since we focus on the $\sigma_8$ parameter\footnote{$\sigma_8$ is the parameter often used in the literature for the normalization of the linear matter power spectrum, corresponding to the rms linear mass density fluctuations within a top-hat sphere of radius $8\,h^{-1}{\rm Mpc}$.}, i.e., the present-day normalization of the linear matter power spectrum instead of the amplitude of the primordial fluctuations, this approximate treatment has little impact on our primary constraints from the HSC-Y3 and SDSS data. 

The particle mass for the fiducial {\it Planck} cosmology is $m=1.02\times 10^{10}~h^{-1}M_\odot$ for the higher resolution simulations used as the basis for {\tt Dark Emulator}. The emulator uses halos with mass greater than $10^{12}~h^{-1}M_\odot$, corresponding to about 100 simulation particles.

For each $N$-body simulation realization (each redshift output) for a given cosmological model, \citet{2018arXiv181109504N} constructed a catalog of halos using \texttt{Rockstar} \citep{Behroozi:2013}, which identifies halos and subhalos based on clustering of $N$-body particles in position and velocity space. Then they constructed the catalog of central halos at each output. In this step, halo mass is defined using the spherical overdensity with respect to the halo center (defined as the position with the maximum mass density): $M\equiv M_{\rm 200 m}=(4\pi/3)R_{\rm 200m}^3\times (200\bar{\rho}_{\rm m0})$, where $R_{\rm 200m}$ is the spherical halo boundary radius within which the mean mass density is 200 times $\bar{\rho}_{\rm m0}$, where $\bar{\rho}_{\rm m0}$ is the present-day mean matter density. By combining the outputs of $N$-body simulations and the halo catalogs at multiple redshifts in the range $z=[0,1.48]$, they built an emulator, dubbed {\tt Dark Emulator}, which enables fast and accurate computations of the following quantities:
\begin{itemize}
\item $\frac{\mathrm{d}n_{\rm h}}{\mathrm{d}M}(M; z,{\bf p})$: the halo mass function for halos in the mass range $[M,M+\mathrm{d}M]$,
\item $\xi_{\rm hm}(r; M, z, {\bf p})$: the halo-matter cross-correlation function for a sample of halos in the mass range $[M,M+\mathrm{d}M]$, and
\item $\xi_{\rm hh}(r; M, M', z,{\bf p})$: the halo-halo auto-correlation function for two samples of halos with masses $[M,M+\mathrm{d}M]$ 
and $[M',M'+\mathrm{d}M']$, respectively, 
\end{itemize}
for an input set of parameters, halo mass $M$ (and $M'$ for the cross-correlation function between two halo samples), redshift $z$, and cosmological parameters ${\bf p}$. 

Fig.~2 of \citet{more2023} showed that the LOWZ, CMASS1 and CMASS2 galaxies in our samples likely reside on host halos with typical masses greater than $10^{13}h^{-1}M_\odot$, assuming a concordance flat $\Lambda$CDM model consistent with the CMB and current large-scale structure data. Hence {\tt Dark Emulator} can be safely used to compute the model predictions of $\dSigma$ and $\wproj$ for these SDSS galaxies.

In addition, {\tt Dark Emulator} outputs ancillary quantities, such as the linear halo bias (the large-scale limit of the halo bias), the Tinker model of the linear halo bias \citep{2010ApJ...724..878T} (see below), the linear matter power spectrum, the linear rms mass fluctuations of halo mass scale $M$ ($\sigma^L_{\rm m}(M)$), and $\sigma_8$. 

The supported range of each cosmological parameter for {\tt Dark Emulator} is given in Table~\ref{tab:cosmological_parameters_supportingrange}. These ranges are sufficiently broad that they cover the range of cosmological constraints from current state-of-the-art large-scale structure probes such as the Subaru HSC cosmic shear results \cite{2019PASJ...71...43H,2020PASJ...72...16H}. Since $\sigma_8$ and $\Omega_{\rm m}$ are the primary parameters to which large-scale structure probes are sensitive, we also quote the supported ranges of these \textit{derived} parameters: $0.55\lesssim \sigma_8\lesssim1.2$ and $0.17\lesssim \Omega_{\rm m}\lesssim0.45$, as shown in Fig.~2 of \citet{2018arXiv181109504N}. In this paper we use {\tt Dark Emulator} to perform cosmological parameter inference in a multi-dimensional parameter space by comparing the model templates of $\dSigma$ and $\wproj$ with the signals measured from the SDSS and HSC-Y3 data.

The Bayesian parameter inference method we use might
occasionally sample models that are outside the supported range of $\Lambda$CDM 
models in {\tt Dark Emulator}. In this case, we make the following simple extrapolation of the model predictions: 
\begin{align}
\xi_{\rm hm}(r; {\bf p}_{\notin})&\rightarrow 
\frac{b^{\rm Tinker}({\bf p}_{\notin})}{b^{\rm Tinker}({\bf p}_{\rm edge})}\frac{\xi^L_{\rm mm}(r;{\bf p}_{\notin})}{\xi^L_{\rm mm}(r;{\bf p}_{\rm edge})}
\xi^{\rm DE}_{\rm hm}(r; {\bf p}_{\rm edge}),\nonumber\\
\xi_{\rm hh}(r; {\bf p}_{\notin})&\rightarrow \left(
\frac{b^{\rm Tinker}({\bf p}_{\notin})}{b^{\rm Tinker}({\bf p}_{\rm edge})}\right)^2\nonumber\\
&\hspace{5em}\times
\frac{\xi^L_{\rm mm}(r;{\bf p}_{\notin})}{\xi^L_{\rm mm}(r;{\bf p}_{\rm edge})}
\xi^{\rm DE}_{\rm hh}(r; {\bf p}_{\rm edge}),
\end{align}
where ${\bf p}_{\notin}$ is a set of six cosmological parameters at a $\Lambda$CDM model that is outside the supported range (Table~\ref{tab:cosmological_parameters_supportingrange}), ${\bf p}_{\rm edge}$ is a set of parameters at the edge $\Lambda$CDM model inside the supported range, $b^{\rm Tinker}(\bf{p}_{\notin})$ and $b^{\rm Tinker}({\bf p}_{\rm edge})$ are the linear bias parameters at the two $\Lambda$CDM models with ${\bf p}_{\notin}$ and ${\bf p}_{\rm edge}$ that are computed based on the fitting formula of \citet{2010ApJ...724..878T}, $\xi^L_{\rm mm}$ is the linear-theory prediction for the matter two-point correlation function at the respective model, and $\xi_{\rm hh}^{\rm DE}$ and $\xi^{\rm DE}_{\rm hm}$ are the {\tt Dark Emulator} outputs at the edge model. Here we use an emulator built on {\tt CLASS} \citep{Lesgourgues:2011, Blas2011} to compute the linear-theory matter correlation, $\xi^{L}_{\rm mm}(r)$, for models outside the supported range \citep[see Appendix~A in Ref.][for details]{2018arXiv181109504N}. We define ${\bf p}_{\rm edge}$ by replacing only the parameter(s) outside the supported range with their value(s) at the edge of the supported range, while keeping the other parameter(s) at their input value(s). In the above extrapolation, we simply assume that the halo-matter cross-correlation and the halo auto-correlation follow the linear theory predictions ($\xi_{\rm hm}\simeq b\xi_{\rm mm}$ and $\xi_{\rm hh}\simeq b^2\xi_{\rm mm}$), and that the ratio of $\xi_{\rm hh}^{\rm L}({\bf p}_{\notin})$ and $\xi_{\rm hh}({\bf p}_{\notin})$ can be accurately captured by a similar ratio between $\xi_{\rm hh}^{\rm L}({\bf p}_{\rm edge})$ and $\xi_{\rm hh}({\bf p}_{\rm edge})$. For the extrapolation we can adopt any input value for $A_{\rm s}$, but need to adopt values in the  specific ranges for $\omega_{\rm c}$ and $\Omega_{\rm de}$, as we will explain in Table~\ref{tab:parameters} in Section~\ref{sec:parameter_estimaton_method}.

Our code outputs the model predictions regardless of whether the cosmological parameters are inside or outside the supported range. This treatment is important, because we perform a blinded cosmological analysis of the HSC and SDSS data. If {\tt Dark Emulator} were to return an error message indicating that an outside model has been sampled, we could unintentionally and prematurely unblind our analysis.

After unblinding our cosmology analysis, we confirmed that all models within the 95\% credible interval of $S_8$ in the chains for our baseline analysis are within the emulator supported range for $\ln (10^{10}A_{\rm s})$ and $\Omega_{\rm de}$\footnote{This means that models outside of the 95\% credible interval of $S_8$ are occasionally computed with the extrapolation.}, the most important parameters that are sensitive to $S_8$\footnote{For this discussion, we used the chains for the models that have $\omega_{\rm c}$ within a $\pm5\sigma$ range of the {\it Planck} constraint, because $\omega_{\rm c}$ is not well-constrained by the observables used in this paper.}. 

\subsubsection{Galaxy-galaxy weak lensing: $\dSigma(R)$}
\label{sec:dsigma_theory}
The details of the galaxy-galaxy weak lensing measurements are presented in \citet{more2023}, but in this section we briefly review the measurement method that we will later use to introduce a residual photo-$z$ error parameter. In particular, we will give the concept of our method to calibrate a residual systematic error in the mean source redshift that affects the weak lensing observables. 

Cross-correlating the positions of {\it spectroscopic} SDSS galaxies with shapes of background {\it photometric} HSC galaxies enables us to probe the average mass distribution around the lens SDSS galaxies -- galaxy-galaxy weak lensing \citep{Mandelbaumetal:05}. Throughout this paper we use the average excess surface mass density profile, $\dSigma(R)$, as the  galaxy-galaxy weak lensing observable, where $\dSigma$ has units
of $[hM_\odot~{\rm pc}^{-2}]$ and is given as a function of the projected comoving separation $R$ with units of $[\mpch]$. An estimator of $\dSigma(R_i)$ for the $i$-th radial bin $R_i$ is given \citep[e.g. see Ref.][]{Miyatakeetal:15}, roughly 
by the following form: 
\begin{align}
\widehat{\dSigma}(R_i)\simeq 
\left.\frac{1}{2{\cal R}\sum_{{\rm ls}}w_{\rm ls}}\sum_{{\rm ls}\in R_i}
w_{\rm ls}\avrg{\Sigma_{\rm cr}^{-1}}_{\rm ls}^{-1}\epsilon_{t,{\rm ls}}\right|_{R_i=\chi(z_{\rm l})\Delta\theta_{\rm ls}},
\label{eq:lens_measurement}
\end{align}
where the summation ``${\rm ls}$'' runs over all lens-source pairs that lie in the $i$-th radial bin $R_i\equiv \chi(z_{\rm l})\Delta \theta_{\rm ls}$, $\chi(z_{\rm l})$ is the comoving angular diameter distance to the l-th SDSS lens galaxy at the spectroscopic redshift $z_{\rm l}$, and $\Delta\theta_{\rm ls}$ is the angular separation between the lens and source in each pair; $\epsilon_{t, {\rm ls}}$ is the tangential component of ellipticity of the s-th HSC source galaxy\footnote{Here we denote the dependence of each lens-source pair, i.e. ``${\rm ls}$'' in the subscript, because the tangential shear component of the s-th HSC source galaxy shape is defined with respect to the line connecting the source and lens galaxies on the sky.}; ${\cal R}$ is the shear responsivity \cite{BernsteinJarvis:02,Mandelbaumetal:05} which accounts for conversion of ``distortion'' ($[a^2-b^2]/([a^2+b^2]$) to ``shear'' ($[a-b]/[a+b]$); $w_{\rm ls}$ is the weight, for which we employ an inverse variance weighting that is nearly optimal in the shape-noise dominated regime, following \citep{Mandelbaumetal:13} \citep[also see][]{Shirasakietal:17}. Additionally, we need to subtract the lensing signal around random points, correct for the additive and multiplicative shear calibration factors \citep{Li2021}, and correct for the multiplicative and additive selection bias. Details of the estimator which we used for actual measurements can be found in \citet{more2023}.

The measured weak lensing signal in Eq.~(\ref{eq:lens_measurement}) depends on the {\it true} redshift distribution of source galaxies. Hence, to obtain an {\it unbiased} estimate of $\dSigma$ for the lens sample, we need the average surface mass density to convert shear to $\dSigma$, in the {\it ensemble average sense}: 
\begin{align}
\left\langle \Sigma_{\rm cr}^{-1} \right\rangle_{\rm ls}
&=\frac{\int_0^\infty\!\mathrm{d}z_{\rm s} p_{\rm s}(z_{\rm s})\Sigma_{\rm cr}^{-1}(z_l, z_{\rm s}) }{\int_0^\infty\!\mathrm{d}z_{\rm s}~  p_{\rm s}(z_{\rm s})} \nonumber\\
&\hspace{-2em}= 4\pi G (1+z_{\rm l})\chi(z_{\rm l})\left[1-\chi(z_{\rm l})\left\langle\frac{1}{\chi(z_{\rm s})}\right\rangle_{p_{\rm s}(z_{\rm s})}\right]\, ,
\label{eq:ave_Sigmacr_def}
\end{align}
for a flat-geometry universe, where $p_{\rm s}(z_{\rm s})$ is the {\it true} redshift distribution of source galaxies in the sample. The factor $(1+z_{\rm l})$ arises from our use of comoving coordinates in the projected separation and we set $\Sigma_{\rm cr}^{-1}=0$ when $z_{\rm s}<z_{\rm l}$ in Eq.~(\ref{eq:ave_Sigmacr_def}). In the second equality on the r.h.s. of Eq.~\ref{eq:ave_Sigmacr_def}), we explicitly show that the dependence of source galaxy redshifts enters only into the average of the inverse of the comoving angular diameter distances to source galaxies over the true redshift distribution of source galaxies: $\avrg{1/\chi(z_{\rm s})}_{p_{\rm s}(z_{\rm s})}$. On the other hand, there is no uncertainty in the dependence of  lens redshifts on an evaluation of $\avrg{\Sigma^{-1}_{\rm cr}}$, because we use the spectroscopic galaxy subsamples (LOWZ,CMASS1 and CMASS2) as the lens sample. Hence we stress  that, as long as a correct value of $\avrg{1/\chi(z_{\rm s})}_{p_{\rm s}(z_{\rm s})}$ is evaluated, the shape of the redshift distribution of source galaxies, such as a high-redshift tail or an outlier redshift population of source galaxies, does not cause a bias in $\dSigma$.

However, the true redshift of individual sources is not available, and we have to use photo-$z$ estimates. To estimate $\widehat{\dSigma}(R)$ in Eq.~(\ref{eq:lens_measurement}), we use, in our baseline method, the posterior distribution of photo-$z$'s for source galaxies to compute $\langle \Sigma_{\rm cr}^{-1} \rangle$, where the photo-$z$ distribution is generally different from the true redshift distribution $p_{\rm s}(z_{\rm s})$, even in the average sense. We will later introduce a nuisance parameter to model the effect of residual systematic error in the mean source redshift, or equivalently a residual error in the estimate of $\langle \Sigma_{\rm cr}^{-1} \rangle$. The nuisance parameter can be calibrated from the relative amplitudes between $\dSigma$ signals for the three lens subsamples and the cosmic shear signals, because the average $\avrg{\Sigma_{\rm cr}^{-1}}$ has characteristic dependences on the lens redshifts, as proposed in \citet{OguriTakada:11}. More exactly speaking, we will implement the self-calibration method along with the estimator used in the measurement, properly taking into  account the weight ($w_{\rm ls}$) for each lens-source pair (see around Eq.~15 in \citet{more2023}), as we will explain below in detail. 

As seen in Eq.~(\ref{eq:lens_measurement}), the estimation of $\dSigma(R)$ involves conversion of the observed angular separation between source and lens, $\Delta\theta$, to the comoving separation $R$ and the multiplicative factor of $\langle\Sigma_{\rm cr}^{-1}\rangle_{\rm ls}$. To do this, we need to assume a ``reference'' cosmology, which generally differs from the underlying true cosmology. In Section~\ref{subsubsec:meascorr} we will describe how to include the effect of varying cosmological models on parameter inference. 

Given an unbiased estimate of $\dSigma$ for a lens sample, we need the theoretical template in cosmology inference. We employ the following two-component model for $\dSigma(R)$: 
\begin{align}
\dSigma(R;z_{\rm l})= \dSigma_{\rm gG}(R;z_{\rm l})+\dSigma_{\rm mag}(R;z_{\rm l}). 
\label{eq:dSigma_two_contributions}
\end{align}
The first term on the right-hand side is the standard contribution to the galaxy-galaxy weak lensing signal, which we refer to as the cross-correlation of the lens galaxies (``g'') and gravitational-lens (``G'') inferred mass in the large scale structure containing the lens sample. Note that $\dSigma_{\rm gG}$ is the {\it standard} excess surface mass density profile of lens galaxies, used in galaxy-galaxy weak lensing. The second term is the contribution caused by the lensing magnification effect, which arises from correlations between shapes of source galaxies and the mass distribution in the foreground structures of lens galaxies along the line-of-sight to source galaxies due to the fact that lens galaxies can preferentially reside in overdensity regions \citep{2020A&A...638A..96U}. Below we describe our models for each contribution within the $\Lambda$CDM model framework. Throughout this paper, we model the clustering observables of each SDSS galaxy sample using the theoretical model prediction at a representative redshift, denoted as $z_{\rm l}$: ${z}_{\rm l}\simeq 0.26, 0.51$ and $0.63$ for the LOWZ, CMASS1 and CMASS2 samples, respectively. That is, we ignore the possible redshift evolution of the clustering observables within each redshift bin for simplicity. In \citet{more2023}, we confirm that this is a reasonable approximation by looking at variations in the measured clustering and lensing signals within each redshift bin for each of the three subsamples.

The excess surface mass density profile $\dSigma$ for a given sample of lens galaxies is expressed as \citep[e.g.][]{Mandelbaumetal:13,2013MNRAS.435.2345H}:
\begin{align}
\dSigma_{\rm gG}(R;z_{\rm l}) &=\bar{\rho}_{\rm m0}\int\!\!\frac{k\mathrm{d}k}{2\pi}~P_{\rm gm}(k;z_{\rm l})J_2(kR),
\label{eq:dSigma_def}
\end{align}
where $J_2(x)$ is the second-order Bessel function and $P_{\rm gm}(k;z_{\rm l})$ is the cross-power spectrum between galaxies and matter at redshift $z_{\rm l}$. Hereafter we omit $z_{\rm l}$ in the argument for notational simplicity.

As described above, {\tt Dark Emulator} outputs halo clustering properties for an input cosmology. To obtain the model predictions for the observable quantities for SDSS galaxies, we need a model for the galaxy-halo connection. For this, we use the halo occupation distribution \cite[HOD][]{1998ApJ...494....1J,Zhengetal:05} \citep[also see][]{Miyatake:2022a,Miyatake:2022b}. Under this setup we can compute $P_{\rm gm}(k)$ for a given model:
\begin{align}
P_{\rm gm}(k)&=\frac{1}{\bar{n}_{\rm g}}
\int\!\mathrm{d}M\frac{\mathrm{d}n_{\rm h}}{\mathrm{d}M}
\avrg{N_{\rm c}}\!(M)\left[1+\lambda_{\rm s}\!(M)\tilde{u}_{\rm s}(k;M,z)\right]\nonumber\\
&\hspace{5em}\times P_{\rm hm}(k; M)\, ,
\label{eq:Pgm_def}
\end{align}
where the mean number density of galaxies is given by
\begin{align}
\bar{n}_{\rm g}=\int\!\mathrm{d}M~ 
\frac{\mathrm{d}n_{\rm h}}{\mathrm{d}M}
\avrg{N_{\rm c}}\!(M)\left[1+\lambda_{\rm s}\!(M)\right],
\label{eq:ng}
\end{align}
$\avrg{N_{\rm c}}\!(M)$ is the HOD of central galaxies, $\avrg{N_{\rm c}}\!(M)\lambda_{\rm s}\!(M)$ is the HOD of satellite galaxies, and $\tilde{u}_{\rm s}(k;M)$ is the Fourier transform of the average radial profile of satellite galaxies in a host halo with mass $M$. All the quantities are evaluated at a representative redshift $z_{\rm l}$ of the lens galaxies in the LOWZ, CMASS1 or CMASS2 subsample. The impact of using representative redshifts instead of integrating over the lens redshift range is less than $\sim6$\% of the square root of the diagonal element of covariance, according to the discussion in \citet{Sugiyama:2022}. Here we use {\tt Dark Emulator}  to compute the halo mass function $\mathrm{d}n_{\rm h}/\mathrm{d}M$ and the halo-matter cross power spectrum, $P_{\rm hm}(k; M)$, for an input cosmological model, where $P_{\rm hm}(k;M)$ is obtained from the Fourier transform of the {\tt Dark Emulator} output, $\xi_{\rm hm}(r;M)$.

We employ the following models for the central and 
satellite HODs in our baseline analysis: 
\begin{align}
\avrg{N_{\rm c}}(M)&=\frac{1}{2}\left[1+{\rm erf}\left(\frac{\log M-\log M_{\rm min}}{\sigma_{\log M}}\right)
\right],\nonumber\\
\avrg{N_{\rm s}}(M)&\equiv \avrg{N_{\rm c}}(M)\lambda_{\rm s}(M)\nonumber\\
&= \avrg{N_{\rm c}}(M)\left(\frac{M-\kappa M_{\rm min}}{M_1}\right)^{\alpha}\, ,
\end{align}
where ${\rm erf}(x)$ is the error function. For our fiducial prescription, we assume that satellite galaxies reside only in a halo that already hosts a central galaxy. Our fiducial HOD model is specified by the five parameters $\{M_{\rm min},\sigma_{\log M},\kappa, M_1,\alpha\}$. 

For $\tilde{u}_{\rm s}(k;M)$ in Eq.~(\ref{eq:Pgm_def}), 
throughout this paper, we assume that satellite galaxies follow a Navarro-Frenk-White (NFW) profile \citep{Navarroetal:97}. To compute the NFW profile as a function of halo mass and redshift for a given cosmological model, we use the halo mass-concentration relation computed using the publicly-available code {\tt Colossus} \footnote{\url{http://www.benediktdiemer.com/code/colossus/}} \citep{2018ApJS..239...35D}. 

For an extended cosmological analysis, we include parameters to model the effect of off-centered ``central'' galaxies or the ``incompleteness'' of central galaxies \citep{2012MNRAS.419.3457H,2013MNRAS.435.2345H}, where the incompleteness effect models the possibility that some massive halos might not host a central galaxy in the sample due to color and magnitude cuts. We use the model in ~\citet{Miyatake:2022a} to model the effects. 

We model the second term in Eq.~(\ref{eq:dSigma_two_contributions}), following the method in  Ref.~\citep{2020A&A...638A..96U} \citep[also see Eq.~4 in Ref.][]{Sugiyama:2022}, as
\begin{align}
\dSigma_{\rm mag}(R)&=2(\alpha_{\rm mag}-1)\int_0^{z_{\rm max}}\!\mathrm{d}z_{\rm l}
p_{\rm l}(z_{\rm l})\int_0^{z_{\rm max}}\!\mathrm{d}z_{\rm s}p_{\rm s}(z_{\rm s})\nonumber\\
&\hspace{2em}\times 
\int\!\frac{\ell\mathrm{d}\ell}{2\pi}
\Sigma_{\rm cr}(z_{\rm l},z_{\rm s})C_\kappa(\ell;z_{\rm l},z_{\rm s})J_2\!\!\left(
\frac{\ell R}{\chi}
\right)\, ,
\label{eq:dSigma_mag_def}
\end{align}
where $C_\kappa(\ell)$ is the cosmic shear convergence power spectrum for source galaxies at redshifts $z_{\rm l}$ and $z_{\rm s}$, defined as 
\begin{align}
C_\kappa(\ell; z_{\rm l}, z_{\rm s})
&\equiv \int_0^{\chi_H}\!\!\mathrm{d}\chi~\frac{W(\chi,\chi_{\rm l})W(\chi,\chi_{\rm s})}{\chi^2}\nonumber\\
&\hspace{2em}\times 
P^{\rm NL}_{\rm mm}\!\left(
k=\frac{\ell+1/2}{\chi}; z\right),
\label{eq:ckapppa_zl_zs}
\end{align}
with the lensing efficiency function, $W(\chi,\chi_{\rm s})$, for lens and source at distances $\chi$ and $\chi_{\rm s}$: 
\begin{align}
W(\chi,\chi_{\rm s})&\equiv 
\frac{3\Omega_{\rm m}}{2}H_0^2(1+z)\chi\left(1-\frac{\chi}{\chi_{\rm s}}\right)\, .
\label{eq:lensefficiency_w}
\end{align}
Here we used the relation between redshift and comoving distance, via relations $\chi=\chi(z)$, for a given cosmological model; $p_{\rm l}(z_{\rm l})$ in Eq.~(\ref{eq:dSigma_mag_def}) denotes the redshift distribution of lens galaxies (LOWZ, CMASS1 or CMASS2), normalized as $\int_0^{z_{\rm max}}\!\mathrm{d}z_{\rm l}~p_{\rm l}(z_{\rm l})=1$; $\alpha_{\rm mag}$ is the power-law slope of number counts of the lens galaxies around a magnitude cut in each sample \citep[see Eq.~10 and Fig.~2 in Ref.][for the estimated value and error]{Miyatake:2022b}; $P^{\rm NL}_{\rm mm}(k)$ is the nonlinear matter power spectrum for which we use {\tt halofit} \citep{Takahashi_2012}  for a given cosmological model. Note that $\dSigma_{\rm mag}$ does not depend on the models for galaxy bias or galaxy-halo connection. In Eq.~(\ref{eq:dSigma_mag_def}) we take into account the redshift distribution of both the lens (SDSS) and source (HSC) galaxies, which is different from our treatment in the HSC-Y1 cosmology analyses \citep{Sugiyama:2022,Miyatake:2022b}. As shown in \citet{Miyatake:2022b}, $\dSigma_{\rm mag}$ leads to about 1\%, 7\% and 10\% contributions to the total power of $\dSigma$ for the LOWZ, CMASS1 and CMASS2 subsamples, respectively, for the {\it Planck} cosmology \cite{PlanckCosmology:16}. Including the $\dSigma_{\rm mag}$ contribution in the theoretical template adds some cosmological information. In our analysis we treat the magnitude slope $\alpha_{\rm mag}$ as a nuisance parameter, with a Gaussian prior with width $\sigma(\alpha_{\rm mag})=0.5$ around the central value taken from the measurement value (see Fig.~2 of Ref.~\cite{Miyatake:2022b}). Note that  $\alpha_{\rm mag}$ is different from $\alpha$, which is a parameter of the satellite HOD. 

\subsubsection{Projected auto-correlation function: $\wproj(R)$}
\label{sec:wproj_theory}

As a second clustering observable of the LOWZ, CMASS1 and CMASS2 galaxy subsamples used in the galaxy-galaxy weak lensing measurements, we use the projected spatial correlation function, denoted as $\wproj(R)$. We model $\wproj(R)$ as
\begin{align}
\wproj(R;z_{\rm l})\equiv 2f^{\rm RSD}_{\rm corr}\!(R;z_{\rm l})\int^{\Pi_{\rm max}}_{0}\!\mathrm{d}\Pi~ \xi_{\rm gg}\!\left(\sqrt{R^2+\Pi^2};z_{\rm l}\right),
\label{eq:wproj_def}
\end{align} 
where we take $\Pi_{\rm max}=100\,h^{-1}{\rm Mpc}$ as our fiducial choice and $\xi_{\rm gg}(r)$ is the real-space, three-dimensional correlation function of galaxies. To compute the radial and projected separations, $\Pi$ and $R$, between galaxies in each pair from their observed redshifts and angular positions, we assume the reference cosmological model as done in our $\dSigma$ analysis above; the flat-geometry model with  $\Omega_{\rm m}^{\rm ref}=0.279$. The prefactor $f^{\rm RSD}_{\rm corr}(R)$ is a correction factor that accounts for the effect of redshift-space distortion (RSD); we assume the linear Kaiser RSD \cite{Kaiser:1984} to compute $f^{\rm RSD}_{\rm corr}$ following the method in \citet{vandenBosch:2013} (see Eq.~48 in the paper) \citep[also see Ref.][]{Miyatake:2022a}.

To use Eq.~(\ref{eq:wproj_def}), we must first compute the three-dimensional correlation function of galaxies for a given set of model parameters. The three-dimensional correlation function $\xi_{\rm gg}$ is given as
\begin{align}
\xi_{\rm gg}(r;z_{\rm l})=\int_0^\infty\!\frac{k^2\mathrm{d}k}{2\pi^2}~P_{\rm gg}(k;z_{\rm l})j_0(kr),
\label{eq:xi_gg}
\end{align}
where $j_0(x)$ is the zero-th order spherical Bessel function, and $P_{\rm gg}(k)$ is the auto-power spectrum of galaxies. Once the power spectrum $P_{\rm gg}(k)$ is given for an input of model parameters, we can compute the model prediction of $\wproj(R)$ according to Eq.~(\ref{eq:wproj_def}). 

In the halo model, $P_{\rm gg}$ can be divided into two contributions, i.e., the 1- and 2-halo terms, as
\begin{align}
P_{\rm gg}(k)=P_{\rm gg}^{\rm 1h}(k)+P_{\rm gg}^{\rm 2h}(k),
\label{eq:pgg_def}
\end{align}
where the 1-halo term describes correlations between galaxies within the same host halo, and the 2-halo term describes correlations between galaxies residing in different halos. In our method, we compute the auto-power spectrum as
\begin{widetext}
\begin{align}
P^{\rm 1h}_{\rm gg}(k)&= \frac{1}{\bar{n}_{\rm g}^2}
\int\!\mathrm{d}M~\frac{\mathrm{d}n_{\rm h}}{\mathrm{d}M}\avrg{N_{\rm c}}\!(M)
\left[
2\lambda_{\rm s}(M)\tilde{u}_{\rm s}(k;M)+\lambda_{\rm s}(M)^2\tilde{u}_{\rm s}(k;M)^2\right], \nonumber\\
P^{\rm 2h}_{\rm gg}(k)&= \frac{1}{\bar{n}_{\rm g}^2}
\left[\int\!\mathrm{d}M~\frac{\mathrm{d}n_{\rm h}}{\mathrm{d}M}
\avrg{N_{\rm c}}\!(M)\left\{1+\lambda_{\rm s}(M)\tilde{u}_{\rm s}(k;M)
\right\}
\right]\nonumber\\
&\hspace{10em}\times
\left[\int\!\mathrm{d}M'~\frac{\mathrm{d}n_{\rm h}}{\mathrm{d}M'}
\avrg{N_{\rm c}}\!(M')\left\{1+\lambda_{\rm s}(M')\tilde{u}_{\rm s}(k;M')
\right\}
\right]P_{\rm hh}(k;M,M').
\label{eq:Pgg_1h2h}
\end{align}
\end{widetext}
Here we use {\tt Dark Emulator} to compute $\mathrm{d}n_{\rm h}/\mathrm{d}M$ and $P_{\rm hh}(k;M,M')$, the power spectrum between halos with masses $M$ and $M'$ for an input cosmological model. Note that in our fiducial model we assume that satellite galaxies reside in halos that host a central galaxy in our sample. \citet{Miyatake:2022a} confirmed that fitting the model to mock observables computed for the case that satellite galaxies are populated in halos irrespective of whether the halos host central galaxies in the sample resulted in a negligible shift in $S_8$, for our baseline analysis setup.

\subsubsection{Cosmic shear correlation functions: $\xi_{\pm}(\vartheta)$}
\label{sec:cosmic shear}
As the third clustering observable in our cosmology analysis, we use the measured two-point correlation functions of galaxy shapes in the HSC source sample used in the $\dSigma$ measurement, denoted as $\xi_{\pm}(\vartheta)$. We model $\xi_{\pm}(\vartheta)$ as a sum of the following three contributions, taking into account contamination of intrinsic alignments (IA): 
\begin{align}
\xi_{\pm}(\vartheta)=\xi_{{\rm GG},\pm}(\vartheta)+\xi_{{\rm GI},\pm}(\vartheta)+\xi_{{\rm II},\pm}(\vartheta)\, .
\label{eq:sourceshape_2pt}
\end{align}
The ``$+$'' and ``$-$'' correlation functions are measured from different combinations of the correlations of the two ellipticity components of source galaxy shapes in each pair, $\xi_{\pm}\leftarrow \langle \epsilon_+\epsilon_+\pm \epsilon_\times\epsilon_\times\rangle$, where $\epsilon_+$ is the ellipticity component along the R.A. or Dec coordinate direction, and $\epsilon_\times$ is its 45~degree rotated component. The first term is the ``gravitational-gravitational'' term (i.e. cosmic shear, ``GG''), the third term is the ``intrinsic-intrinsic'' (``II'') IA contribution \citep{Heavensetal:00,2000ApJ...545..561C,2000ApJ...532L...5L,2001MNRAS.320L...7C}, and the second term is the ``gravitational-intrinsic'' correlation (``GI'') \citep{Hirata:2004} that arises in pairs of galaxies for which common large-scale structure in the line of sight affects the intrinsic shapes of one of the galaxies and the gravitational lensing shear on the other.

The GG term in Eq.~(\ref{eq:sourceshape_2pt}) is given in terms of the cosmic shear convergence power spectrum, $C_\kappa(\ell)$, as
\begin{align}
\xi_{{\rm GG},\pm}(\vartheta)\equiv \int\!\frac{\ell \mathrm{d}\ell}{2\pi}~C_\kappa(\ell)J_{0,4}(\ell\vartheta),
\label{eq:cosmic_shear_hankel}
\end{align}
where $J_{0,4}(x)$ is the zero-th order (for $\xi_+$) or fourth-th order (for $\xi_-$) Bessel function. Using the flat-sky approximation and Limber's approximation \citep{Limber:54}, $C_\kappa(\ell)$ is computed from the line-of-sight integral of the nonlinear matter power spectrum as
\begin{align}
C_\kappa(\ell)=\int_0^{\chi_H}\!\mathrm{d}\chi~\frac{q(\chi)^2}{\chi}P^{\rm NL}_{\rm mm}\!\!\left(
k=\frac{\ell+1/2}{\chi},z\right),
\label{eq:P_kappa}
\end{align}
where $\chi_H$ is the comoving horizon radius, and $z$ is given by the inverse of $\chi=\chi(z)$. To model $P^{\rm NL}_{\rm mm}$ for a given cosmological model, we employ {\tt halofit} \citep{Takahashi_2012} in the same way as used in $\dSigma_{\rm mag}$ in Eq.~(\ref{eq:dSigma_two_contributions}). The lensing efficiency function $q(\chi)$ \citep[also see][]{TakadaJain:04} is defined as
\begin{align}
q(\chi)\equiv \int_{z=z(\chi)}^{z_{\rm max}}\!\mathrm{d}z_{\rm s}~ p_{\rm s}(z_{\rm s})W(\chi,\chi_{\rm s}).
\end{align}
where $W(\chi,\chi_{\rm s})$ is defined by Eq.~(\ref{eq:lensefficiency_w}). We note that we use the same redshift distribution of source galaxies, $p_{\rm s}(z_{\rm s})$, as used in the $\dSigma$ measurement. Adding the cosmic shear information in parameter inference further helps the self-calibration of the residual photo-$z$ errors of the HSC source galaxies, as we will show later. 

To model the IA correlation functions, in this paper we adopt the NLA model \citep{2007NJPh....9..444B} in our baseline model. In this model, the II and GI correlation functions are given by
\begin{align}
\xi_{{\rm II/GI},\pm}(\vartheta)&=
\int\!\!\frac{\ell\mathrm{d}\ell}{2\pi}~ C_{{\rm II/GI}}(\ell)J_{0,4}(\ell\vartheta)\, ,
\label{eq:xi_ii_gi_def}
\end{align}
with
\begin{align}
C_{{\rm II}}(\ell)&=\int^{\chi_H}\!\!\mathrm{d}\chi~ F^2(\chi)\frac{p_{\rm s}(\chi)p_{\rm s}(\chi)}{\chi^2}
P^{\rm NL}_{\rm mm}\!\left(k=\frac{\ell+1/2}{\chi},z\right), \nonumber\\
C_{{\rm GI}}(\ell)&= 
2\int^{\chi_H}_0\!\!\mathrm{d}\chi~F(\chi)\frac{q(\chi)p_{\rm s}(\chi)}{\chi^2}
P^{\rm NL}_{\rm mm}\!\left(k=\frac{\ell+1/2}{\chi},z\right).
\label{eq:IA_ps}
\end{align}
Here the redshift- and cosmology-dependent factor, $F(\chi)$, relating the galaxy ellipticity and the gravitational tidal field is parametrized as
\begin{align}
F(z)=-A_{\rm IA}C_1\rho_{\rm cr,0}\frac{\Omega_{\rm m}}{D(z)}
\left(\frac{1+z}{1+z_0}\right)^{\eta_{\rm IA}},
\end{align}
where $A_{\rm IA}$ is a dimensionless amplitude parameter, $\rho_{\rm cr,0}$ is the critical density of the Universe at $z=0$, and $D(z)$ is the linear growth factor normalized to unity at $z=0$. The additional redshift dependence is assumed to have a power-law form, with power-law index parameter $\eta_{\rm IA}$. We use a single parameter model of IA by fixing $\eta_{\rm IA}=0$ as our cosmic shear data is only for a single redshift bin, that is, does not contain the redshift information of the IA effect. The normalization constant factor $C_1$ is set to $5\times 10^{-14}h^{-2}M_\odot^{-1}{\rm Mpc}^3$ at $z_0=0.62$ by convention, which is motivated by the observed ellipticity variance in SuperCOSMOS \citep{2002MNRAS.333..501B}. This model has previously been used in cosmic shear cosmology analyses
\citep{2019PASJ...71...43H,2017MNRAS.465.1454H}.
While this is merely a phenomenological model of the IA effect, our cosmological constraints are from the joint information of $\wproj$, $\dSigma$ and $\xi_\pm$, we expect its effect is small. Indeed, we will find that the cosmological constraints are changed very little even if we ignore the IA contamination in the model template. 

We note that galaxy-galaxy weak lensing is not affected by IA contamination, as long as the redshifts of source galaxies do not overlap with those of lens galaxies, which we believe is the case for our source galaxy selection. 

\subsection{Modeling Residual Systematic Errors}
\label{sec:systematic_errors}
In this section, we present a method to account for the effects of residual systematic errors 
on our cosmology analysis. 
In what follows, we include the systematic effects in the theoretical templates rather than in the signals 
to keep the data vector and the covariance matrix invariant. 

\subsubsection{Residual systematic photo-$z$ uncertainty: $\Delta z_{\rm ph}$}
\label{subsubsec:photo-z}

Photo-$z$ uncertainty is one of the most important systematic effects in weak lensing measurements, i.e. $\dSigma$ and $\xi_{\pm}(\vartheta)$ in our data vector. As detailed in \citet{more2023}, the redshift distribution of HSC source galaxies was inferred by combining the individual photo-$z$ posteriors with the cross-correlation clustering measurement of HSC galaxies with the {\tt CAMIRA} sample of Luminous Red Galaxies (LRGs) that have accurate photo-$z$ estimates (typically a few per cent in $\sigma(z)/(1+z)$), based on the method in \citet{Rau:2022wrq}. However, we were not able to fully calibrate the redshift distribution due to the lack of a calibration sample of CAMIRA LRGs at $z\gtrsim 1$ (more exactly speaking, the photo-$z$ accuracies of LRGs at $1\lesssim z\lesssim 1.2$ are degraded, and there are no LRGs available at $z\gtrsim 1.2$). Hence we take into account the possibility that there is an unknown residual systematic error in the mean redshift of source galaxies. To study the impact of such residual photo-$z$ calibrated uncertainty, we introduce a nuisance parameter, denoted as $\deltapz$, to model a systematic shift in the mean source redshift by shifting the posterior distribution of source redshifts, given as $z^{\rm est}=z^{\rm true} + \Delta z_{\rm ph}$ \citep{Hutereretal:06,OguriTakada:11,Miyatake:2022b}. That is, we use the shifted photo-$z$ distribution to model the true distribution as
\begin{align}
p^{\rm true}(z)=p^{\rm est}(z+\Delta z_{\rm ph}).
\end{align}
A positive $\Delta z_{\rm ph}$ corresponds to the true mean redshift being  lower than what is inferred from the photo-$z$ posterior, and vice-versa. Note that the discussion around Eq.~(\ref{eq:ave_Sigmacr_def}) gives a justification of this shifted model for the galaxy-galaxy weak lensign ($\dSigma$) and \citet{Zhang:2023} gave a quantitative justification for the HSC-Y3 analyses  on the cosmic shear signals.

For $\dSigma$ (Eq.~\ref{eq:dSigma_two_contributions}), we first need to recompute the averaged lensing efficiency $\avrg{\Sigma_{\rm cr}^{-1}}$ and the weight $w_{\rm ls}$ using the shifted redshift distribution (Eqs.~\ref{eq:lens_measurement} and \ref{eq:ave_Sigmacr_def}): we define the correction factor as
\begin{align}
    f_{\dSigma}(\Delta z_{\rm ph}) \equiv  
    \frac{\sum_{\rm ls}w_{\rm ls}\langle\Sigma_{\rm c}^{-1}\rangle^{\rm true}_{\rm ls}/\langle\Sigma_{\rm c}^{-1}\rangle^{\rm est}_{\rm ls}}{\sum_{\rm ls}w_{\rm ls}}. \label{eq:photo-z-corr}
\end{align}
We compute the correction factor for each of the three lens subsamples, LOWZ, CMASS1 and CMASS2. In our method, we multiply the correction factor by the model template of $\dSigma$, rather than varying the signal, as 
\begin{align}
    \dSigma^{\rm corr}(R, \Delta z_{\rm ph}; z_{\rm l}) 
    = f_{\dSigma}(\Delta z_{\rm ph}; z_{\rm l})\dSigma(R; z_{\rm l}).
\end{align}
Note that $\dSigma$ includes both the galaxy-galaxy weak lensing and the magnification term in Eq.~(\ref{eq:dSigma_two_contributions}): $\dSigma=\dSigma_{\rm gG}+\dSigma_{\rm mag}$, since the correction factor is an overall factor that is applied to the estimator of $\widehat{\dSigma}$ (Eq.~\ref{eq:lens_measurement}). In the theoretical template, in addition to the overall factor, we properly use the shifted redshift distribution of source galaxies to re-compute the magnification bias term, $\dSigma_{\rm mag}$. Also note that the definition of $f_{\dSigma}$ is the inverse of the similar correction factor 
$f_{\rm ph}$ used in the HSC-Y1 papers \citep{Miyatake:2022b,Sugiyama:2022}.

Similarly, we recompute the model prediction for the cosmic shear correlation functions 
$\xi_\pm(\vartheta)$ 
using the shifted redshift distribution of source galaxies. 

\subsubsection{Correction for the reference cosmology used in our measurement}
\label{subsubsec:meascorr}
In the measurements of $\wproj$ and $\dSigma$, we need to assume a ``reference'' cosmology to convert the angular separation between galaxies in each pair to the projected separation $R$, and the redshift difference to the radial separation, $\Pi$. For $\dSigma$, we also need the reference cosmology to convert the shear to $\dSigma$. Throughout our series of papers, the reference cosmology is a flat $\Lambda$CDM model with $\Omega_{\rm m}^{\rm ref}=0.279$. However, the reference cosmology generally differs from the true underlying cosmology, and we need to correct for this discrepancy in our cosmology analysis. We denote a cosmology taken in the parameter inference as $\mathbb{C}$ and the reference cosmology as $\mathbb{C}^{\rm ref}$. The corrections for $R$ and $\Pi$ are obtained as 
\begin{align}
    R &= \frac{\chi(z_{\rm l}; \mathbb{C})}{\chi(z_{\rm l}; \mathbb{C}^{\rm ref})}R^{\rm ref},\nonumber\\
    \Pi &= \frac{E(z_{\rm l};\mathbb{C}^{\rm ref})}{E(z_{\rm l};\mathbb{C})} \Pi^{\rm ref}, \label{eq:meascorr-pimax}
\end{align}
where $E(z)$ is the normalized, dimension-less Hubble rate at redshift $z$, defined as $E(z) \equiv H(z)/H_0$. 
Thus we include the measurement corrections in the theoretical templates of $\dSigma$ and $\wproj$ as
\begin{align}
&\dSigma^{\rm ref}\!(R^{\rm ref}; \Delta z_{\rm ph}, \mathbb{C}, z_{\rm l})=
f_{\dSigma}(\Delta z_{\rm ph}; \mathbb{C}, z_{\rm l})
\dSigma(R; \mathbb{C}, z_{\rm l}), 
\label{eq:dsigma_varyingcosmology}
\\
&\wproj^{\rm ref}\!(R^{\rm ref};\mathbb{C},z_{\rm l})= 2f^{\rm RSD}_{\rm corr}\!(R;\mathbb{C}, z_{\rm l})\frac{E(z_{\rm l};\mathbb{C})}{E(z_{\rm l};\mathbb{C}^{\rm ref})}\nonumber\\
&\hspace{6em}\times
\int_0^{\Pi_{\rm max}}\!\mathrm{d}\Pi~\xi_{\rm gg}\!\left(\sqrt{R^2+\Pi^2}; \mathbb{C},z_{\rm l}\right),
\label{eq:wp_varyingcosmology}
\end{align} 
where $R$ and $\Pi$ are given by $R^{\rm ref}$ or $\Pi^{\rm ref}$ and the cosmological parameters ($\Omega^{\rm ref}_{\rm m}$ and $\Omega_{\rm m}$ for a flat model) in the  $\mathbb{C}^{\rm ref}$ and $\mathbb{C}$ models (Eq.~\ref{eq:meascorr-pimax}). Note that we adopt $\Pi_{\rm max}=[E(\mathbb{C}^{\rm ref})/E(\mathbb{C})]\Pi_{\rm max}^{\rm ref}=[E(\mathbb{C}^{\rm ref})/E(\mathbb{C})]\times 100\,h^{-1}{\rm Mpc}$, as we use the fixed $\Pi_{\rm max}^{\rm ref}=100\,h^{-1}{\rm Mpc}$ in the measurement. Also note that $\dSigma(R)$ and $\xi_{\rm gg}(r)$ on the r.h.s. of the above equations are computed from theory ({\tt Dark Emulator} in our method) for a given cosmological model ($\mathbb{C}$). The overall correction factor for $\dSigma$ is defined as
\begin{align}
    f_{\dSigma}(\Delta z_{\rm ph}; \mathbb{C},z_{\rm l})\equiv 
    \frac{\sum_{\rm ls}w_{\rm ls}\langle\Sigma_{\rm c}^{-1}\rangle^{{\rm true}, \mathbb{C}}_{\rm ls}/\langle\Sigma_{\rm c}^{-1}\rangle^{{\rm est}, \mathbb{C}^{\rm ref}}_{\rm ls}}{\sum_{\rm ls}w_{\rm ls}}. \label{eq:photo-z-corr-cosmo}
\end{align}
Thus this correction factor accounts for both the effects of residual photo-$z$ errors ($\Delta z_{\rm ph}$) and the measurement correction ($\mathbb{C}$). We evaluate the model templates, $\dSigma^{\rm ref}$ and $\wproj^{\rm ref}$, at the discrete sampling points of $R^{\rm ref}$ as used in the data vector in \citet{more2023}.

\subsubsection{Residual multiplicative shear error}
\label{subsubsec:mbias}

In order to account for possible residual errors in the shape calibration, we introduce a nuisance parameter which quantifies the residual multiplicative bias $\Delta m$ and shifts the theoretical templates of the lensing observables: 
\begin{align}
    \dSigma^{\rm corr}(R,\Delta m; z_{\rm l}) &= (1+\Delta m)\dSigma(R; z_{\rm l}),\\
    \xi_\pm^{\rm corr}(\vartheta, \Delta m) &= (1+\Delta m)^2\xi_\pm(\vartheta).
\end{align}
Since we use a single source sample for both the galaxy-galaxy lensing and cosmic shear, we use the same residual multiplicative bias parameter for the theoretical templates of $\dSigma$ for the three lens subsamples and for $\xi_{\pm}$. Hence comparing these data vectors allows us to calibrate the $\Delta m$ parameter, simultaneously with the calibration of the photo-$z$ error parameter $\Delta z_{\rm ph}$.

\subsubsection{PSF systematics}
\label{subsubsec:psf-residual}
As discussed in the HSC-Y3 shape catalog paper \cite{Li2021} \citep[also see][]{2022arXiv221203257Z}, PSF leakage and residual PSF modeling error contaminate the measured cosmic shear correlation functions. Such residual PSF systematic errors could produce artificial two-point correlations and hence bias the cosmic shear measurements. Here we examine the impact of these systematics in our cosmic shear measurements, following the method used for the Year 1 analyses \citet{2019PASJ...71...43H,2020PASJ...72...16H} \citep[also see][]{2018PhRvD..98d3528T},  where we assume that the measured galaxy shapes have an additional additive bias given by
\begin{align}
\epsilon^{(\rm sys)}=\alpha_{\rm psf}\epsilon^{\rm p}+\beta_{\rm psf}\epsilon^{\rm q}.
\end{align}
The first term, referred to as PSF leakage, represents a systematic error proportional to the PSF model ellipticity $\epsilon^{\rm p}$ due to the imperfect PSF correction. The second term represents the systematic error associated with the difference between the model PSF ellipticity, $\epsilon^{\rm p}$, and the true PSF ellipticity estimated from individual ``reserved'' stars $\epsilon^{\rm star}$, i.e. $\epsilon^{\rm q}\equiv \epsilon^{\rm p}-\epsilon^{\rm star}$ \citep{2018PhRvD..98d3528T}. Non-zero residual PSF ellipticities $\epsilon^{\rm q}$ indicate an imperfect PSF estimate, which will propagate to estimates of galaxy shears. Note that the above PSF systematics causes additive shear bias, so does not affect the galaxy-galaxy weak lensing.

When the observed galaxy ellipticity is contaminated by $\epsilon^{(\rm sys)}$, these systematic terms cause an additional contamination to the measured cosmic shear correlation functions as 
\begin{align}
\xi_{{\rm psf},\pm}(\vartheta)=\alpha_{\rm psf}^2\hat{\xi}^{\rm pp}_{\pm}(\vartheta)
+2\alpha_{\rm psf}\beta_{\rm psf}\hat{\xi}^{\rm pq}_{\pm}(\vartheta)
+\beta_{\rm psf}^2\hat{\xi}^{\rm qq}_{\pm}(\vartheta)\, ,
\label{eq:psf_sys_cosmicshear}
\end{align}
where $\hat{\xi}^{\rm pp}_{\pm}$, $\hat{\xi}^{\rm qq}_{\pm}$ and $\hat{\xi}^{\rm pq}_{\pm}$ represent the auto-correlation of the model PSF ellipticity $\epsilon^{\rm p}_{\pm}$, the auto-correlation of the residual PSF ellipticity $\epsilon^{\rm q}_{\pm}$, and the cross-correlation of $\epsilon^{\rm p}_{\pm}$ and $\epsilon^{\rm q}_{\pm}$, respectively. The hat notation, ``$\hat{\hspace{1em}}$'', denotes the correlation function measured from the HSC data using the model PSF and the reserved stars (see \citep{more2023}). The coefficients $\alpha_{\rm psf}$ and $\beta_{\rm psf}$ are estimated by cross-correlating $\epsilon^{\rm p}_{\pm}$ and $\epsilon^{\rm q}_{\pm}$ with the observed galaxy ellipticities, as
\begin{align}
&\hat{\xi}^{\rm gp}_{\pm}(\vartheta)
=\alpha_{\rm psf}\hat{\xi}^{\rm pp}_{\pm}(\vartheta)+
\beta_{\rm psf}\hat{\xi}^{\rm pq}_{\pm}(\vartheta),\nonumber\\
&\hat{\xi}^{\rm gq}_{\pm}(\vartheta)
=\alpha_{\rm psf}\hat{\xi}^{\rm pq}(\vartheta)+\beta_{\rm psf}\hat{\xi}^{\rm qq}_{\pm}(\vartheta),
\label{eq:galaxy_psf_correlation}
\end{align}
where $\hat{\xi}_{\pm}^{\rm gp}$ and $\hat{\xi}_{\pm}^{\rm gq}$ are the measured cross-correlations between the galaxy ellipticities, used for the cosmic shear data vector, and $\epsilon^{\rm p}_{\pm}$ and $\epsilon^{\rm q}_{\pm}$. As discussed in \citet{more2023} (also see \citet{li2023,2022arXiv221203257Z}), we used the measurements of mock galaxy shape catalogs and the real star catalog to estimate the statistical errors of the $\hat{\xi}^{\rm gp}$ and $\hat{\xi}^{\rm gq}$ measurements, where the errors take into account the cosmic variance. By comparing the measured $\hat{\xi}^{\rm gp}$ and $\hat{\xi}^{\rm gq}$ with Eq.~(\ref{eq:galaxy_psf_correlation}) using the measured $\hat{\xi}^{\rm pp}$, $\hat{\xi}^{\rm pg}$ and $\hat{\xi}^{\rm qq}$, we found $\alpha_{\rm psf}=-0.0292\pm 0.0129$ and $\beta_{\rm psf}=-2.59\pm 1.65$ for our fiducial source catalog (see Fig.~16 of \citet{more2023}).
 
To take into account the impact of the residual PSF modeling errors on parameter inference, we add the PSF error contamination $\xi_{{\rm psf},\pm}$ (Eq.~\ref{eq:psf_sys_cosmicshear}) to the model cosmic correlation function $\xi_{\pm}$ in Eq.~(\ref{eq:sourceshape_2pt}) and then estimate parameters by varying the parameters $\alpha_{\rm psf}$ and $\beta_{\rm psf}$ using Gaussian priors with widths inferred from the above errors.

The above method of PSF systematics takes into account the PSF systematics up to the second-order moment of PSF. The HSC-Y3 cosmic shear cosmology papers, \citet{li2023} and \citet{dalal2023}, used the more sophisticated, accurate method developed in \citet{2022arXiv221203257Z}, which accounts for the effects up to fourth-moment PSF leakage and fourth-moment PSF modeling error on cosmic shear correlations. Using the same method, we also measured up to the $\alpha$ and $\beta$ coefficients of the fourth-moments of PSF for the HSC source galaxy sample used in this paper. We then generated synthetic cosmic shear data vectors including the measured PSF systematic effects up to the fourth-order moment and checked that the estimated $S_8$ remains almost unchanged compared to our baseline analysis method using the $\alpha$ and $\beta$ coefficients of the PSF second-moment with the priors described above. The main reason for this is that most of the constraining power is from the galaxy clustering information of SDSS galaxies. The details are given in Appendix~\ref{sec:validation} \citep[also see][]{sugiyama2023}.

\subsection{Summary: Model Templates}
\label{sec:summary_theoreticaltemplates}
For convenience, here we write down the model templates used in cosmology inference where we explicitly show which parameters 
are used in the templates of each observable: 
\begin{widetext}
\begin{align}
\dSigma^{\rm t}\!(R^{\rm ref},z_{\rm l}|\mathbb{C},\boldsymbol{\theta}_{\rm g},\Delta z_{\rm ph}, \Delta m,\alpha_{\rm mag}(z_{\rm l}))&=
(1+\Delta m)\Delta\!\Sigma^{\rm ref}\!(R^{\rm ref},z_{\rm l}|\mathbb{C},\boldsymbol{\theta}_{\rm g},\Delta z_{\rm ph},\alpha_{\rm mag}(z_{\rm l})), \nonumber\\
\wproj^{\rm t}\!(R^{\rm ref},z_{\rm l}|\mathbb{C},\boldsymbol{\theta}_{\rm g}):&\hspace{1em} \mbox{Eqs.~(\ref{eq:xi_gg})--(\ref{eq:Pgg_1h2h}),
(\ref{eq:wp_varyingcosmology})}\, ,\nonumber\\
\xi^{\rm t}_{\pm}(\vartheta|\mathbb{C},\Delta z_{\rm ph},A_{\rm IA},\eta_{\rm IA},
\alpha_{\rm psf},\beta_{\rm psf})
&=(1+\Delta m)^2\xi_{\pm}(\vartheta|\mathbb{C},\Delta z_{\rm ph},A_{\rm IA},\eta_{\rm IA})
+\xi_{{\rm psf},\pm}(\vartheta|\alpha_{\rm psf},\beta_{\rm psf}), 
\label{eq:summary_theoretical_templates}
\end{align}
\end{widetext}
where $\mathbb{C}$ denotes a cosmological model sampled in  parameter inference within the flat-geometry $\Lambda$CDM model characterized by five cosmological parameters, $\boldsymbol{\theta}_{\rm g}$ is a set of parameters to model the galaxy-halo connection (five parameters for each of LOWZ, CMASS1 and CMASS2 in our baseline model), $\dSigma^{\rm ref}$ is given by Eq.~(\ref{eq:dsigma_varyingcosmology}), and others are nuisance parameters to model the residual systematic errors. For our baseline model, we have 28~parameters in total: $5~(\mathbb{C})+3\times 5~(\boldsymbol{\theta}_{\rm g})+8~(\Delta z_{\rm ph},\Delta m,\alpha_{\rm mag}(z_{\rm l}),A_{\rm IA},\alpha_{\rm psf},\beta_{\rm psf})$.

\subsection{Computation Time}
\label{sec:computation_time}
We use {\tt Dark Emulator}  to compute the model predictions, $\dSigma^{\rm t}(R)$  and $\wproj^{\rm t}(R)$, for an input model. We use the publicly available {\tt FFTLog} code \citep{Hamilton00} to perform the Hankel transforms in Eqs.~(\ref{eq:dSigma_def}), (\ref{eq:dSigma_mag_def}), \ref{eq:xi_gg}, and (\ref{eq:cosmic_shear_hankel}); for our analysis we use the updated code in \citet{2020MNRAS.497.2699F}. Since our data vector is given by discrete bins of $R^{\rm ref}$ or $\vartheta$, we properly take into account the weighted average of the model predictions within the bin width, more precisely $\Delta \ln R^{\rm fid}=0.246$ for $\dSigma^{\rm t}$, $\Delta\ln R^{\rm fid}=0.169$ for $\wproj^{\rm t}$, and $\Delta\ln\vartheta=0.242$ for $\xi_{\pm}(\vartheta)$, respectively. With our current analysis pipeline, we can compute the model predictions of $\dSigma^{\rm t}$ for all three lens samples (LOWZ, CMASS1 and CMASS2) in about 2~CPU seconds in total, those of $\wproj^{\rm t}$ in about 2~seconds total, and those of $\xi_{\pm}$ in about 0.15~seconds for a given model. This is fast enough to enable cosmological parameter inference in a high dimensional parameter space (28~parameters in our baseline setup). 

\begin{table}
\caption{Model parameters and priors used in our cosmological inference. The label ${\cal U}(a,b)$  denotes a uniform (or equivalently flat) prior with minimum $a$ and maximum $b$, while ${\cal N}(\mu,\sigma)$ denotes a normal (or Gaussian) prior with mean $\mu$ and width $\sigma$. The parameters used in our baseline analysis are listed above the horizontal double lines: five cosmological parameters, five HOD parameters for each of the LOWZ, CMASS1 and CMASS2 subsamples, two nuisance parameters to model residual photo-$z$ and multiplicative shear biases, three parameters to model the magnitude slope of the galaxy number counts that characterizes the magnification bias on $\dSigma$ for each of the LOWZ, CMASS1 and CMASS2 subsamples, two parameters to model residual PSF modeling errors in the cosmic shear 2pt functions, and one parameter to model the IA contamination to cosmic shear: $28=5+3\times 5+2+3+2+1$ in total. The parameters below the double lines are used in the extended models.}
\label{tab:parameters}
\setlength{\tabcolsep}{15pt}
\begin{center}
\input{param_prior_b2.tex}
\end{center}
\end{table}

\subsection{Parameter Estimation Method}
\label{sec:parameter_estimaton_method}
We assume that the likelihood of data for a given model follows a multivariate Gaussian distribution: 
\begin{align}
\ln{\cal L}({\bf d}|\boldsymbol{\theta})=-\frac{1}{2}\sum_{i,j} \left[
d_i - t_i(\boldsymbol{\theta})\right]
{\bf C}^{-1}_{ij} 
\left[
d_j - t_j(\boldsymbol{\theta})\right],
\label{eq:likelihood}
\end{align}
where ${\bf d}$ is the data vector, ${\bf t}$ is the model prediction for the data vector given the model parameters $\boldsymbol{\theta}$, ${\bf C}^{-1}$ is the inverse of the covariance matrix, and the summation runs over indices corresponding to the dimension of the data vector. Note that non-Gaussianity in the likelihood might affect our results as indicated by \citet{2020MNRAS.499.2977L}, although they showed that the non-Gaussianity does not cause a significant bias in the parameter value, and rather changes the size of the confidence region. We will leave this question to future studies. Please see \citet{more2023} for our method to construct the covariance matrix using the mock catalogs of SDSS and HSC galaxies. In our baseline analysis, the data vector consists of $\dSigma(R)$ in 9 logarithmically-spaced radial bins  within $3\le R/[h^{-1}{\rm Mpc}]\le 30$, and $\wproj\!(R)$ in 16 logarithmically-spaced radial bins within $2\le R/[h^{-1}{\rm Mpc}]\le 30$, for each galaxy subsample (LOWZ, CMASS1 and CMASS2), and 8 and 7 logarithmically-spaced angular bins within $8\lesssim \vartheta/[{\rm arcmin}]\lesssim 50$ and $30\lesssim \vartheta/[{\rm arcmin}]\lesssim 150$ for $\xi_+$ and $\xi_-$, respectively. Thus, we use $90(=3\times (9+16)+8+7)$ data points in total. When we use the data vector with different scale cuts from the baseline analysis, we use the submatrix of the full covariance matrix computed in \citet{more2023}, corresponding to that range of scales, and then invert the matrix to obtain the inverse of the covariance submatrix.

Our analysis uses a set of parameters and priors summarized in Table~\ref{tab:parameters}. The parameters include five cosmological parameters denoted by $\mathbb{C}=\{\Omega_{\rm de},\ln(10^{10}A_{\rm s}),\omega_{\rm b\}, \omega_{\rm c},n_{\rm s}}$ for the flat $\Lambda$CDM model, as well as five HOD parameters for each of the LOWZ, CMASS1, and CMASS2 samples. For for $\omega_b$, a Gaussian prior with a mean and width inferred from Big Bang nucleosynthesis (BBN) constraints is employed. For $n_{\rm s}$, a Gaussian prior, which was inferred from the {\it Planck}2018 ``TT,EE,TE+lowE'' constraints, with a mean value of $0.9649\pm (3\times 0.0042)$ and a Gaussian width three times wider than the $1\sigma$ uncertainty ($0.0042$) of the {\it Planck} constraint is used. We employ these priors because the clustering observables $\dSigma$ and $w_{\rm p}$ are insensitive to $\omega_b$ and $n_{\rm s}$. Broad, flat priors are adopted for $\Omega_{\rm de}$ and $\omega_{\rm c}$, with ranges corresponding to about $\pm 30\sigma$ and $\pm 15\sigma$, respectively, compared to the $1\sigma$ error of the {\it Planck} constraints for the flat $\Lambda$CDM model. These ranges correspond to the supported range of the extrapolation of {\tt Dark Emulator} (for more information, see Section\ref{sec:dark_emulator}). Additionally, we use a broad and uninformative flat prior for $\ln{(10^{10} A_{\rm s})}$, as there is no limitation on its extrapolation.

To account for possible uncertainty in the magnitude slope of the number counts when modeling the magnification bias for each lens sample, we incorporate $\alpha_{\rm mag}(z_i)$ into our analysis. We use the measured value of $\alpha_{\rm mag}$ for the central value (see Section IIIA and Fig. 2 in \citet{Miyatake:2022b}) and adopt a Gaussian prior with a width of $\sigma(\alpha_{\rm mag})=0.5$. Our choice of Gaussian width is conservative, since it is much wider than the measurement error on $\alpha_{\rm mag}$. However, we demonstrate that the results remain largely unchanged even when $\alpha_{\rm mag}$ is fixed to the central (measured) value.

We account for residual uncertainties in the source photo-$z$ error and the multiplicative shear bias by including nuisance parameters $\Delta z_{\rm ph}$ and $\Delta m$. Since we use a single population of source galaxies, we needed to adopt only one $\Delta z_{\rm ph}$ and one $\Delta m$ parameter to model the impact on the galaxy-galaxy weak lensing signals
for all three lens galaxy samples and the cosmic shear correlation functions. In Section~\ref{subsubsec:photo-z}, we discussed the uncertainty in $\Delta z_{\rm ph}$ and hence chose to use an uninformative flat, wide prior of ${\cal U}(-1.0,1.0)$ in our baseline setup as the most conservative option.

We will demonstrate that our method enables 
a self-calibration of $\Delta z_{\rm ph}$. We made this choice while the analysis was still blinded, i.e. before the unblinding. We also consider a Gaussian prior with width $\sigma(\Delta z_{\rm ph})=0.1$ and mean $\Delta z_{\rm ph}=0$, as done in the Year 1 analysis \citep{Miyatake:2022b}. This allows us to study how the cosmological parameter inference is altered by this informative prior. This prior is still wider than the width of a few $O(10^{-2})$ that is inferred from the photo-$z$ method in \citet{Rau:2022wrq}. If we use only a subset of the observables, either only the 2$\times$2pt or the cosmic shear correlations, the analysis cannot constrain $\Delta z_{\rm ph}$. Hence, for analyses aimed at internal consistency tests of the data, we used a Gaussian prior whose width and mean are given by the posterior of $\Delta z_{\rm ph}$ obtained from the baseline 3$\times$2pt analysis. This prior is denoted by the superscript ``$^\ast$'' in Table~\ref{tab:parameters} and hereafter. 

For $\Delta m$, we employ a prior range that corresponds to about $1\sigma$ statistical uncertainties in the shape measurement calibration \cite{2018MNRAS.481.3170M} \citep[also see Table~6 in Ref.][]{2019PASJ...71...43H}. We will discuss the case where the prior range of $\Delta m$ is broadened in Section~\ref{sec:lcdm}. 

For $\alpha_{\rm psf}$ and $\beta_{\rm psf}$, which model the residual PSF modeling errors, we  
use the measured values 
for the central values and employ a Gaussian prior with width given by the $1\sigma$ measurement uncertainty for each of $\alpha_{\rm psf}$
and $\beta_{\rm psf}$. 
The details of estimation of these parameters are given in \citet{more2023}.

The parameters we described above are the model parameters used for our baseline analysis. We also employ the extended halo model to check how the cosmological parameters obtained from the baseline analysis are robust against possible variations in the model template. For the extended model, we consider the effects of off-centered central galaxies and the incompeleteness of central galaxies \citep[also see][for details]{Miyatake:2022a,Miyatake:2022b}. Table~\ref{tab:parameters} gives the parameters to model these effects.

We then obtain the posterior distribution ${\cal P}(\boldsymbol{\theta}|{\bf d})$ of our parameters $\boldsymbol{\theta}$ given the data ${\bf d}$, by performing Bayesian inference:
\begin{align}
{\cal P}(\boldsymbol{\theta}|{\bf d})&\propto {\cal L}({\bf d}|\boldsymbol{\theta})\Pi(\boldsymbol{\theta}),
\label{eq:bayes}
\end{align}
where $\Pi(\boldsymbol{\theta})$ is the prior distribution of the parameters. 
The marginalized posterior distributions of the derived parameters $\Omega_{\rm m}$, $\sigma_8$, and $S_8\equiv \sigma_8(\Omega_{\rm m}/0.3)^{0.5}$, where $\Omega_{\rm m} = 1-\Omega_{\rm de}$ for a flat $\Lambda$CDM model, are the main focus of this paper. While $\ln{(10^{10} A_{\rm s})}$ is sampled in logarithmic space with a flat prior, we effectively produce a flat prior in linear space of $\sigma_8$ when obtaining the posterior distribution of $\sigma_8$ as a derived parameter by taking into account Jacobian as weights (see Section~IV~A in \citet{Sugiyama:2020} and Section~V~G in \citet{dalal2023} for a detailed discussion). However, the effect is negligible  because the Jacobian is nearly constant in the range of our credible interval on $\sigma_8$. 

To obtain the posterior distribution of parameters in our multi-dimensional parameter space, we use the importance nested sampling algorithm implemented in the publicly-available software package {\tt MultiNest} \cite{Feroz:2008,Feroz:2009,Feroz:2019} and its python wrapper, \texttt{PyMultiNest} \citep{2014A&A...564A.125B}. We use ${\rm nlive}=600$, ${\rm tol}=0.1$ for the hyper parameters of {\tt MultiNest}. However, we found that {\tt MultiNest} tends to underestimate the credible interval, e.g., that of $S_8$ by $\sim 10\%$. This is because {\tt MultiNest} samples a parameter with a Gaussian prior in Table~\ref{tab:parameters}, in a limited volume that is specified by another hyper parameter ${\rm efr}$. We checked that we can avoid this inaccuracy by treating the Gaussian prior as an additive term to the likelihood of our obeservables, rather than injecting the prior to the prior interface in {\tt MultiNest}. We use this implementation for our baseline 3$\times$2pt analysis. On the other hand, we use the standard {\tt MultiNest} implementation with ${\rm nlive}=600$, ${\rm tol}=0.1$, and ${\rm efr}=0.5$ for model/method validations and internal consistency tests described in Section~\ref{sec:consistency_test_main}, since we need to save computing time to run chains for each setup listed in Table~\ref{tab:analysis_setups}. Still, we note that the central value of cosmological parameter is stable for both the implementations (typically only a few percent difference). We describe detailed investigations, such as a convergence of {\tt MultiNest} chains and comparison with the Metropolis 
algorithm, in Appendix~\ref{sec:chain_convergence}.
\begin{table*}
\caption{A summary of the analysis setups. The first column 
denotes each analysis setup. The scale cuts ``$(X,Y)$'' denote the lower scale cuts applied to $\wproj(R)$ and $\dSigma(R)$, respectively, which means that we use $\wproj$ and $\dSigma$ for $X\le R/[h^{-1}{\rm Mpc}]\le 30$ and $Y\le R/[h^{-1}{\rm Mpc}]\le 30$, respectively, in the cosmology analysis. The symbol ${}^\ast$ at the end of the setup label indicates an analysis using a prior of photo-$z$ shift parameter derived from the baseline 3$\times$2pt analysis, given by $\Pi(\Delta z_{\rm ph})={\cal N}(-0.06,0.08)$, while ${}^\dagger$ denotes the analysis with a Gaussian prior $\Pi(\Delta z_{\rm ph})={\cal N}(0,0.1)$. ${\cal D}(\boldsymbol{\theta})$ and ${\cal D}({\bf d})$ denote the dimension of parameters and data vector, respectively, in each analysis.
 \label{tab:analysis_setups}}
\renewcommand{\arraystretch}{1.2}
\setlength{\tabcolsep}{2pt}
\begin{center}
\begin{tabular}{llc}\hline\hline
setup label & description & $\mathcal{D}(\bm{\theta})$, $\mathcal{D}({\bf d})$ \\
\hline 
3$\times$2pt                                               & {\it baseline analysis} $\dSigma+\wproj+\xi_{\pm}$, with (2,3)$h^{-1}{\rm Mpc}$  scale cuts for $\wproj$ and $\dSigma$                                  & 28, 90\\
2$\times$2pt${}^{\ast}$                                   & 2$\times$2pt ($\dSigma+\wproj$), w/o $\xi_{\pm}$, using $\Delta z_{\rm ph}$ posterior from 3$\times$2pt analysis as a prior                             & 25, 75\\
cosmic shear${}^{\ast}$                                   & $\xi_{\pm}$ alone , using $\Delta z_{\rm ph}$ posterior from 3$\times$2pt analysis as a prior                                                           & 10, 15\\
\hline
3$\times$2pt, $R_{\rm min}=(4,6)~h^{-1}{\rm Mpc}$${}^\ast$  & 3$\times$2pt, with the minimum scale cuts $R_{\rm min}=(4,6)~h^{-1}{\rm Mpc}$ for $\wproj$ and $\dSigma$                                                & 28, 72\\
3$\times$2pt, $R_{\rm min}=(8,12)~h^{-1}{\rm Mpc}$${}^\ast$ & 3$\times$2pt, with the minimum scale cuts $R_{\rm min}=(8,12)~h^{-1}{\rm Mpc}$ for $\wproj$ and $\dSigma$                                               & 28, 51\\
\hline
3$\times$2pt, w/o LOWZ                                     & 3$\times$2pt w/o LOWZ                                                                                                                                   & 22, 65\\
3$\times$2pt, w/o CMASS1                                   & 3$\times$2pt w/o CMASS1                                                                                                                                 & 22, 65\\
3$\times$2pt, w/o CMASS2                                   & 3$\times$2pt w/o CMASS2                                                                                                                                 & 22, 65\\
\hline
2$\times$2pt, w/o LOWZ${}^{\ast}$                                  & 2$\times$2pt w/o LOWZ                                                                                                                                   & 18, 51\\
2$\times$2pt, w/o CMASS1${}^{\ast}$                                  & 2$\times$2pt w/o CMASS1                                                                                                                                 & 18, 51\\
2$\times$2pt, w/o CMASS2${}^{\ast}$                                   & 2$\times$2pt w/o CMASS2                                                                                                                                 & 18, 51\\
\hline
no photo-$z$ error                                         & 3$\times$2pt, but fixing $\Delta z_{\rm ph}=0$                                                                                                          & 27, 90\\
no shear error                                             & 3$\times$2pt, but fixing $\Delta m=0$                                                                                                                   & 27, 90\\
fix mag. bias                                              & 3$\times$2pt, but fixing $\alpha_{\rm mag}=\mu$                                                                                                         & 25, 90\\
no PSF error                                               & 3$\times$2pt, but fixing $\alpha_{\rm psf}=\beta_{\rm psf}=0$                                                                                           & 26, 90\\
no IA                                                      & 3$\times$2pt, but fixing $A_{\rm IA}=0$                                                                                                                 & 27, 90\\
extreme IA                                                 & 3$\times$2pt, but fixing $A_{\rm IA}=5$                                                                                                                 & 27, 90\\
\hline
3$\times$2pt${}^{\dagger}$                                & 3$\times$2pt with a prior $\Pi(\Delta z_{\rm ph})={\cal N}(0, 0.1)$                                                                                     & 28, 90\\
2$\times$2pt${}^{\dagger}$                                & 2$\times$2pt with a prior $\Pi(\Delta z_{\rm ph})={\cal N}(0, 0.1)$                                                                                     & 24, 75\\
cosmic shear${}^{\dagger}$                                & cosmic shear with a prior $\Pi(\Delta z_{\rm ph})={\cal N}(0, 0.1)$                                                                                     & 11, 15\\
\hline
2$\times$2pt                                               & 2$\times$2pt with a prior $\Pi(\Delta z_{\rm ph})={\cal U}(-1,1)$                                                                                       & 25, 75\\
cosmic shear                                               & $\xi_{\pm}$ with a prior $\Pi(\Delta z_{\rm ph})={\cal U}(-1,1)$                                                                                        & 10, 15\\
\hline
XMM  ($\sim33$~deg$^2$)${}^{\ast}$                                 & 3$\times$2pt, but using the signals of the XMM field alone                                                                                              & 28, 90\\
GAMA15H ($\sim41$~deg$^2$)${}^{\ast}$                                & 3$\times$2pt, but using the signals of the GAMA15H field alone                                                                                          & 28, 90\\
HECTOMAP ($\sim43$~deg$^2$)${}^{\ast}$                               & 3$\times$2pt, but using the signals of the HECTOMAP field alone                                                                                         & 28, 90\\
GAMA09H ($\sim78$~deg$^2$)${}^{\ast}$                                & 3$\times$2pt, but using the signals of the GAMA09H field alone                                                                                          & 28, 90\\
VVDS ($\sim96$~deg$^2$)${}^{\ast}$                                 & 3$\times$2pt, but using the signals of the VVDS field alone                                                                                             & 28, 90\\
WIDE12H ($\sim121$~deg$^2$)${}^{\ast}$                              & 3$\times$2pt, but using the signals of the WIDE12H field alone                                                                                          & 28, 90\\
\hline
\dempz\&WX                                                 & 
\dempz\&WX is used to infer the source redshift distribution 
and  for $\dSigma$ measurement.                                                            
& 28, 90\\
\mizuki                                                    & \mizuki is used for source sample selection and $\dSigma$ measurement                                                                                   & 28, 90\\
\dnnz                                                      & \dnnz is used for source sample selection and $\dSigma$ measurement                                                                                     & 28, 90\\
\hline
w/o star weight                                           & 3$\times$2pt, but without using star weight when computing $w_{\rm p}$ & 28, 90\\
\hline
offcentering                                               & 3$\times$2pt, but including the off-centering effect in galaxy-halo connection                                                                          & 34, 90\\
incompleteness                                             & 3$\times$2pt, but including the incompleteness effect in galaxy-halo connection                                                                         & 34, 90\\
\hline
{\bf Additional tests after unblinding}&&\\
$\sigma(\Delta m)=0.1$ prior                                               & 3$\times$2pt, but using prior $\Pi(\Delta m)={\cal N}(0, 0.1)$                                                                      
& 28, 90\\
$\sigma(\Delta z_{\rm ph})=0.2$ prior                                                 & 3$\times$2pt, but using prior 
$\Pi(\Delta z_{\rm ph})={\cal N}(0, 0.2)$                                                                              
    & 28, 90\\
2 cosmo paras                                              & 3$\times$2pt, but varying only $(\Omega_{\rm de},\ln (10^{10}A_{\rm s}))$ and fixing $(\omega_{\rm c},\omega_{\rm },n_{\rm s})$ to their {\it Planck}
values& 25, 90\\
\hline\hline
\end{tabular}
\end{center}
\renewcommand{\arraystretch}{1}
\end{table*}

\subsection{Analysis Setups}
\label{sec:analysis_setups}

To carry out the inference of cosmological parameters, we need to define the analysis setups, which include the range of scales and combinations of observables to be used. The setups employed in this paper are summarized in Table~\ref{tab:analysis_setups}.

The setup labeled ``3$\times$2pt'' is our baseline setup that serves as a reference; if we identify any internal consistency test that significantly deviates from the cosmological parameters obtained from this reference setup, we should consider changing the baseline setup. 

If we use either $\dSigma$ or $\wproj$ alone, the parameter inference suffers from severe degeneracies, especially between the galaxy bias (and therefore the HOD model parameters) and the cosmological parameters that encode information about the power spectrum amplitude, as shown in our validation paper \cite{Miyatake:2022a} (see Fig.~9). Hence, in the following we show only the results of the joint analysis of $\dSigma$ and $\wproj$, even when not combined with $\xi_{\pm}$. The ``2$\times$2pt${}^\ast$'' setup is such a case, but we set the prior on the residual photo-$z$ parameter $\Delta z_{\rm ph}$ to the one obtained from the baseline 3$\times$2pt analysis because without the prior cosmological constraints are quite weak. We also run the cosmic shear-only analysis with the same prior which is labeled as ``cosmic shear${}^\ast$''. \footnote{Note that the Gaissian prior employed for internal consistency tests, i.e., $\Pi(\Delta z_{\rm ph})={\cal N}(-0.06,0.08)$ is slightly different from what we show as the baseline result, $\deltapz=-0.05\pm0.09$, in Section~\ref{sec:lcdm}. This is because we use the $\Delta z_{\rm ph}$ constraint from the standard {\tt MultiNest} implementation for this Gaussian prior rather than the setup used for the baseline 3$\times$2pt analysis described in Section~\ref{sec:parameter_estimaton_method}. This is the case for the Gaussian prior on $\Delta z_{\rm ph}$ used in \citet{sugiyama2023}.}

An important aspect of the analysis is determining the appropriate ``scale cuts'', which refers to the range of scales ($R$) in $\dSigma(R)$ and $\wproj(R)$ used in the cosmological analysis. There are two competing factors to consider: on one hand, including information from $\dSigma(R)$ and $\wproj(R)$ down to smaller scales can increase the statistical power to constrain cosmological parameters. On the other hand, observables at small scales may be more affected by physical systematic effects inherent in galaxy formation/physics, which are difficult to accurately model. Our validation paper \cite{Miyatake:2022a} describes our choice of $(2,3)~h^{-1}{\rm Mpc}$ scale cuts for $\wproj$ and $\dSigma$ as reasonable choices for obtaining unbiased estimates of the cosmological parameters, with reasonably small credible intervals given the statistical power of HSC-Y1 and SDSS data. These scale cuts are larger than the virial radii of massive halos, so we exclude information from scales that are in the deeply 1-halo term regime in our cosmology analysis. However, we note that the galaxy-galaxy weak lensing signal around the scale cut is sensitive to the {\it interior} mass inside that radius. Thus, we can extract the average mass of halos hosting the SDSS galaxies, which in turn helps constrain the large-scale bias of SDSS galaxies via the scaling relation of halo bias with halo mass, encoded in {\tt Dark Emulator}, when combined with the measurement of $\wproj$. We also examine the results for scale cuts of $(4,6)$ and ($8,12$)$~h^{-1}{\rm Mpc}$, respectively, to investigate the impact of the scale cut choice. 

For the scale cuts of $\xi_\pm$ we follow the same scale cuts in $\xi_\pm(\vartheta)$ as those in \citet{li2023}, because we did not find any significant residual $B$-mode signal in the range. For this paper, we use the cosmic shear signals in the rane of $8\lesssim\vartheta/{\rm arcmin}\lesssim 50$ and $30\lesssim\vartheta/{\rm arcmin}\lesssim150$
for $\xi_+$ and $\xi_-$, respectively.  

As an internal consistency test, we perform various analyses to assess the robustness of our results to various splits of our data and the sensitivity of our results to the different analysis choices. We perform the analyses excluding one of the LOWZ, CMASS1 or CMASS2 subsample for both ``3$\times$2pt'' and ``2$\times$2pt${}^\ast$'' analysis. To study the impact of the nuisance parameters on our results, we perform the analysis by fixing either one of the nuisance parameters, rather than varying it, in the parameter inference: the residual photo-$z$ error $\Delta z_{\rm ph}=0$, the multiplicative error $\Delta m=0$, the magnification bias parameters $\alpha_{\rm mag}$ to their measured values (see Table~\ref{tab:parameters}), the PSF systematics parameters $\alpha_{\rm psf}=\beta_{\rm psf}=0$, intrinsic alignment $A_{\rm IA}=0$, or $A_{\rm IA}=5$, respectively.

To check the impact of the prior choice of the residual photo-$z$ error parameter, which is a key parameter in our analysis, we perform the analysis using a Gaussian prior given by $\Pi(\deltapz)={\cal N}(0,0.1)$, for the 3$\times$2pt analysis or the 2$\times$2pt- or the cosmic shear-only analysis. We also perform the analysis using an uninformative flat prior, $\Pi(\Delta z_{\rm ph})={\cal U}(-1,1)$, for the 2$\times$2pt- or the cosmic shear-only analysis.

We also perform field-by-field analyses which are labeled by the field name, e.g. ``XMM'' in Table~\ref{tab:analysis_setups}, which is the result when using the HSC-Y3 data only in the XMM region \citep{Li2021}. Note that the signals measured in different fields are almost independent.

To check for possible systematic biases arising from the different photo-$z$ estimate methods, we perform the analyses using the lensing signals that are measured using the different photo-$z$ method instead of our fiducial method ({\sc DEmPz}). More exactly, we use the different method to {\it select} source galaxies based on the same selection cut (Eq.~\ref{eq:source-selection}) and then use the inferred source redshift distribution in the weak lensing observables (the signal  of $\dSigma$ and the theory of $\xi_{\pm}$). For ``{\sc DEmPz\&Wx}'', we use the source redshift distribution inferred by the  ``{\sc DEmPz\&WX}'' method  that is obtaine by combining the {\sc DEmPz} photo-$z$ estimates and the clustering with the {\sc CAMIRA} LRGs in \citet{Rau:2022wrq}. We also use the \mizuki photo-$z$ method or the \dnnz photo-$z$ method, respectively. We use the uninformative flat prior, $\Pi(\deltapz)={\cal U}(-1,1)$ for all these tests as in our baseline analysis. Note that the {\sc DEmPz\&Wx} analysis uses the same source sample as that in the baseline analysis, but the other methods use the different source samples.

The analysis labeled ``w/o star weight'' is for testing the impact of the star weight used in the clustering measurement of $\wproj$ (see \citet{more2023} for details) on our results, where the star weight is designed to account for the systematic effect related to seeing and stellar density in the target selection of SDSS galaxies. We use the $\wproj$ data measured without the star weight in \cite{more2023}, instead of our default $\wproj$, for the baseline 3$\times$2pt analysis. 

We also perform the analyses using the extended models that are more flexible models for the galaxy-halo connection; we use the extended models including the effects of off-centered central galaxies or the incompleteness effect of central galaxies \citep{Miyatake:2022a}. For both extended models, we introduce two additional model parameters, as indicated in the rows  ``{off-centering}'' or ``{incompleteness}''.

In addition, after unblinding our cosmology results (see next section), we further decided to perform the analyses labeled ``$\sigma(\Delta m)=0.1$ prior'' and ``$\sigma(\Delta z_{\rm ph})=0.2$ prior''. For these, we use Gaussian priors, $\Pi(\Delta m)={\cal N}(0,0.1)$ or $\Pi(\deltapz)={\cal N}(0,0.2)$, to study the impact of the prior choices on the cosmological parameters. Here $\sigma(\Delta m)=0.1$ is much wider than the fiducial prior, $\sigma(\Delta m)=0.01$ (see Table~\ref{tab:parameters}), and this test is for assessing the self-calibration power of the residual shear calibration factor with our 3$\times$2pt method (therefore, we consider the overwhelmingly pessimistic shear calibration error). Furthermore, we an analysis with the cosmological parameters  except $(\Omega_{\rm m}, \ln(10^{10}A_{\rm s}))$ fixed to the {\it Planck}~2015 ``TT,TE,EE+lowP'' constraints \cite{PlanckCosmology:16} to check how the parameters are poorly constrained by our data vector affect our cosmological constraints. This setup is labeled ``2 cosmo'' in Table~\ref{tab:analysis_setups}.

\subsection{Validation of Modeling and Analysis Choices}
\label{sec:validation_method}

To check the robustness of our results to our modeling and analysis choices, we perform various validation tests. This includes tests of different samplers, different models of galaxy-halo connection, different models for baryonic feedback, as well as different models of observational systematics including residual systematic error in the mean source redshift and PSF systematics. Please also see \citet{Miyatake:2022a} for  the detailed validation tests, where the scale cuts of $\wproj$ and $\dSigma$, $R=2$ and $3~h^{-1}$Mpc, were validated in the sense that estimated parameters, such as $S_8$, do not have significant biases compared to their statistical errors. For the tests in this paper, we also include the synthetic data of cosmic shear correlation functions $\xi_{\pm}$ and perform the validation tests using the covariance matrix for the HSC-Y3 data. The details of the validation tests are given in Appendix~\ref{sec:validation}, and here we give a summary of the important points. 

For $\wproj$ and $\dSigma$ we have to study the impact of uncertainties in the galaxy-halo connection on the cosmological constraints. Since HOD is an empirical prescription for the galaxy-halo connection, our philosophy is that we should include a sufficient number of the HOD parameters and then extract the cosmological information from the halo clustering quantities, accurately modeled by {\tt Dark Emulator}, after marginalizing over the galaxy-halo connection parameters. For the validation tests of our analysis method, we generate various types of mock SDSS galaxies, where we employ different ways to populate galaxies into halos in $N$-body simulations and then generate the synthetic data vectors of $\wproj$ and $\dSigma$ from the mock catalogs \citep[also see][]{Miyatake:2022a}. We then apply our baseline analysis pipeline using the fiducial HOD model to the synthetic data vector to assess whether our method can recover the input cosmological parameters used in the mock catalogs. As shown in Appendix~\ref{sec:validation}, our method can recover the $S_8$ value with an accuracy better than $\sim 0.5\sigma$ for most of the mock SDSS galaxy catalogs. 

Nevertheless, reflecting on the fact that there is no established theory of the halo-galaxy connection, we also consider the ``worst-case'' scenario in order for us to be ready for surprises due to unknown systematic effects.  We prepare extreme mock catalogs of SDSS galaxies, where we implement a non-standard prescription of the galaxy-halo connection, e.g. the mock catalogs including the overwhelmingly large assembly bias effect and the off-centering effect of central galaxies. These worst-case scenario mocks change $\wproj$ and $\dSigma$ at scales around the scale cuts, $R=2$ and $3~h^{-1}$Mpc, and could cause a significant bias in $S_8$ (more than $0.5\sigma$). Even for these worst-case scenarios, we have a useful diagnostic to flag such an extreme systematic effect on the cosmological parameters. Since the halo model based theory includes the linear-theory prediction at large scales, which can be applied to any types of galaxies \citep{Nishimichi:2020}, the method can recover the underlying cosmological parameters {\it if} employing sufficiently large scale cuts, such as $R=8$ and $12~h^{-1}$Mpc \citep{Sugiyama:2020,Miyatake:2022a}. Hence, if a significant systematic effect exists in the actual SDSS galaxies, the estimated central value of $S_8$ would have a systematic shift with changing the scale cuts. Based on these findings, we can employ the different scale cuts for the actual cosmology analysis to monitor a change in cosmological parameters as an indicator of potential systematic effects. 

For the systematic effects on the cosmic shear signal $\xi_{\pm}$, our modeling method is very similar to that of \citet{li2023}. We employ exactly the same scale cuts in $\xi_{\pm}$ as those in \citet{li2023}. All the validation tests are passed in the sense that the $S_8$ value is recovered using synthetic data vectors of $\xi_{\pm}$ where different models of systematic effects (baryonic feedbacks and PSF systematics) are implemented. We did not find any flag in our analysis method arising from modeling inaccuracy and systematic effects in cosmic shear. 

One of the most important validation tests is to assess the sensitivity of our method to a residual systematic error in the mean source redshift of HSC galaxies, $\deltapz$. As described in \citet{sugiyama2023} in detail, we implement a non-zero shift (non-zero $\deltapz$) in the source redshift distribution to generate synthetic data vectors of $\dSigma$ and $\xi_{\pm}$ and then assess whether our analysis method can recover the input $\deltapz$ as well as the $S_8$ parameter. As shown in Appendix~\ref{sec:validation}, we find that, if a non-zero shift of $\deltapz$ by more than $|\deltapz|\sim 0.1$ exists in the synthetic data, the use of the uninformative flat prior of $\deltapz$ allows us to recover the underlying true $S_8$ value to within the credible interval. In other words, if we employ an informative prior on $\deltapz$, such as ${\cal N}(0,0.1)$, it could yield a significant bias in $S_8$ larger than the $1\sigma$ statistical error. Thus, the HSC-Y3 data has the statistical power to calibrate $\deltapz$ to the precision of $\sigma(\deltapz)\simeq 0.1$ and then recover $S_8$. This gives a validation of our analysis method using an uninformative prior of $\deltapz$, ${\cal U}(-1,1)$, even if the method gives a large credible interval in exchange. In this sense, our cosmology analysis can be considered robust and conservative. 

\section{Blinding scheme and Unblinding criteria}
\label{sec:blinding}
To avoid confirmation bias we perform our cosmological analysis in a blind fashion. To prevent inadvertent unblinding during our cosmological analysis, we implemented a two-tier blinding approach. The two tiers are outlined below:
\begin{itemize}
\item {\it Catalog level}: During the cosmological analysis, the analysis team performs teh cosmological analysis using three different weak lensing shape catalogs, with only one of them being the true catalog and the other two being fake catalogs. The team members conducting the analysis are unaware of which is the true catalog.
\item {\it Analysis level}: When the analysis team makes plots comparing the measurements with theoretical models, the $y$-axis values (e.g., the amplitudes of $\dSigma$) are hidden and the analysis team is not allowed to see the values of cosmological parameters used in the theoretical models. 
When the analysis team makes plots showing the credible intervals of cosmological parameters (i.e., posterior distribution), the central value(s) of the parameter(s) are shifted by the mode value of the baseline 3$\times$2pt analysis, without any inspection of the mode value. The plotted figures only display the range of the credible interval(s). With this blinding method, we can evaluate how the cosmological parameters change by each analysis setup compared to those of the baseline analysis. Before unblinding the results, the analysis team is not allowed to compare the posterior of cosmological parameter(s) or the best-fit model the predictions to external findings such as the {\it Planck} CMB cosmology.
\end{itemize}

Please refer to Section~II~B of \citet{more2023} for more details regarding the creation of the fake catalogs in a manner that minimizes the possibility of the analysis team inadvertently unblinding the results. Although using these catalogs necessitates the analysis team performing the same analyses three times, this approach avoids the need for a reanalysis upon unblinding the catalogs.

The set of the three shape catalogs used in this paper is shared with the two companion papers, \citet{more2023} and \citet{sugiyama2023}. The first of these presents the details of the measurements of clustering observables used in this paper. The companion paper \citet{sugiyama2023} presents the cosmological parameter estimation from the same signals as that of this paper, but using the perturbation theory based model template that is compared to the signals on scales with the larger scale cuts of $(8,12)h^{-1}{\rm Mpc}$ for $\wproj$ and $\dSigma$, respectively.

We imposed the following criteria for deciding to unblind our results: 
\begin{itemize}
    \setlength{\itemsep}{0em}
    \item Analysis pipeline codes are made available to collaboration members and some specific members are 
    assigned to review each part of the code.
    \item Various analysis setups, described in Table~\ref{tab:analysis_setups}, are tested to ensure that the cosmological constraints are robust to the different setups. We check whether a significant shift in $S_8$, $>0.5\sigma$, (where $\sigma$ is from the baseline 3$\times$2pt analysis), is found from any of the analysis setups. If such a shift is found, we check whether it is consistent with the statistical scatter by comparing the shift with the distribution of $S_8$ values estimated from a set of hypothetical analyses using 100 realizations of the synthetic noisy data vector. In particular we check how the $S_8$ value changes for different scale cuts as a flag for the assembly bias effect (see Section~\ref{sec:assembly_bias} for detailed discussion).
    \item 
    Internal consistency tests to check whether an estimation of the key cosmological parameter changes, compared to that from the baseline analysis method, using subsets of data vector and/or different analysis methods, where the different methods are based on more flexible models of galaxy-halo connection including the incompleteness and off-centering effects of central galaxies (see Table~\ref{tab:parameters}). 
    \item Quantify the goodness of fit of the best-fit model predictions to the data vector in each of the three blind catalogs.
\end{itemize}

Once the collaboration agrees to unblind the analysis, the analysis-level unblinding is first removed by the analysis team. The final catalog-level unblinding happens a few hours later. The analysis team resolved that the results would be published regardless of the outcome, once the results are unblinded. Furthermore, the analysis method could not be changed or modified after unblinding. Following these agreed rules, we unblinded the results at our regular telecon on Dec 3, 2022 in JST. The figures shown in this paper, unless otherwise noted, were made prior to unblinding, with only the axes changed after unblinding to show true values.

\section{Results}
\label{sec:results}

In this section we show the main results of this paper, the cosmological parameters estimated from 
the joint measurements of $\dSigma(R)$, $\wproj(R)$ and $\xi_{\pm}(\vartheta)$ for the HSC-Y3 and SDSS catalogs. 
All the analyses in this section are done before
unblinding, and the results are presented without any change after unblinding.
\subsection{$\Lambda$CDM Constraints}
\label{sec:lcdm}
\begin{figure*}
    \includegraphics[width=1.8\columnwidth]{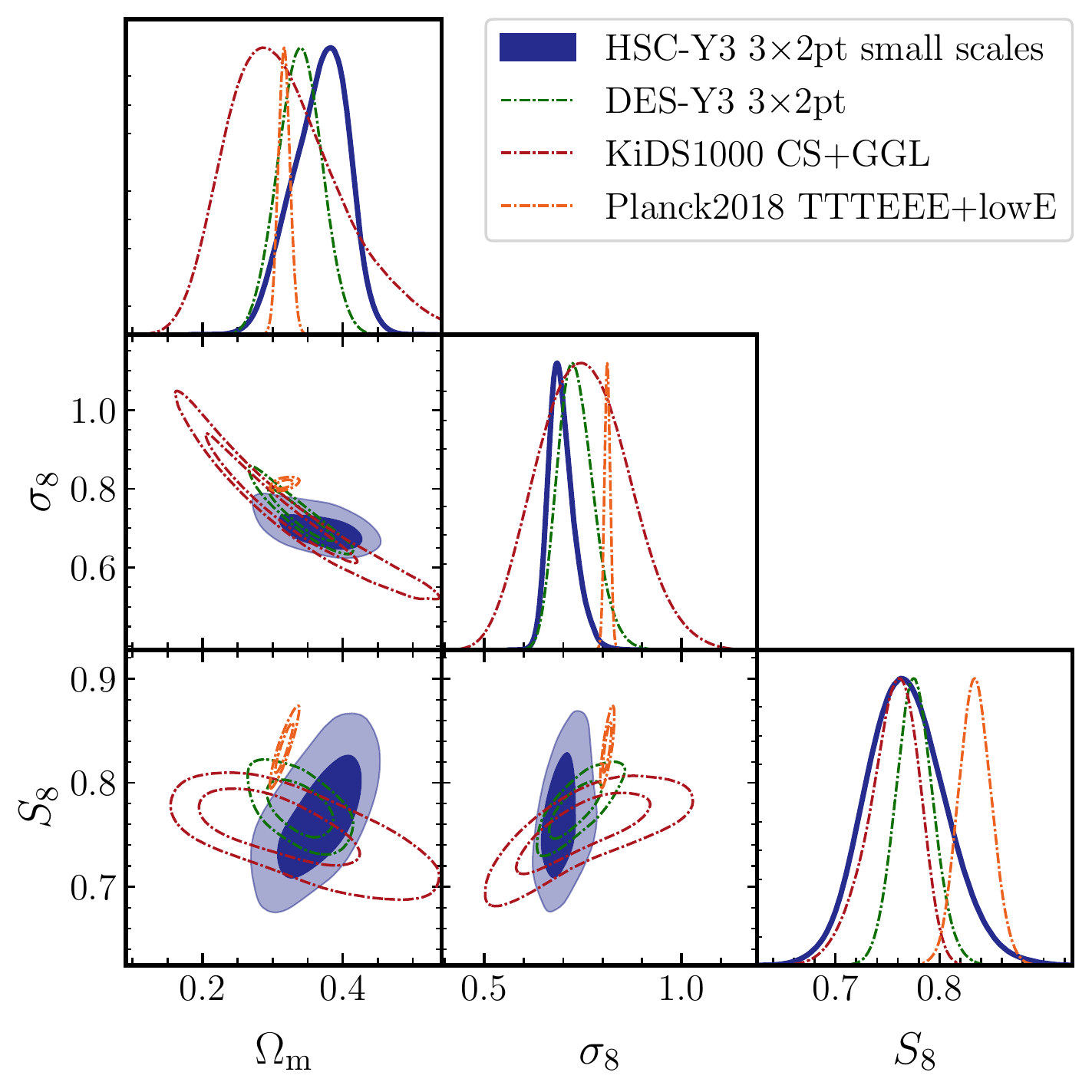}
    \caption{The 1d and 2d posterior distributions in the sub-space of $S_8$, $\sigma_8$ and $\Omega_{\rm m}$ for the flat $\Lambda$CDM cosmology. The blue dark (light) shaded regions denote the 68\% (95\%) credible interval, respectively, for our HSC-Y3 3$\times$2pt baseline analysis in Table~\ref{tab:analysis_setups}. For comparison we show the results for other recent cosmological analyses. The red contours are from the DES-Y3 3$\times$2pt analysis \citep{DES-Y3}. The blue contours are from the KiDS-1000 analysis \citep{2020arXiv200701844J} with cosmic shear (``CS'') and galaxy-galaxy weak lensing (``GGL'') (see text for details). The green contours are the {\it Planck}~2018 results using the primary CMB anisotropy information (``TT,TE,EE+lowE'') \citep{2020A&A...641A...6P}. Note that the degeneracy direction of the HSC-Y3 result in each 2d subspace of the parameters are different from those of DES-Y3 and KiDS-1000, since the relative constraining powers of the cosmological parameters for different observables are different.}
    \label{fig:contour-main}
\end{figure*}
\begin{figure*}
    \includegraphics[width=2\columnwidth]{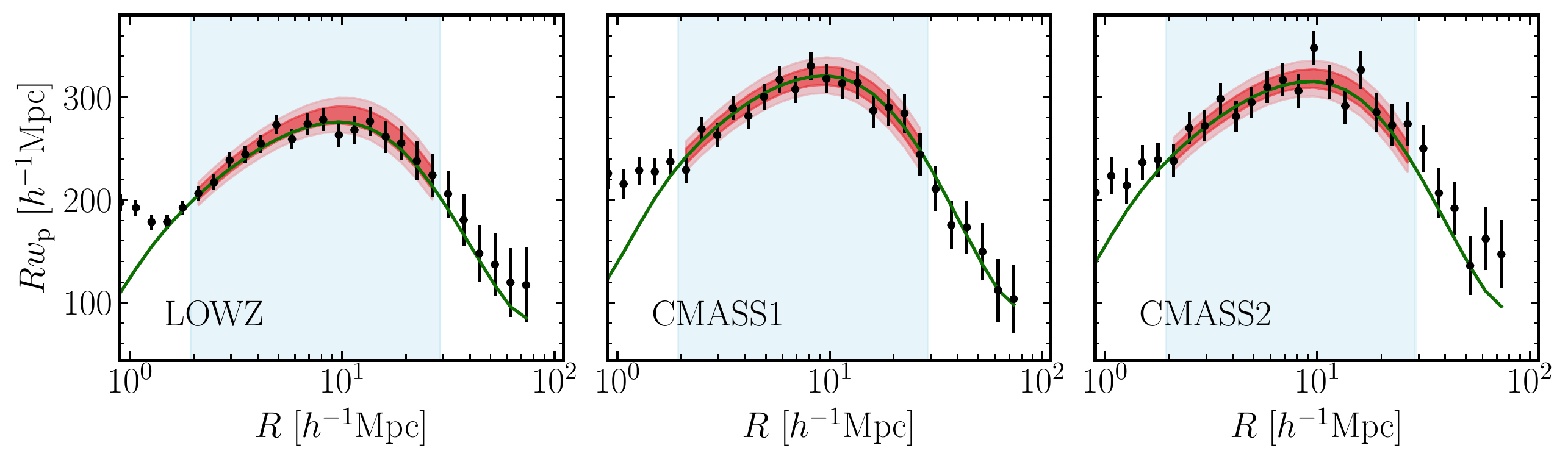}
    \includegraphics[width=2\columnwidth]{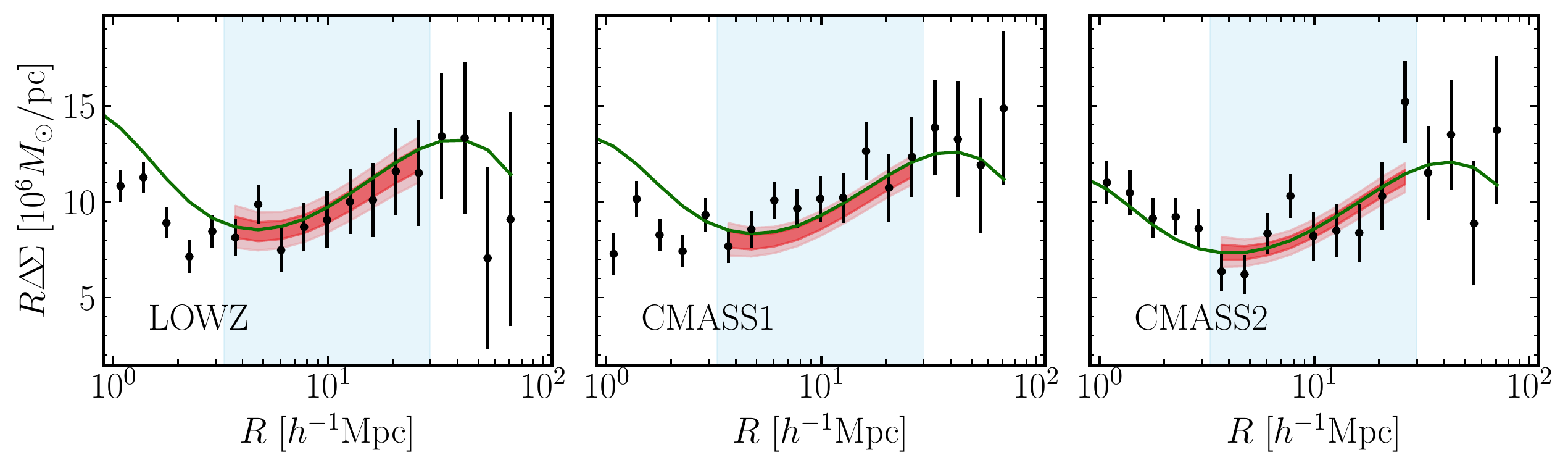}
    \includegraphics[width=1.4\columnwidth]{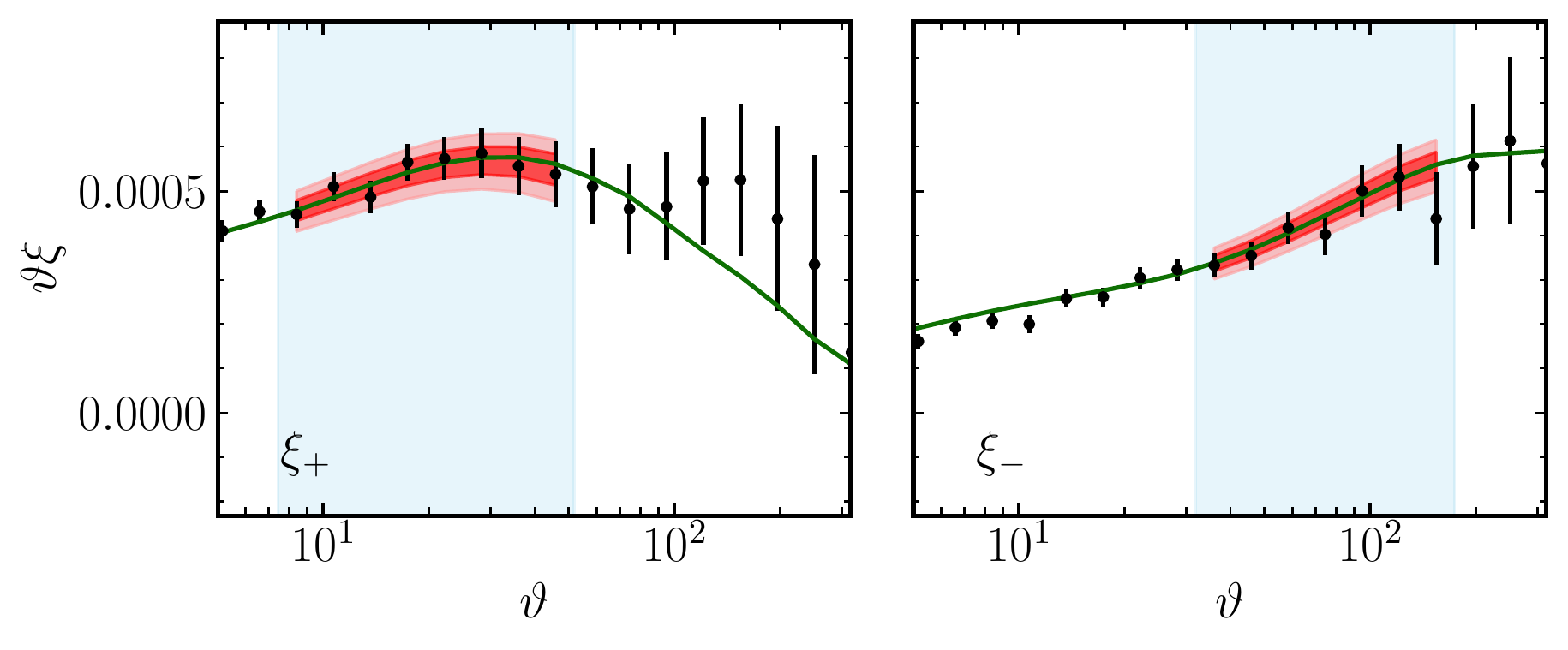}
    \caption{The green solid line in each panel denotes the model prediction at the {\it maximum a posteriori} (MAP) for the baseline analysis in Fig.~\ref{fig:contour-main}, while the data points with error bars are the measured signals. The upper-row three panels are for the projected correlation functions of galaxies, $\wproj(R)$, for the LOWZ, CMASS1 and CMASS2 samples in the redshift ranges $z=[0.15,0.35]$, $[0.43,0.55]$ and $[0.55,0.70]$, respectively.The middle-row three panels are for the galaxy-galaxy weak lensing using the HSC galaxies as source sample, $\dSigma(R)$, for the same LOWZ, CMASS1, and CMASS2 samples as lens samples, respectively. The bottom-row two panels are for the cosmic shear correlation functions, $\xi_{\pm}(\vartheta)$. For illustration purpose, we show $R\times \wproj(R)$, $R\times \dSigma(R)$ and $\vartheta\times \xi_{\pm}(\vartheta)$. The red shaded regions around the green line denote the 68\% and 95\% credible intervals of the model predictions in each separation bin, which are computed from the posterior distributions in the Bayesian cosmology inference. Note that the errors are computed from the diagonal components of the covariance matrix. The blue-color shaded region in each panel denotes 
    the range of projected or angular separation bins that is used for the cosmology analysis. }
    \label{fig:signal-fitting}
\end{figure*}
\begin{figure}
    \includegraphics[width=\columnwidth]{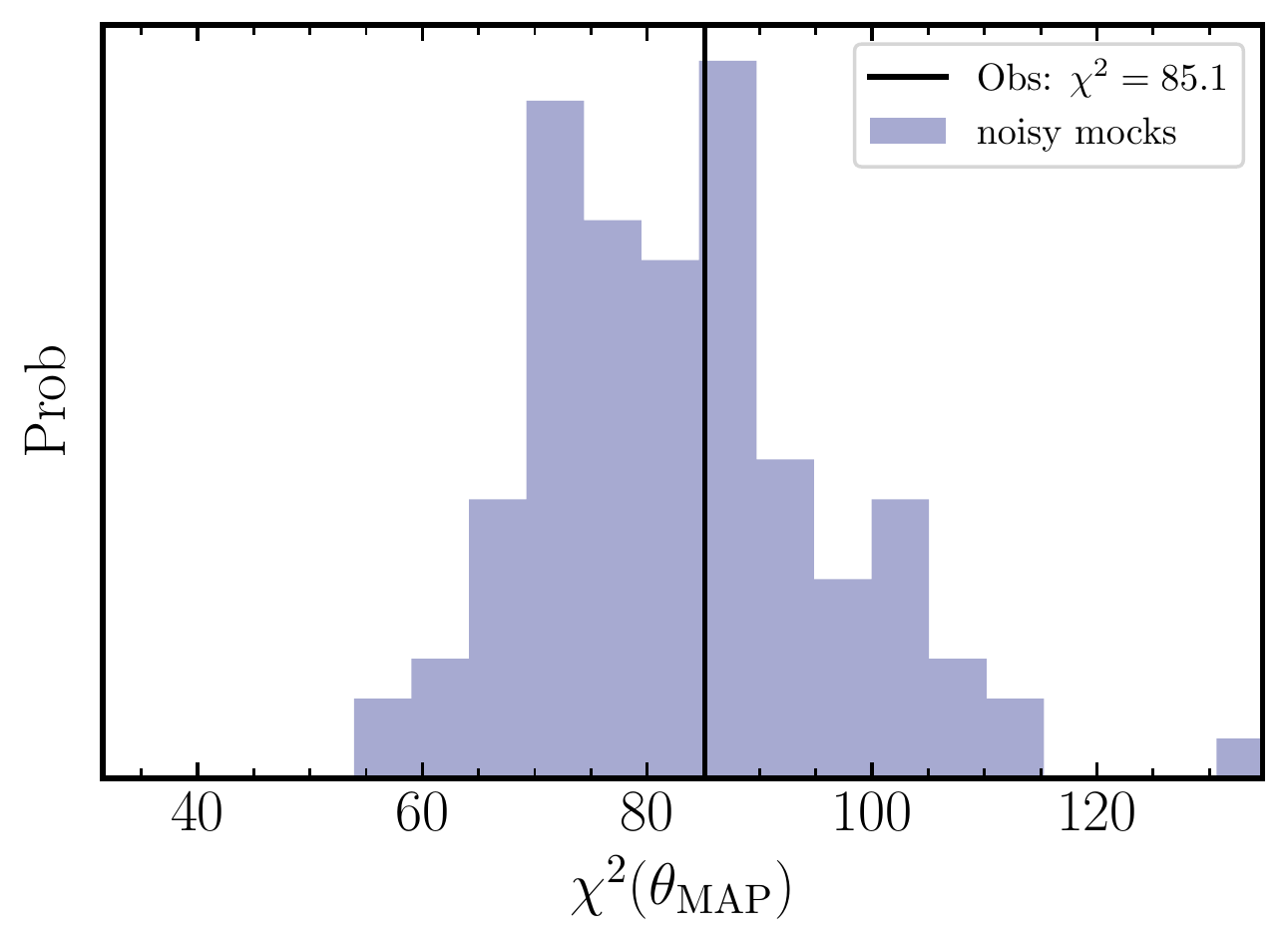}
    \caption{An evaluation of the goodness-of-fit of the best-fit (MAP) model in Fig.~\ref{fig:contour-main}. The histogram shows the distribution of the $\chi^2$ values at the MAP model, obtained by applying the same baseline analysis to 100 noisy mock datasets (see text for details). The vertical black line denotes the measured $\chi^2$-value ($\chi^2=85.1$)  at MAP for the actual analysis of HSC-Y3 and SDSS data. The probability of finding the $\chi^2$ value larger than the observed value ($p$ value)
    is about 41\%.}
    \label{fig:goodness-of-fit}
\end{figure}
The shaded contours in 
Fig.~\ref{fig:contour-main} are the 1d and 2d posterior distributions of the key parameters, $S_8$, $\sigma_8$ and $\Omega_{\rm m}$ for flat $\Lambda$CDM model, obtained from the baseline 3$\times$2pt analysis setup of the HSC-Y3 data as given in Table~\ref{tab:analysis_setups}. The central value and credible interval for each parameter are given as 
\begin{align}
\input{3x2pt_result_b2_dempz.tex},
\label{eq:cosmological_parameters_baseline}
\end{align}
where the number in parentheses denotes the value for the MAP ({\it maximum a posteriori}) model in the sampled chains. The HSC-Y3 3$\times$2pt analysis achieves about 5\% fractional accuracy in the $S_8$ parameter. 

Fig.~\ref{fig:signal-fitting} shows that the best-fit (MAP) model fits all the measured quantities simultaneously over the range of radial or angular separations that are used in the cosmology analysis. We note that our cosmology analysis does not include the information in the deeply nonlinear regime such as the 1-halo term regime, e.g. $R\lesssim \mbox{a few}~h^{-1}$Mpc corresponding to the virial radii of massive halos. Nevertheless we stress that the use of {\tt Dark Emulator} is critical to accurately model the clustering observables on scales in the mildly nonlinear regime. If we use the simpler perturbation-theory based model (\citet{sugiyama2023}), it cannot describe the signals well in the range of scales we use in this paper \citep[also see][]{Sugiyama:2022}. 

In Fig~\ref{fig:goodness-of-fit} we evaluate the goodness-of-fit of the best-fit  model to the measured signal. To do this evaluation, we generate 100 realizations of noisy mock data vectors using the ``full'' covariance matrix; the full covariance includes the elements in radial or angular bins outside those used in our cosmology analysis and the cross-covariance terms that describe correlated scatter between the observables, i.e., the galaxy-galaxy weak lensing signals for the different lens subsamples and cosmic shear as shown in Fig.~6 in \citet{more2023}. The histogram in Fig.~\ref{fig:goodness-of-fit} shows the distribution of the $\chi^2$-value of the MAP model prediction for each of the 100 realizations. We find that the $\chi^2$ values tend to exceed that expected from the degrees of freedom, $\nu=90-28=62$ (see Table~\ref{tab:analysis_setups}). We ascribe this excess to severe parameter degeneracies; some of the model parameters, especially the HOD parameters, are not well constrained by the data vector. The histogram can be compared to the $\chi^2$ value of the actual HSC-Y3 and SDSS analysis (solid black line), showing that the observed $\chi^2$ value is near the middle of the distribution. Hence, we conclude that the best-fit model is quite acceptable.

\subsection{Internal Consistency Tests}
\label{sec:consistency_test_main}
\begin{figure*}
    \includegraphics[width=2\columnwidth]{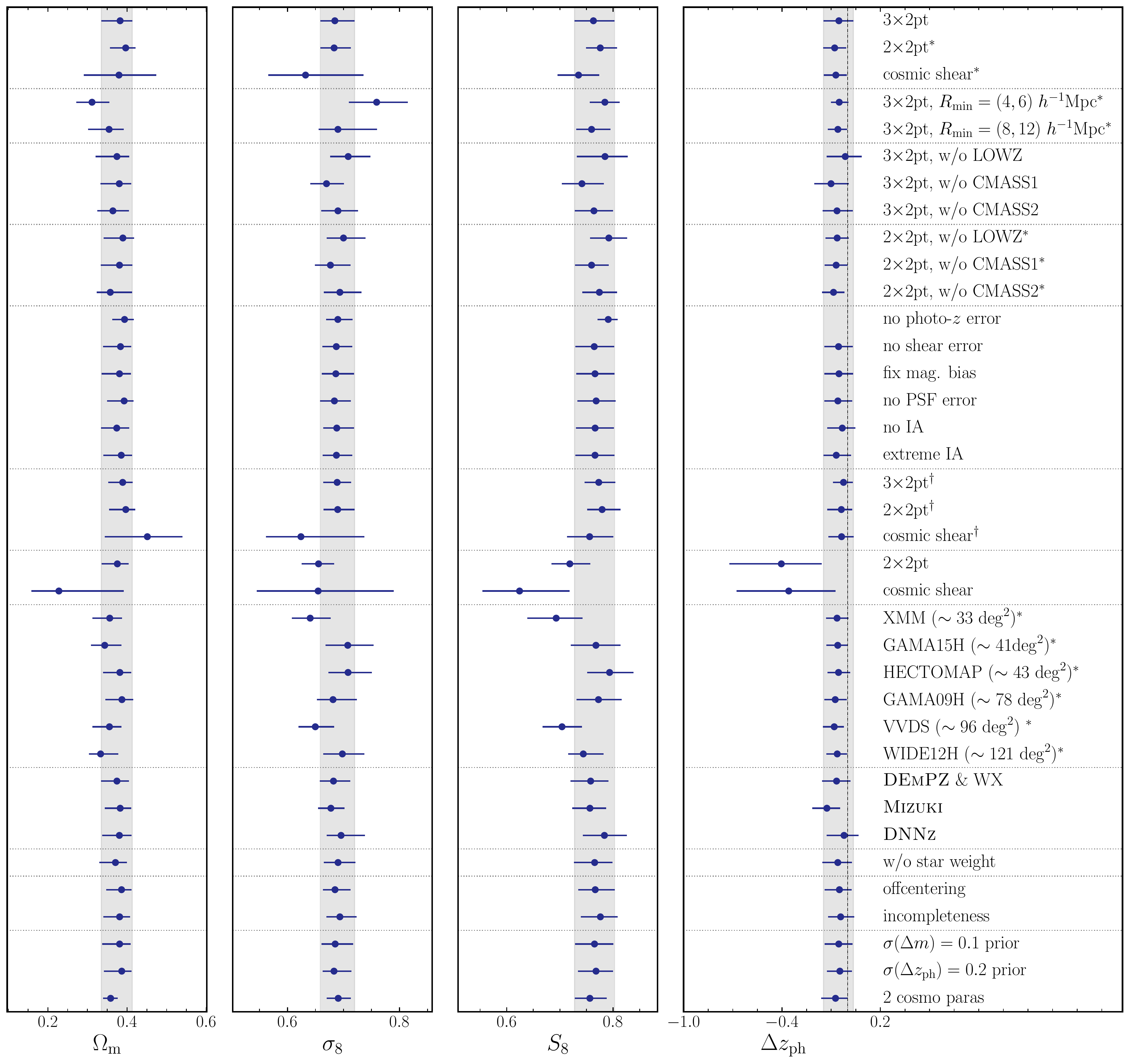}
    \caption{A summary of the cosmological parameters and the residual photo-$z$ error parameter     $\Delta z_{\rm ph}$, estimated from each of the different analysis setups in Table~\ref{tab:analysis_setups}. The vertical dashed line in the panel of $\Delta z_{\rm ph}$ denotes $\Delta z_{\rm ph}=0$, i.e. the case of no residual photo-$z$ error or equivalently the case that the mean redshift estimate of HSC source galaxies inferred from the photo-$z$ estimates is perfect. As our default choice, we employ the uninformative flat prior on the residual photo-$z$ error, $\Pi(\Delta z_{\rm ph})={\cal U}(-1,1)$. As explained in Table~\ref{tab:analysis_setups}, the analysis with superscript ``$^{}\ast$'' denotes an analysis using the informative Gaussian prior on the photo-$z$ error, given by ${\cal N}=(-0.06,0.08)$, which is inferred from the posterior of the baseline 3$\times$2pt analysis. The analysis with superscript ``$^\dagger$'' denotes an analysis using the Gaussian prior on $\deltapz$ with mean around $\deltapz=0$, given by $\Pi(\deltapz)={\cal N}(0,0.1)$. The analysis with label ``$\sigma(\deltapz)=0.2$ prior'' denotes the result 
    using the Gaussian prior, $\Pi(\deltapz)={\cal N}(0,0.2)$.}
    \label{fig:summary-real-data}
\end{figure*}
\begin{figure}
    \includegraphics[width=\columnwidth]{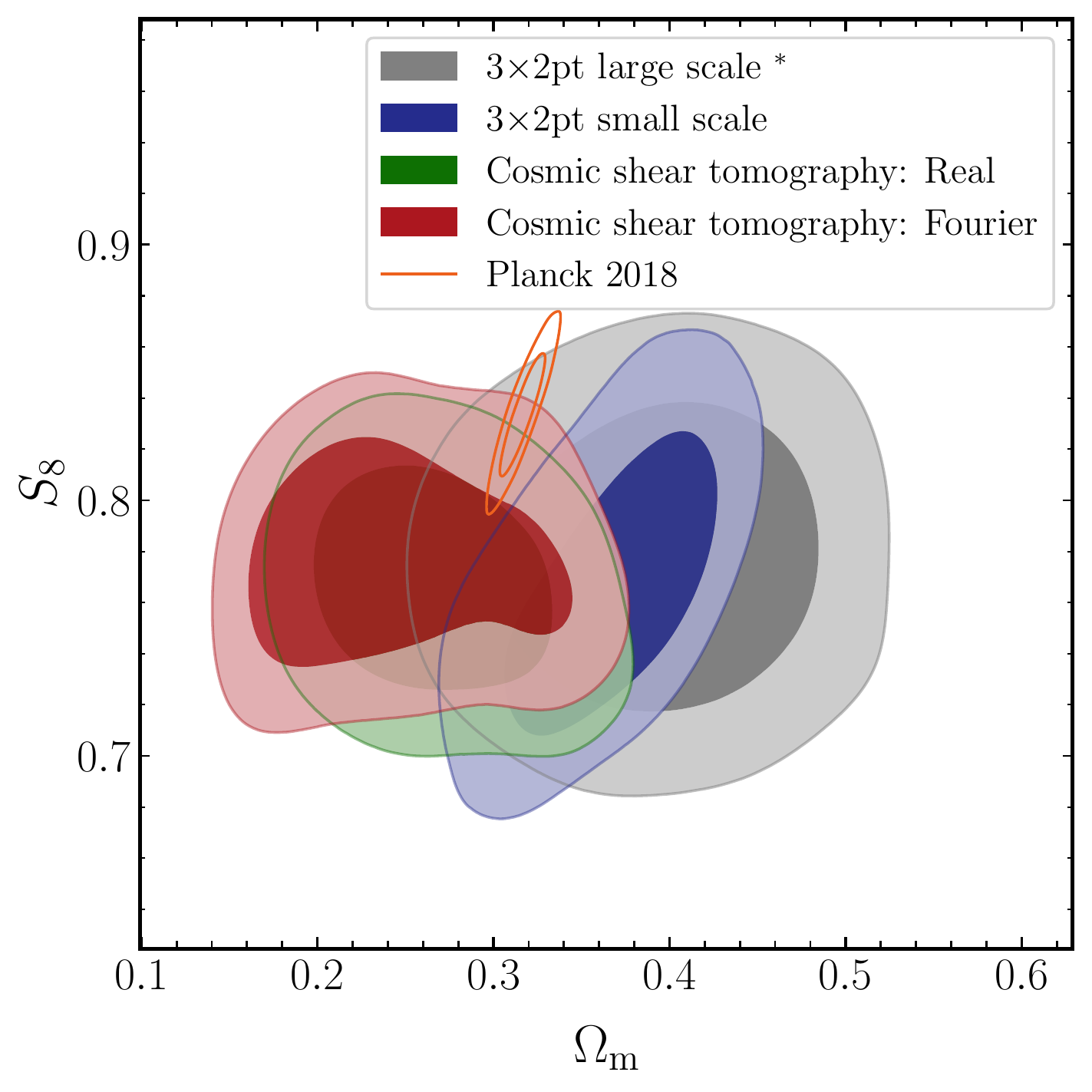}
    \caption{Comparison of the cosmological constraints from the different cosmological analyses of the HSC-Y3 data. The result labeled as ``3$\times$2pt large scale'' uses the same data vector as that in this paper, but uses the perturbation theory based model template when comparing the predictions to the measurements on large scales (\citet{sugiyama2023}). ``Cosmic shear tomography: Real'' is the result from the cosmic shear tomography analysis using the real-space cosmic shear correlation functions with 4 tomographic redshift bins (\citet{li2023}). ``Cosmic shear tomography: Fourier'' is from the cosmic shear tomography using the cosmic shear power spectra (\citet{dalal2023}). The two cosmic shear results use the different scale cuts. 
    \label{fig:internal_HSCy3}
    }
\end{figure}
As self-consistency tests, we performed the cosmological parameter estimation for each of the different setups listed in Table~\ref{tab:analysis_setups}, before unblinding. The cosmological constraints for each setup are shown in Fig.~\ref{fig:summary-real-data} and the mode, credible interval, and MAP value of each parameter are summarized in Table~\ref{tab:internal_consistency} of Appendix~\ref{sec:consistency_tests}. We find that the $S_8$ parameter is robust to these different tests, changing by $<1\sigma$ in each case. The exception is the test in which the residual photo-$z$ error parameter is fixed to $\Delta z_{\rm ph}=0$, i.e. the case in which the mean redshift of HSC source galaxies is assumed to be perfectly estimated based on their photo-$z$'s. The test using the fixed $\Delta z_{\rm ph}=0$ gives a smaller error bar in $S_8$ and gives a sizable shift in the central value of $S_8$ ($1.5\sigma$ where $\sigma$ is taken from the test with fixed $\Delta z_{\rm ph}$). This indicates the existence of a non-zero residual photo-$z$ error as discussed below. Some tests show a scattered shift in the values of $\Omega_{\rm m}$ and $\sigma_8$, but such shifts are also seen in the validation tests using the mock catalogs. We did not find any significant shift of $S_8$ compared to the tests using the mock data, or did not find any evidence of unknown systematic effects in our results. We also note that the shifts in these parameters are likely due to projection effects of the non-Gaussian posterior distribution in the full-dimensional parameter space. In Appendix~\ref{sec:consistency_tests} we give a detailed discussion of the internal consistency tests.  

In Fig.~\ref{fig:internal_HSCy3} we compare the cosmological constraints from the different cosmology analyses using the HSC-Y3 data. Although the result ``3$\times$2pt large-scale'' uses the same data vector as that in this paper, the analysis uses the perturbation theory based theoretical template to compare with the measurements at $R>8$ and $12~h^{-1}$Mpc for $\wproj$ and $\dSigma$, respectively, where the perturbation theory model is valid (\citet{sugiyama2023}) The other two results are from the cosmic shear tomography analyses using the cosmic shear two-point correlations \citep{li2023} and the power spectra \citep{dalal2023}. The three sets of analyses (both 3$\times$2pt analyses, the real-space cosmic shear analysis and the Fourier-space cosmic shear analysis), performed blinded cosmology analyses using different blinded catalogs, and we made this comparison plot after the unblinding. The cosmological results from all the four analyses, especially the $S_8$ results, are in good agreement with each other. This is quite encouraging, because the constraining power of the 3$\times$2pt analyses is mainly from the clustering ($\wproj$) information of SDSS galaxies and the two cosmic shear constraints are sensitive to different scales in the cosmic shear information of HSC-Y3 data due to the different scale cuts in the real- and Fourier-space. We also emphasize that the two cosmic shear analyses adopted the uninformative prior on the residual photo-$z$ error parameters for the two high-redshift tomographic bins, $z_3$ and $z_4$, that correspond to redshifts at $z\gtrsim 0.9$. If the cosmic shear analyses employ informative priors on the photo-$z$ error parameters, the $S_8$ parameter is shifted and the agreement in Fig.~\ref{fig:internal_HSCy3} cannot be realized. Hence all the cosmological analyses indicate a non-zero residual systematic error in the photo-$z$ estimates of such high-redshift HSC galaxies. 

\subsection{Comparison with External Data and $S_8$ Tension}
\label{sec:tension}
In this section, we discuss the comparison of our HSC-Y3 result with external cosmology results (see Fig.~\ref{fig:contour-main}). Note that, as we described in Section~\ref{sec:blinding}, we never compared the HSC-Y3 results with external cosmology results during the blind analysis stage, and made plots like Fig.~\ref{fig:contour-main} only after unblinding. For the CMB constraints, we consider the ``Planck2018'' results \citep{2020A&A...641A...6P}, from the analysis where the primary CMB temperature and $E$-mode polarization anisotropy information (``TT, EE, TE+lowE'') are used and the neutrino mass is fixed to $0.06$~eV as we did in our setup. To obtain the posterior distribution we used the public chain ``base/plikHM\_TTTEEE\_lowl\_lowE'' available from the website\footnote{\url{https://pla.esac.esa.int/pla/aio/product-action?COSMOLOGY.FILE_ID=COM_CosmoParams_fullGrid_R3.01.zip}}. For the posterior distribution of the DES Year 3 (``DES-Y3'') result, we used the public chain\footnote{\url{https://des.ncsa.illinois.edu/releases/y3a2/Y3key-products}}, which is the result obtained from the 3$\times$2pt cosmological analysis using the photometric ``MagLim'' samples for both lens and source galaxies \citep{DES-Y3}. For the ``KiDS-1000'' result, we used the public chain\footnote{\url{ https://kids.strw.leidenuniv.nl/DR4/KiDS-1000_3x2pt_Cosmology.php}} to exhibit the result from a joint analysis of cosmic shear (``CS''in the legend) and galaxy-galaxy (``GGL'') weak lensing in \citet{2020arXiv200701844J}, where the {\it spectroscopic} samples in BOSS and the 2-degree Field Lensing Survey (2dFLenS) were used as the lens samples in the galaxy-galaxy weak lensing measurements. Note that the 3$\times$2pt results from KiDS-1000 include BAO information in the BOSS galaxies that can give a tighter constraint on $\Omega_{\rm m}$, so we instead refer the above $2\times$2pt CS$\times$GGL result as KiDS-1000. For both the DES-Y3 and KiDS-1000 analyses, the weak lensing observables are angular correlation functions -- $\gamma_T(\theta)$ and/or $w(\theta)$ -- rather than $\dSigma(R)$ and $\wproj(R)$.

Fig.~\ref{fig:contour-main} shows that the HSC-Y3 result is generally consistent with the DES-Y3 and KiDS-1000 results within the credible intervals. However, the degeneracy direction of the HSC-Y3 result in each 2d subspace of the parameters are different from those of DES-Y3 and KiDS-1000, as the relative constraining powers of different observables for the cosmological parameters (after marginalizing over other parameters) are different. For our case, the galaxy-galaxy clustering of BOSS galaxies has the most constraining power, after lifting the parameter degeneracies between the galaxy bias and the cosmological parameters with the weak lensing information as we will discuss below. 

Extending the definition of $S_8$ to $S'_8\equiv \sigma_8(\Omega_{\rm m}/0.3)^\alpha$ with $\alpha$ being a free parameter, we find that the best-constrained parameter is $\alpha\simeq 0.22$: with this value, we find $S_8'\simeq 0.721\pm0.0279$, the fractional precision is 4\%. This precision is 30\% smaller than the $\sigma(S_8)\simeq 0.040$ for  the standard $S_8$ in Eq.~(\ref{eq:cosmological_parameters_baseline}). If we compare the width of the 2d contour along the narrowest direction in the ($\Omega_{\rm m},S_8$) sub-space in Fig.~\ref{fig:contour-main}, the HSC-Y3 result is comparable with the KiDS-1000 result, but is somewhat larger than the DES-Y3 result. This is partly due to our use of an uninformative prior on the residual photo-$z$ error parameter ($\Delta z_{\rm ph}$). If we employ a tighter prior on $\Delta z_{\rm ph}$ such as the prior width inferred from the original photo-$z$ estimate, $\sigma(\Delta z_{\rm ph})\simeq O(10^{-2})$, we can obtain a tighter credible interval, however, the central value of $S_8$ shows a non-negligible shift. Hence $\Delta z_{\rm ph}$ is a key parameter in our analysis to obtain a robust estimate of the cosmological parameters, and we decided to adopt the uninformative prior of $\Delta z_{\rm ph}$ during the blind analysis stage before revealing the central value of $S_8$. In Section~\ref{sec:photoz_bias}, we will give a more detailed discussion of how different treatments of the residual photo-$z$ error, e.g. informative vs. uninformative prior, alter our cosmological constraints. 

Fig.~\ref{fig:contour-main} displays a $2\sigma$-level tension between the HSC-Y3 3$\times$2pt result and the {\it Planck}~2018 result. To quantify the possible tension, we use the methods developed in \citet{Park:2020} and \citet{2020PhRvD.101j3527R}, which are called {\tt eigentension} and {\tt tensiometer}, respectively \citep[also see][]{2021MNRAS.505.6179L}.

For the {\tt eigentension} method, we start by diagonalizing the covariance matrix of cosmological parameters to find the eigenvectors and eigenvalues. Among the 5 cosmological parameters in our 3$\times$2pt analysis $n_{\rm s}$ and $\omega_{\rm b}$ are prior dominated, so we focus on the parameters, $\sigma_8$, $\Omega_{\rm m}$, and $\omega_{\rm c}$. When we diagonalize the covariance matrix of these parameters, obtained from the chains in our baseline analysis, we find the two eigenvectors, $(e_0, e_1)=(\sigma_8\ \omega_c^{0.19}\ \Omega_{\rm m}^{0.38}, \omega_c\ \Omega_{\rm m}^{0.54}\ \sigma_8^{-0.40})$, are well constrained by the HSC-Y3 3$\times$2pt observables compared to the prior widths, while the third eigenvector is prior-dominated. If we compute the posterior distribution of the eigenvector differences, defined as ($\Delta e_0, \Delta e_1 ) \equiv (e_0,e_1)_{\rm HSC-Y3}-(e_0,e_1)_{\rm Planck} $, from the two chains of the HSC-Y3 baseline analysis and {\it Planck}~2018, we find that the point where the two data sets are consistent with each other, i.e.,  $(\Delta e_0,\Delta e_1)=(0,0)$, is located at $\sim 2.5\sigma$ in the posterior. Note that this method allows us to compute the posterior distributions of these parameter differences from the existing chains of HSC-Y3 and {\it Planck}, as long as the two datasets are independent \cite{2017PhRvD..95l3535C}. Thus, we conclude that the HSC-Y3 result displays a $2.5\sigma$ tension with the {\it Planck}~2018 result. 

To implement the {\tt tensiometer} method, we use the publicly-released code\footnote{\url{https://github.com/mraveri/tensiometer}}. This code allows us to generate the posterior distribution of the three parameter differences, $(\Delta \sigma_8, \Delta \Omega_{\rm m},\Delta \omega_{\rm c})$, from the two chains of the HSC-Y3 baseline analysis and {\it Planck}~2018 using machine learning modeling of the posterior distribution with normalizing flows. It then quantifies a disagreement significance in the full 3d parameter space: we find a $2.7\sigma$ tension between the HSC-Y3 and the {\it Planck}~2018 results, in close agreement with the estimate from {\tt eigentension} above. 

Hence, we conclude that the HSC-Y3 3$\times$2pt result has about 
$2.5\sigma$ tension with the {\it Planck}~2018 constraints within the flat-geometry 
$\Lambda$CDM framework. 

\section{Discussion}
\label{sec:discussion}

\subsection{An Implication of Residual Systematic Photo-$z$ Error}
\begin{figure*}
    \includegraphics[width=\columnwidth]{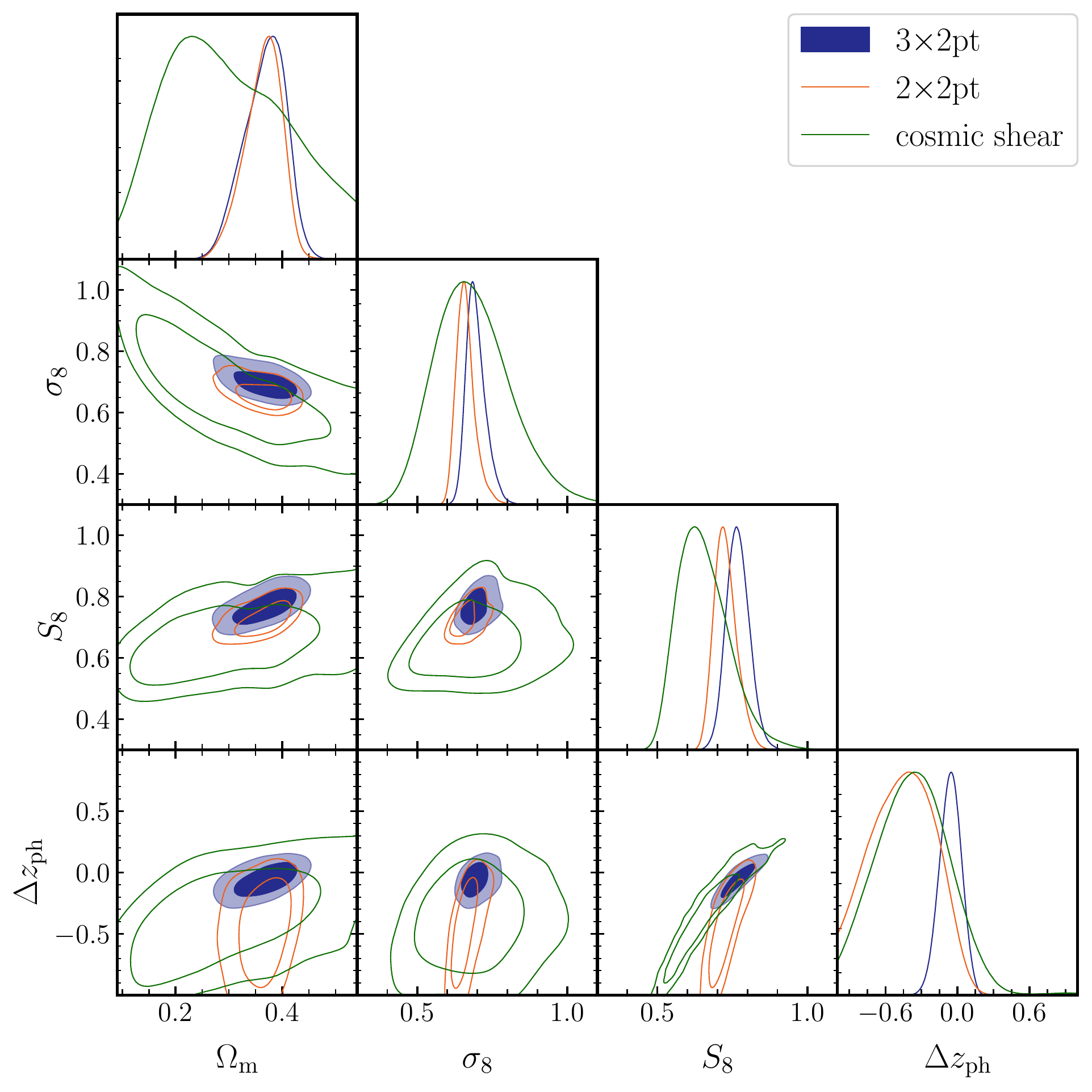}
    \includegraphics[width=\columnwidth]{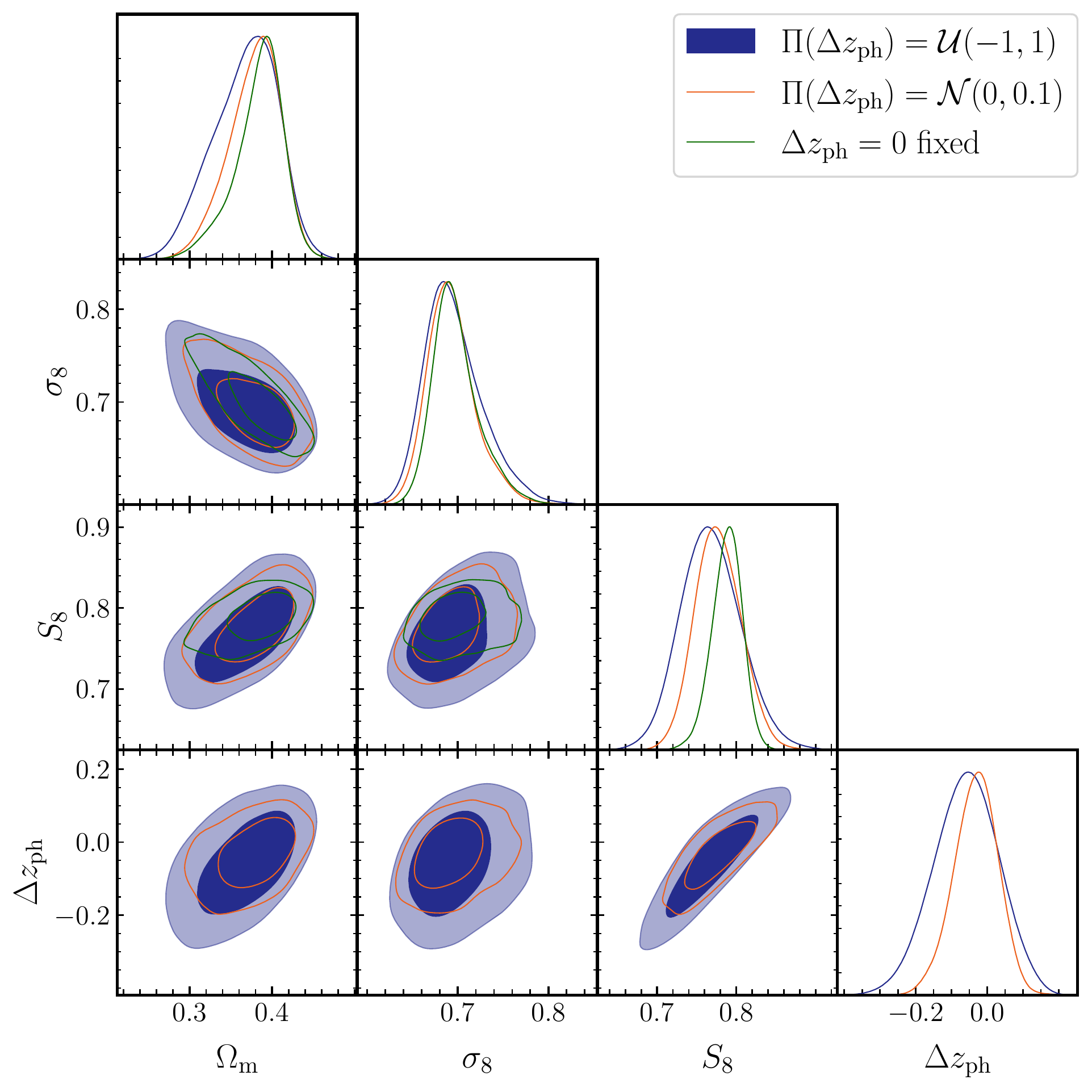}
    \caption{
    The importance of the uniformative prior on the residual photo-$z$ error parameter ($\Delta z_{\rm ph}$) in the cosmology analysis. {\it Left panel}: The 1d and 2d posterior distributions obtained using the different observables: the baseline (3$\times$2pt), the $2\times 2$pt ($\dSigma\times \wproj$), and the cosmic shear correlations. For all the analyses, we employ the flat prior of $\Delta z_{\rm ph}$: ${\cal U}(-1,1)$ as our baseline analysis. {\it Right}: The posterior distributions for the 3$\times$2pt analyses when using the different priors of $\Delta z_{\rm ph}$: the baseline analysis (the flat prior), the Gaussian prior of ${\cal N}(0,0.1)$ and the case fixing $\Delta z_{\rm ph}=0$, respectively. The result of $\Delta z_{\rm p}=0$ approximately corresponds to the case where the redshift distribution of HSC source galaxies is as inferred from the photo-$z$ estimate, because the prior is as informative as $\sigma(\deltapz)\sim 10^{-2}$. The different treatments of $\deltapz$ affect the mode value and the size of the credible interval of the cosmological parameters.}
    \label{fig:photoz_bias}
\end{figure*}
\label{sec:photoz_bias}

A notable aspect of this study, compared to other weak lensing cosmology analyses, is that we estimate the cosmological parameters employing an {\it uninformative} prior on the residual photo-$z$ error parameter of source galaxies; $\Delta z_{\rm ph}: {\cal U}(-1,1)$ (see Table~\ref{tab:parameters}). This is a conservative setup which is equivalent to the case in which we do not adopt any prior knowledge about the mean redshift of HSC source galaxies. In this section, we show how the self-calibration of the photo-$z$ error parameter is achieved by our method, and also study how the cosmological parameters change when using different priors on $\Delta z_{\rm ph}$. 

The left panel of Fig.~\ref{fig:photoz_bias} shows how the residual photo-$z$ error parameter $\Delta z_{\rm ph}$ is calibrated by combining the different observables. For the $2\times$2pt ($\dSigma\times\wproj$) analysis and the cosmic shear, we also employ the flat prior ${\cal U}(-1,1)$ for $\Delta z_{\rm ph}$. The figure shows that the different observables are complementary to each other and $\Delta z_{\rm ph}$ is accurately estimated by combining the three observables. The self-calibration of $\Delta z_{\rm ph}$ is achieved by comparing the galaxy-galaxy weak lensing signals for the three lens subsamples (LOWZ, CMASS1 and CMASS2) at the three different spectroscopic redshifts and the cosmic shear signals for the same source galaxies. The baseline 3$\times$2pt analysis suggests a non-zero value of $\Delta z_{\rm ph}=-0.05\pm0.09$, i.e. a $1\sigma$-level hint of a non-zero residual systematic error in the mean source redshift. This indicates that the true mean  redshift of HSC source galaxies is higher than the photo-$z$ estimate by $|\Delta z_{\rm ph}|=0.05$. Furthermore, if we focus on the $\Omega_{\rm m}=0.3$ cross section of the posterior, where  $\Omega_{\rm m}\simeq 0.3$ is indicated by measurements of BAO or the galaxy clustering  \citep[e.g.][]{Kobayashi:2021}, one fits an even larger bias, $\Delta z_{\rm ph}\sim -0.2$, as we will below study in more detail. This case also implies a lower value of $S_8$, since $\Delta z_{\rm ph}$ and $S_8$ are positively correlated. Hence, employing an uninformative flat prior of $\Delta z_{\rm ph}$ is important to obtain an unbiased estimate of $S_8$, if the non-zero $\Delta z_{\rm ph}$ is genuine. Encouragingly, a similar residual photo-$z$ error for HSC source galaxies at high redshifts is also implied by the real- and Fourier-space cosmology analyses of HSC-Y3 cosmic shear tomography in \citet{li2023} and \citet{dalal2023}, respectively. For these cosmology analyses, the photo-$z$ error parameters for the two high-$z$ bins (corresponding roughly to our source galaxy sample) are calibrated by the cosmic shear signals relative to those in the lower redshift bins that are more reliably estimated by the photo-$z$ and the cross-correlation method \citep{Rau:2022wrq}. Thus the photo-$z$ error calibration by the cosmic shear methods is somewhat independent from the calibration of the 3$\times$2pt method in this paper.

\begin{figure}
    \includegraphics[width=\columnwidth]{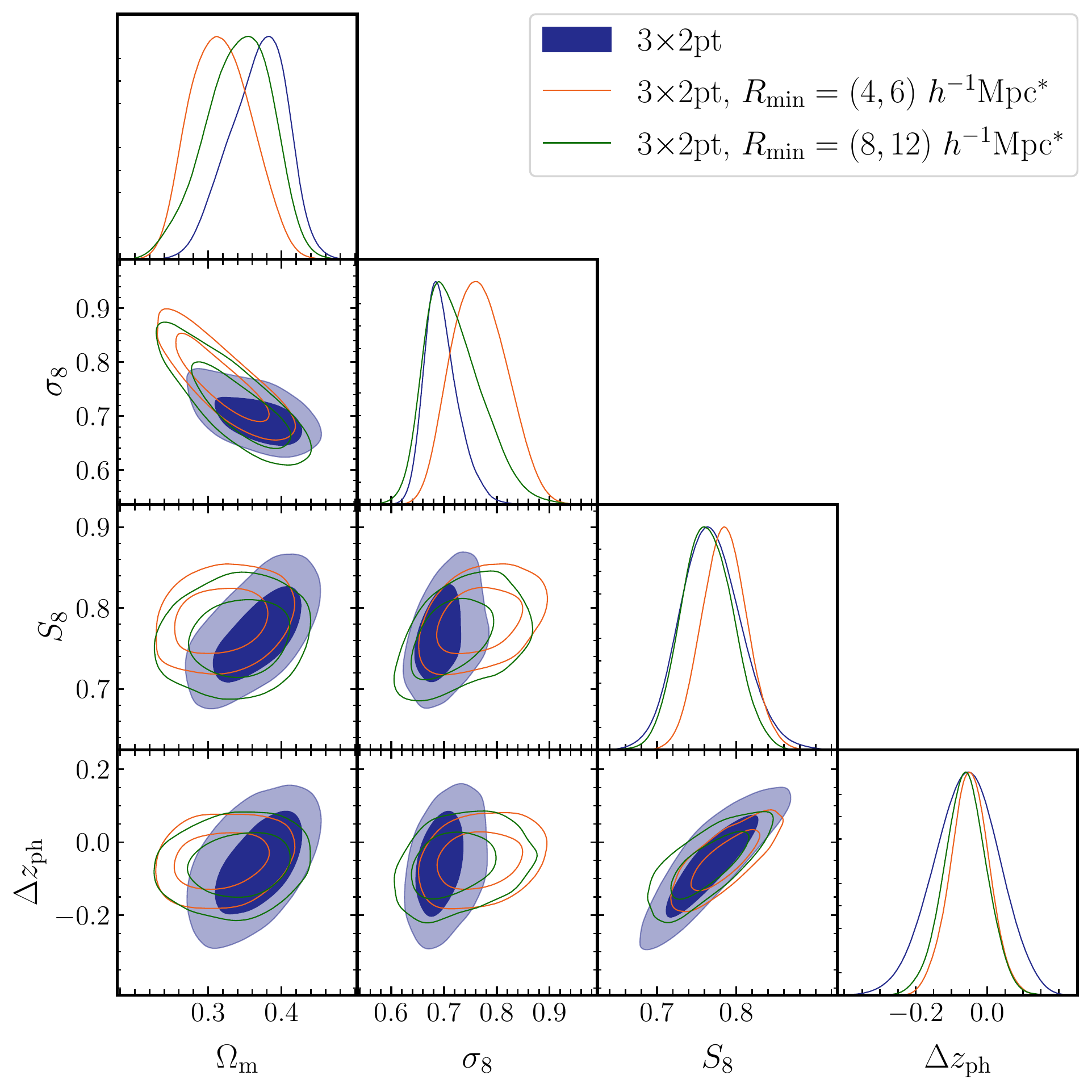}
    \caption{The posterior distributions of the cosmological parameters when using different scale cuts of $R=[4,6]$ or $[8,12]~h^{-1}$Mpc for the $\wproj$ and $\dSigma$ signals, respectively, in the cosmology analysis. The gray contours are the same as those in Fig.~\ref{fig:contour-main}. Note that we used the Gaussian prior of the residual photo-$z$ error parameter, ${\cal N}(-0.06,0.08)$, that is inferred from the baseline 3$\times$2pt analysis and the results include the cosmic shear information too, where we used the same range of the angular separations as in the baseline analysis.}
    \label{fig:contour-scalecuts}
\end{figure}

\subsection{Assembly Bias}
\label{sec:assembly_bias}
\begin{figure}
    \includegraphics[width=\columnwidth]{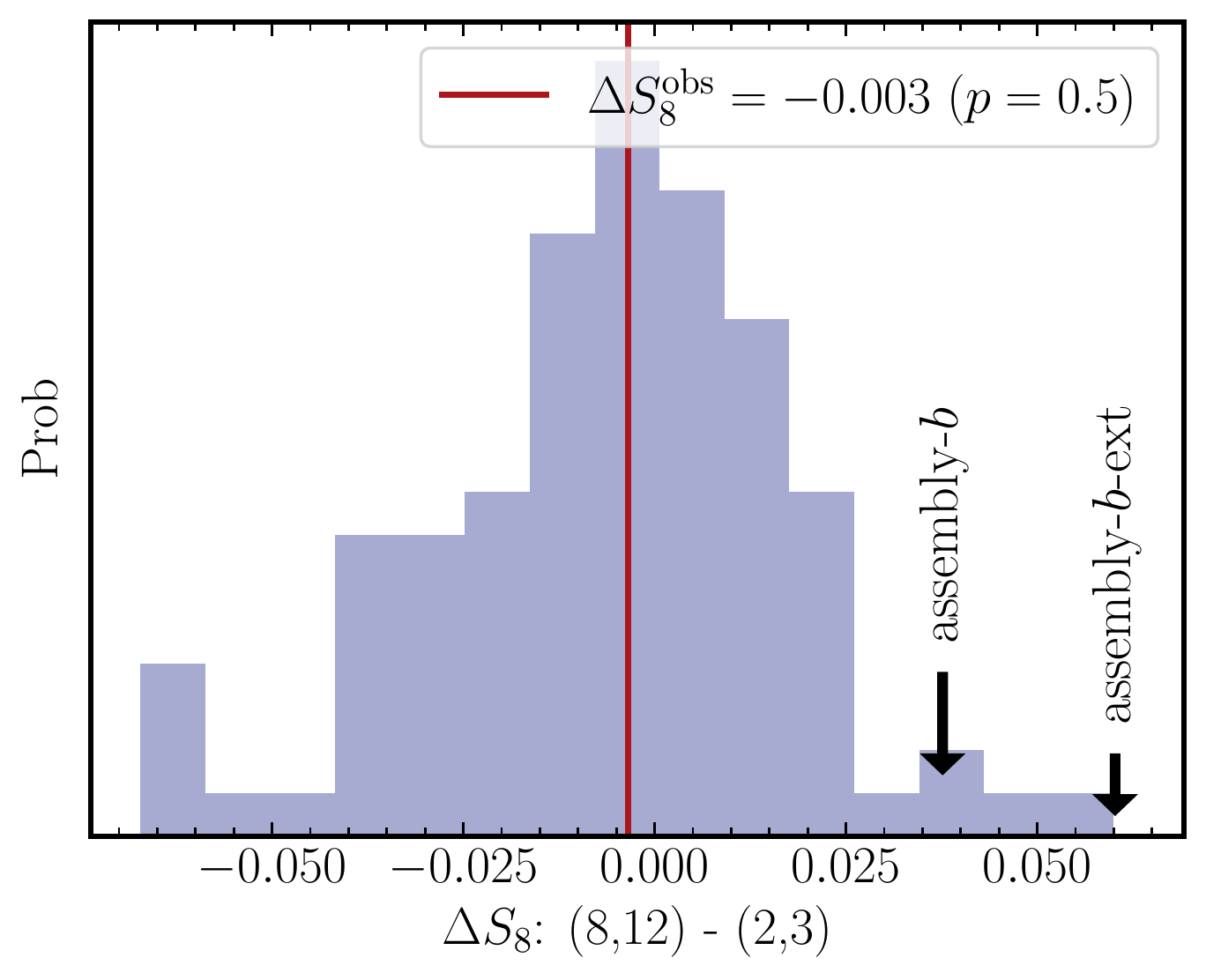}
    \caption{The shaded histogram shows the expected distribution of the differences between the $S_8$ values obtained from the 3$\times$2pt analyses using the different scale cuts of $(2,3)$ and $(8,12)h^{-1}$Mpc for $\wproj$ and $\dSigma$, respectively, assuming that  $\wproj$ and $\dSigma$ are not contaminated by the assembly bias effect. To obtain the distribution, we perform the same cosmology analysis on each of the 100 realizations of the noisy mock data vector. Note that in this inference simulation, we employ the prior on the residual photo-$z$ error, $\Delta z_{\rm ph}$ obtained from the fiducial analysis of (2,3)$h^{-1}$Mpc scale cut to each realization for the (8,12)-analyses, as we did for the actual analysis. When there is no assembly bias effect, the $S_8$ values from the (2,3)- and (8,12)-scale cuts should be consistent with each other, and the actual observed difference of $S_8$, as denoted by the vertical solid line, is consistent with the distribution from the synthetic data vector.     The probability of finding $\Delta S_8$ larger than the observed value ($p$-value) is about 50\%. The two arrows indicated by ``assembly-$b$'' and ``assembly-$b$-ext'' denote the expected difference values of $S_8$ obtained from the simulated synthetic data, where assembly bias effects with different amplitudes are included.}     \label{fig:S8_scalecut_diff} 
\end{figure}
One concern in the halo model based analysis is the effect of possible assembly bias of the SDSS galaxies on the cosmological parameters \citep{Miyatake:2022a,Yuetal:12}. Even if we use the $\wproj$ and $\dSigma$ information on scales greater than the size of most massive halos, $R\gtrsim 2$ and $3\,h^{-1}{\rm Mpc}$, respectively, the galaxy-galaxy lensing ($\dSigma$) contains information on the interior mass of halos hosting the SDSS galaxies, which in turn lifts degeneracies in the galaxy-halo connection in the clustering amplitudes of the 2-halo term regime. If the SDSS galaxies are affected by assembly bias, it could cause a bias in the cosmological parameters, because of a breakdown in the {\it simple} galaxy-halo connection as a function of halo mass. To test the impact of possible assembly bias, we perform the parameter estimation using the different scale cuts, $R=[4,6]$ or $[8,12]~h^{-1}$Mpc, respectively. On sufficiently large scales, galaxy clustering properties are governed by gravity, and the correlation coefficient function of galaxy clustering approaches to the simple relation irrespective of galaxy types including a galaxy sample with assembly bias, given by $\xi_{\rm gm}/[\xi_{\rm mm}\xi_{\rm gg}]^{1/2}\simeq 1$ \citep[see Fig.~6 in Ref.][for the results using the Illustris hydrodynamical simulations]{Hadzhiyska:2021} \citep[also see][]{2018arXiv181109504N,Miyatake:2022a}. As demonstrated in \citet{Miyatake:2022a}, if we adopt the large scale cuts of $[8,12]~h^{-1}$Mpc, the clustering signals are safely in the 2-halo term regime, and the cosmology analysis can recover the cosmological parameters even if the assembly bias effect exists. 

Fig.~\ref{fig:contour-scalecuts} shows the posterior distribution when using the different scale cuts, $R=[4,6]$ or $[8,12]~h^{-1}$Mpc for the $\wproj$ and $\dSigma$ signals, respectively, in the cosmology analysis. Even if we use the largest scale cut $R=[8,12]~h^{-1}$Mpc, the cosmological parameters are almost unchanged. In fact, the cosmological parameters for the baseline analysis are also consistent with the results obtained using the perturbation theory based method in \citet{sugiyama2023} as shown in Fig.~\ref{fig:internal_HSCy3}. Thus, the clustering signals do not exhibit any signature of the assembly bias effect. While the results for $R=[4,6]~h^{-1}$Mpc show a shift in the posterior distribution, we checked that the shift is caused primarily by upward scatter in the data points of $\dSigma$ around the scale cut for the LOWZ and CMASS1 subsamples (see Fig.~\ref{fig:signal-fitting}), which causes the code to prefer unphysical regions of the HOD parameters and then leads to a shift in cosmological parameters. We found that, if we remove the scattered data points, the shift in cosmological parameters do not occur.

As a further sanity check, we run the cosmology analyses for 100 realizations of the mock data vector that do not include the assembly bias effect. Note that for this test, we employ the photo-$z$ prior indicated from the (2,3)$h^{-1}$Mpc scale cut analysis for the $(8,12)~h^{-1}$Mpc scale-cut analysis in each realization as we did in the actual analysis. Fig.~\ref{fig:S8_scalecut_diff} shows in the difference in the $S_8$ values with different scale cuts we found from the real HSC-Y3 data occur with a reasonable chance. The two arrows in the figure show a shift in the $S_8$ values found from the two assembly bias mocks that we use in our validation tests (see Appendix~\ref{sec:validation}), and the $S_8$ difference is located at the tail of the 100 realizations, significantly displaced from the measured difference value. Thus we conclude that our cosmology results are unlikely to be affected by assembly bias. 

\subsection{Post-unblinding Analysis: The Impact of $\Omega_{\rm m}$ Prior}
\label{sec:om_prior}

\begin{figure}
    \includegraphics[width=\columnwidth]{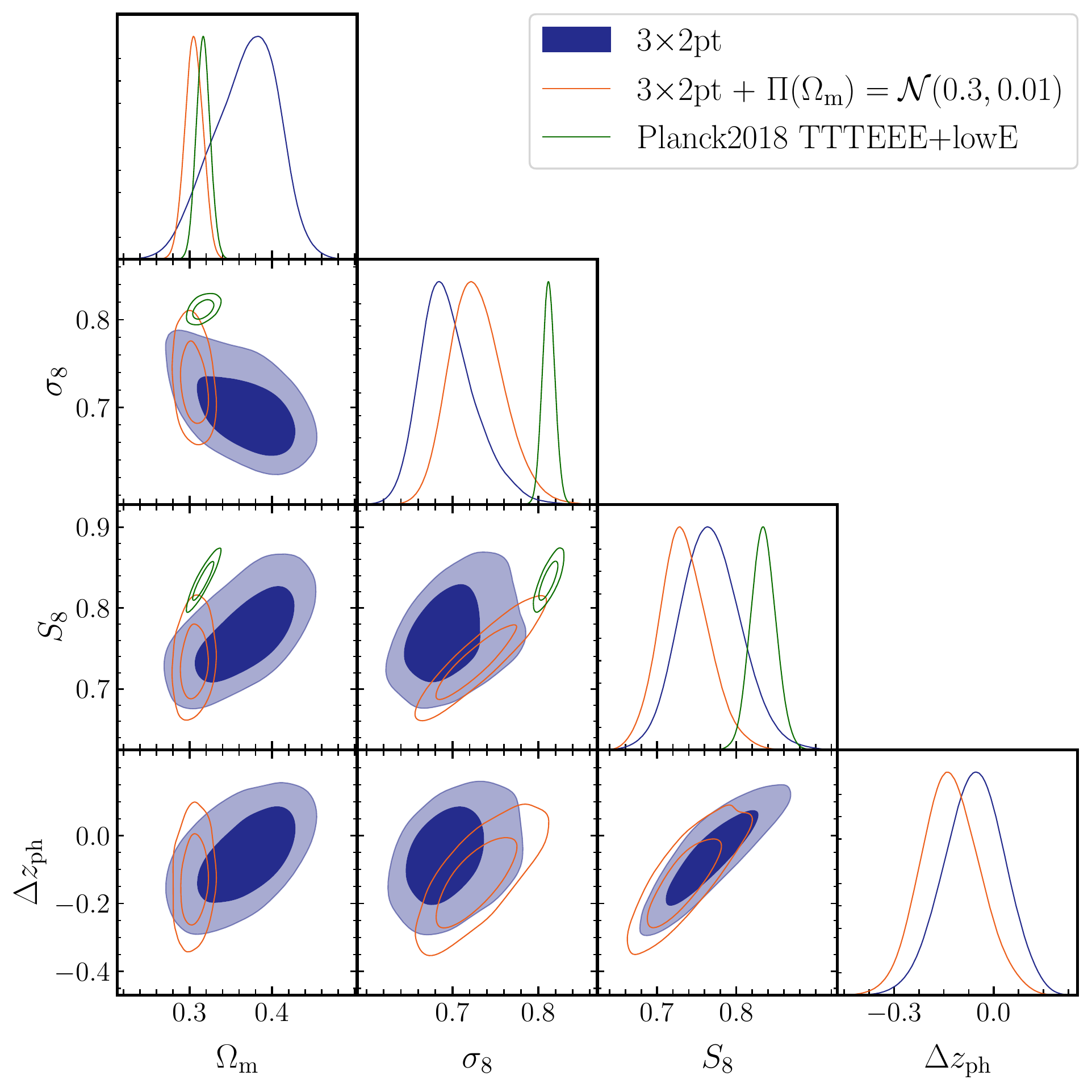}
    \caption{The posterior distribution of the parameters when the Gaussian prior of $\Pi(\Omega_{\rm m})={\cal N}(0.3,0.01)$ as motivated by BAO constraints is added. This analysis was done as a part of the post-unblinding analysis.} 
    \label{fig:contour_Omm_prior}
\end{figure}
During the blind analysis stage (see Section~\ref{sec:blinding}), we did not compare the posterior distribution of the cosmological parameters in our analysis with any external datasets. Our main result of  Fig.~\ref{fig:contour-main} indicates a higher $\Omega_{\rm m}$ than the {\it Planck} constraint. $\Omega_{\rm m}$ is well constrained by the baryon acoustic oscillation information in galaxy clustering \citep{Eisensteinetal:05} for flat-geometry $\Lambda$CDM model. Since we did not intentionally include the BAO information of SDSS galaxy clustering in our analysis, here we study how a BAO-motivated prior of $\Omega_{\rm m}$ affects our results, as part of our post-unblinding analysis. To do this, we employ the Gaussian prior given by $\Omega_{\rm m}: {\cal N}(0.3,0.01)$ in the baseline 3$\times$2pt analysis, where the central value of $\Omega_{\rm m0}$ and the width of $\sigma(\Omega_{\rm m0})=0.01$ are roughly consistent with the constraints obtained from the BAO analyses \citep{2021PhRvD.103h3533A,Kobayashi:2020a} \citep[also see][for the CMB-independent constraint]{2020JCAP...05..032P}.

Fig.~\ref{fig:contour_Omm_prior} shows the 2d posterior distributions of the parameters. We find
\begin{align}
S_8&= 0.732^{+0.027}_{-0.029} (0.738), \nonumber\\
\deltapz&=-0.133^{+0.077}_{-0.084} (-0.132).
\label{eq:S8_Omprior}
\end{align}
The prior of $\Omega_{\rm m} $ sightly lowers the central value of $S_8$, and also indicates a larger central value of $|\deltapz|$ than we found for the fiducial analysis (Eq.~\ref{eq:cosmological_parameters_baseline}). If we apply the {\tt eigentension} method to quantify a tension between the HSC 3$\times$2pt analysis and the {\it Planck} result, we find the tension is at about $2.4\sigma$, almost no change from the result in Section~\ref{sec:tension}. 

\section{Conclusion}
\label{sec:conclusion}
In this paper we have carried out a cosmology analysis combining three clustering observables, the projected correlation function ($\wproj$), galaxy-galaxy weak lensing ($\dSigma$), and cosmic shear correlation functions ($\xi_{\pm}$). These quantitites are measured from the spectroscopic SDSS galaxy samples as lens samples of $\dSigma$ and tracers of the $\wproj$, and the HSC-Y3 photometric galaxy samples for source galaxies in the $\dSigma$ and $\xi_{\pm}$ measurements. One of the most important aspects of our analysis is that we use a {\it single} source sample in the weak lensing measurements, allowing us to self-calibrate the residual error in the mean redshift of source galaxies, which is one of the most important systematic effects in weak lensing cosmology. We do so by comparing the relative $\dSigma$-amplitudes for the three spectroscopic lens subamples and the cosmic shear signal, as suggested in \citet{OguriTakada:11}. We employ a completely {\it uninformative} flat prior, ${\cal U}(-1,1)$, for the residual photo-$z$ error parameter $\deltapz$ in our cosmology analysis. We showed that, with the statistical power of the HSC-Y3 data, we can estimate the cosmological parameters and the photo-$z$ parameter $\deltapz$ simultaneously. We decided on this analysis setup during the blind analysis stage, without looking at the estimated values of cosmological parameters, and we froze the analysis method including the flat prior of $\deltapz$ before unblinding. This allowed us to obtain a {\it robust} estimate of the cosmological parameters, which minimizes the impact of the possible photo-$z$ errors, even at the cost of larger error bars of the cosmological parameters. 

The parameters we obtained for the flat $\Lambda$CDM model are: $S_8=0.763^{+0.040}_{-0.036}$ and $\deltapz=-0.05\pm 0.09$ after marginalizing over a number of other parameters. Thus we have estimated $S_8$ with a fractional precision of 5\%. Extending the $S_8$ definition to $S'_8=\sigma_8(\Omega_{\rm m}/0.3)^\alpha$, we showed that our method gives the best constraint with $\alpha=0.22$; we find $S'_8=0.721\pm0.028$ with $\alpha=0.22$, about 4\% fractional precision. These $S_8$ values are lower than indicated by the {\it Planck} CMB result. Using the tension metric in Refs.~\cite{Park:2020} and \cite{2020PhRvD.101j3527R}, we quantified the tension to be about 2.5$\sigma$. We plan to use extended models such as dark energy models, i.e., $w\neq-1$, or model templates with varying neutrino masses for the HSC-Y3 3$\times$2pt analysis to study whether the $S_8$ tension is relaxed. This requires a joint likelihood analysis of the HSC-Y3 3$\times$2pt and the {\it Planck} data using such extended models.

We also showed that when our HSC-Y3 analysis is combined with the external BAO constraints on $\Omega_{\rm m}$ with ${\cal N}(0.3,0.01)$, the parameters are changed to  $S_8=0.732^{+0.027}_{-0.029}$ and $\deltapz=-0.133^{+0.077}_{-0.084}$. This result indicates a $2\sigma$-level residual photo-$z$ error, implying that the mean redshift of the HSC galaxies at $z\gtrsim 0.7$ is {\it higher} by $|\Delta z|=0.133$ than implied by the photo-$z$ estimates. Interestingly, such a large photo-$z$ bias for the high-redshift HSC galaxies is also indicated in the companion works of the HSC cosmic shear tomography analyses (\citet{li2023} and \citet{dalal2023}). For the cosmic shear analyses, the photo-$z$'s at the high redshifts are calibrated by the cosmic shear amplitudes at different redshift bins, while the photo-$z$ error in our method is calibrated by the combination of the galaxy-galaxy lensing signal of the spectroscopic SDSS galaxies and the cosmic shear. 

We stress that we carried out all our cosmology analyses, the 3$\times 2{\rm pt}$ analyses and the real- and Fourier-space cosmic shear analyses, using different blinded catalogs. We did not compare the cosmological constraints from the different methods during the blind analysis stage. After unblinding we found that all cosmological constraints are in agreement with each other, and also indicate a non-zero residual photo-$z$ error for the high-redshift HSC galaxies. The significance of the non-zero residual photo-$z$ error and the consistency tests of these HSC cosmology results are studied and presented in the upcoming paper, Sugiyama~et al. (in prep.) using a mock analysis of these HSC cosmology analyses taking into account the cross covariances between the different observables. Thus the HSC results might suggest an unknown systematic error in the photo-$z$ estimates for high-redshift galaxies that are not calibrated out by the COSMOS data. The upcoming spectroscopic samples to be delivered from the DESI\footnote{\url{https://www.desi.lbl.gov}} and PFS surveys \cite{Takadaetal:14} will be very powerful samples for calibrating these high-redshift photo-$z$'s using the clustering redshift method to higher redshifts, $z\gtrsim 1$ \cite{Rau:2022wrq}. If we can constrain the photo-$z$ systematics to the precision of $\sigma(\deltapz)\sim {\cal O}(10^{-2})$, we can significantly improve the precision of our $S_8$ constraint even with the current HSC-Y3 data. 

There are various directions to improve the cosmological constraints in this paper. First of all the cosmological analysis in this paper is based on the HSC Year 3 dataset of 416~\sqdeg, which is about one-third of the full HSC dataset covering about 1,100~\sqdeg. Obviously, it is worth pursuing this possible $S_8$-tension with the full HSC dataset. In addition, this paper uses the projected correlation function of the SDSS galaxies for the joint analysis. In this paper we intentionally did not include the BAO information or the redshift-space distortion (RSD) effect that can be measured from the redshift-space three-dimensional correlation function or power spectrum. Since the BAO and RSD information are very powerful probes of cosmological parameters, it would be very interesting to explore a full joint analysis of the galaxy-galaxy weak lensing, the cosmic shear and the redshift-space power spectrum. In our future work, we will do this, using an emulator-based method similar to that in this paper to model the redshift-space power spectrum of galaxies based on the redshift-space halo power spectrum and the halo occupation method \citep{Kobayashi:2021}. Lastly, our method can be applied for the Stage-IV surveys, i.e., ground-based survey: Vera C. Rubin Observatory Legacy Survey of Space and Time \citep[LSST; ][]{LSSTOverviwe2019}), and space-based surveys: Euclid \citep{Euclid2011} and the Nancy Grace Roman space telescope \citep[Roman; ][]{WFIRST15}, by which statistical uncertainties will be significantly improved. 
 
\begin{acknowledgments}
This work was supported in part by World Premier International Research Center Initiative (WPI Initiative), MEXT, Japan, and JSPS KAKENHI Grant Numbers JP18H04350, JP18H04358, JP19H00677, JP19K14767, JP20H00181, JP20H01932, JP20H04723, JP20H05850, JP20H05855, JP20H05856, JP20H05861, JP21J00011, JP21H05456, JP21J10314, JP21H01081, JP21H05456,  JP22K03634, JP22K03655 and JP22K21349 by Japan Science and Technology Agency (JST) CREST JPMHCR1414, by JST AIP Acceleration Research Grant Number JP20317829, Japan, and by Basic Research Grant (Super AI) of Institute for AI and Beyond of the University of Tokyo. SS was supported in part by International Graduate Program for Excellence in Earth-Space Science (IGPEES), WINGS Program, the University of Tokyo. YK is supported in part by the David and Lucile Packard foundation. RM is supported by a grant from the Simons Foundation (Simons Investigator in Astrophysics, Award ID 620789). RD acknowledges support from the NSF Graduate Research Fellowship Program under Grant No.\ DGE-2039656. WL acknowledge the support from the National Key R\&D Program of China (2021YFC2203100), the 111 Project for "Observational and Theoretical Research on Dark Matter and Dark Energy'' (B23042), NSFC(NO. 11833005, 12192224) as well as the Fundamental Research Funds for the Central Universities (WK3440000006).

The Hyper Suprime-Cam (HSC) collaboration includes the astronomical communities of Japan and Taiwan, and Princeton University. The HSC instrumentation and software were developed by the National Astronomical Observatory of Japan (NAOJ), the Kavli Institute for the Physics and Mathematics of the Universe (Kavli IPMU), the University of Tokyo, the High Energy Accelerator Research Organization (KEK), the Academia Sinica Institute for Astronomy and Astrophysics in Taiwan (ASIAA), and Princeton University. Funding was contributed by the FIRST program from Japanese Cabinet Office, the Ministry of Education, Culture, Sports, Science and Technology (MEXT), the Japan Society for the Promotion of Science (JSPS), Japan Science and Technology Agency (JST), the Toray Science Foundation, NAOJ, Kavli IPMU, KEK, ASIAA, and Princeton University. This paper makes use of software developed for the Large Synoptic Survey Telescope. We thank the LSST Project for making their code available as free software at \url{http://dm.lsst.org}

The Pan-STARRS1 Surveys (PS1) have been made possible through contributions of the Institute for Astronomy, the University of Hawaii, the Pan-STARRS Project Office, the Max-Planck Society and its participating institutes, the Max Planck Institute for Astronomy, Heidelberg and the Max Planck Institute for Extraterrestrial Physics, Garching, The Johns Hopkins University, Durham University, the University of Edinburgh, Queen's University Belfast, the Harvard-Smithsonian Center for Astrophysics, the Las Cumbres Observatory Global Telescope Network Incorporated, the National Central University of Taiwan, the Space Telescope Science Institute, the National Aeronautics and Space Administration under Grant No. NNX08AR22G issued through the Planetary Science Division of the NASA Science Mission Directorate, the National Science Foundation under Grant No. AST-1238877, the University of Maryland, and Eotvos Lorand University (ELTE) and the Los Alamos National Laboratory.

Based in part on data collected at the Subaru Telescope and retrieved from the HSC data archive system, which is operated by Subaru Telescope and Astronomy Data Center at National Astronomical Observatory of Japan.
\end{acknowledgments}

\appendix

\section{Model Validation with Mock Galaxy Catalogs}
\label{sec:validation}
In this section we describe the validation tests of our modeling and analysis methods that we perform as one of the unblinding criterion. Table~\ref{tab:validation} summarizes the synthetic data vectors used for validation tests. For $\dSigma$ and $\wproj$ we use the same data vector as described in \citet{Miyatake:2022a} except for the synthetic data labelled as ``$\Delta z_{\rm ph}=-0.2$'' and ``$\Delta z_{\rm ph}=-0.2$${}^\dagger$''. For $\xi_{\pm}$, we use the data vector described in \citet{sugiyama2023}. Note that we take extreme cases for the baryonic effect on  $\xi_{\pm}$ by setting $A_{\rm bary}=1.6$ and $T_{\rm AGN}=7.3$. The detailed procedures to generate the synthetic data vectors for ``$\Delta z_{\rm ph}^{\rm in}=-0.2$'' and ``$\Delta z_{\rm ph}^{\rm in}=-0.2$${}^\dagger$,'' assuming that the estimated redshift distribution of source galaxies is systematically lower than the true distribution by $|\Delta z_{\rm ph}^{\rm in}|=0.2$ are described in Appendix~A of \citet{sugiyama2023}. Here $\Delta z_{\rm ph}^{\rm in}=-0.2$ is about $2\sigma$ away from the central value of $\Delta z_{\rm ph}$ in our fiducial analysis, i.e. $\Delta z_{\rm ph}=-0.05\pm 0.09$. Hence this validation test gives the worst case scenario for the impact of residual photo-$z$ error. Note that the validation test for ``$\Delta z_{\rm ph}^{\rm in}=-0.2$${}^\dagger$'' assumes the informative prior of $\Delta z_{\rm ph}$ given by ${\cal N}(0,0.1)$, aimed at studying how the informative prior gives a biased estimate in the cosmological parameters in the presence of the photo-$z$ bias given by  $\Delta z_{\rm ph}^{\rm in}=-0.2$. We apply the baseline analysis pipeline to each of the synthetic data vector to estimate the cosmological parameters using  the covariance matrix for the HSC-Y3 data.

\begin{figure*}
    \includegraphics[width=2\columnwidth]{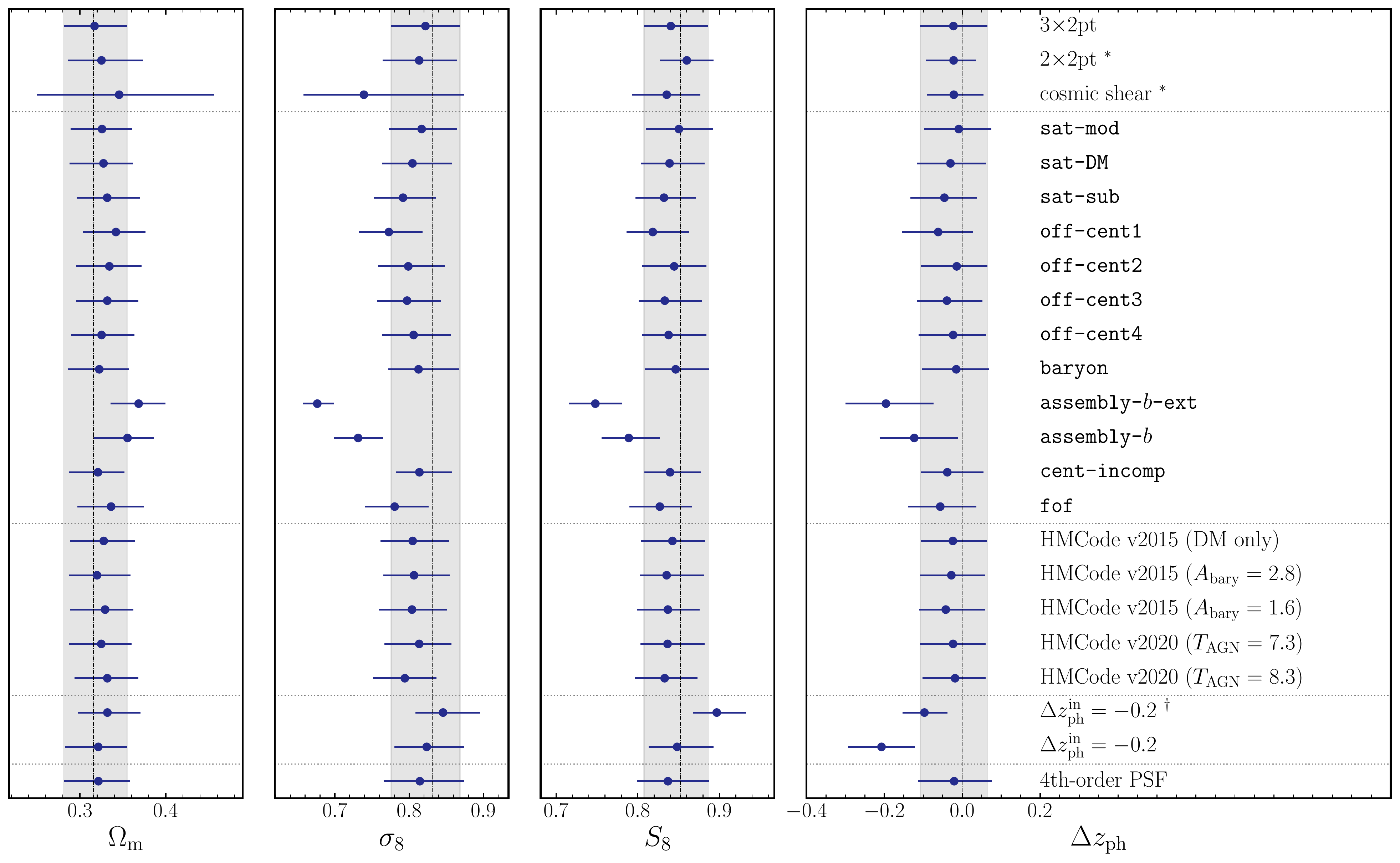}
    \caption{Summary of model validation tests for the 3$\times$2pt analysis. Constraints on cosmological parameters $\Omega_{\rm m}$, $\sigma_8$, and $S_8$ and the photo-$z$ shift parameter $\Delta z_{\rm ph}$ are shown in each panel from left to right. The input cosmological parameters used for making synthetic data vectors are indicated by the vertical dashed lines in the panels. The shaded regions are the 68\% confidence intervals from the constraints in the fiducial``3$\times$2pt'' case.}
    \label{fig:summary-mock-validation_3x2pt}
\end{figure*}

Fig.~\ref{fig:summary-mock-validation_3x2pt} shows the summary of the validation tests. As described in the main text, we do not find any significant deviation from the input cosmological parameters, except for ``assembly-$b$-ext,'' ``assembly-$b$,'' and ``$\Delta z_{\rm ph}=-0.2$ ${}^\dagger$.'' In ``assembly-$b$-ext'' and ``assembly-$b$'', we assume the large assembly bias amplitudes, so these tests give the worst case scenario in our cosmological constraints if the SDSS galaxies are affected by such large assembly bias effects, although there has been no detection of assembly bias for actual SDSS galaxies.  As described in Section~\ref{sec:assembly_bias}, a possible assembly bias signature can be identified from actual data analysis: if the assembly bias effect exists, we expect that using the different scale cuts of $\wproj$ and $\dSigma$ in the cosmology analysis would lead to a systematic shift in $S_8$. For example, if we employ the sufficiently large scale cuts such as $R=(8,12)h^{-1}$Mpc for $\wproj$ and $\dSigma$, where the linear theory or perturbation theory model is valid, the cosmological parameters are safely recovered. For the actual SDSS data, we did not observe such a systematic shift in $S_8$, so we concluded that the SDSS galaxies do not display any evidence of the assembly bias effect. 

The row of ``$\Delta z_{\rm ph}^{\rm in}=-0.2$ ${}^\dagger$'' in Fig.~\ref{fig:summary-mock-validation_3x2pt} shows a significant bias in $S_8$ by $>1\sigma$ if we employ an informative Gaussian prior on $\deltapz$ given by ${\cal N}(0,0.1)$ even when the photo-$z$ bias given by $\Delta z_{\rm ph}=-0.2$ exists. On the other hand, the row of ``$\Delta z_{\rm ph}^{\rm in}=-0.2$$^{\ast}$'' shows that the input cosmological parameters are safely recovered if we employ an uninformative, flat prior $\Pi(\Delta z_{\rm ph})={\cal U}(-1, 1)$ as our baseline analysis. Note that the photo-$z$ bias introduced in the data vector is also recovered in this case. This means that the HSC-Y3 data has a calibration power of $\deltapz$ to the precision of $\sigma(\deltapz)\sim 0.1$. Based on these findings, we decided to implement an analysis setup that uses the uninformative, flat prior of $\deltapz$. 

The row of ``4th-order PSF'' shows the results when including the fourth-moment PSF leakage and fourth-moment PSF modeling error in the synthetic data of $\xi_{\pm}$. As described in \citet{sugiyama2023}, we used the method in \citet{2022arXiv221203257Z} to measure the fourth-moment PSF leakage and fourth-moment PSF modeling errors from the HSC source galaxy sample used in this paper. Then we include the measured PSF systematics contamination in the synthetic data of $\xi_\pm$ and then apply the baseline analysis pipeline to the synthetic data vector including $\wproj$ and $\dSigma$. The result shows that the cosmological constraints are not affected by the PSF systematics. The impact is smaller than that found from the cosmic shear analyses of HSC-Y3 data \citep{li2023,dalal2023}, because the constraining power in our 3$\times$2pt analysis is mainly from the clustering information of SDSS galaxies, not from the cosmic shear signal.

\begin{table*}
\caption{A summary of mock signals used for the validation tests (see \citet{2021arXiv210100113M} for the details). All the mock catalogs, except for ``cent-imcomp.'' and ``FoF-halo'' catalogs, have the same HOD in the average sense, but use the different ways to populate galaxies into halos in $N$-body simulations. The column ``satellite gals.'' denotes a model of the spatial distribution of satellite galaxies in the host halo. In the columns of $\dSigma$, $\wproj$, and $\xi_{\pm}$, ``$\checkmark$'' or ``--'' denote whether they are modified from the fiducial mock or not, respectively. Note that the ``assembly-$b$'' and ``assembly-$b$-ext'' are the worst-case scenarios, where we implemented the overwhelmingly large assembly bias in the sense that the catalogs give the larger clustering amplitudes in $\wproj$ than those of the fiducial mocks (where the halo bias is simply given by the host halo mass) by a factor of 1.3 and 1.5 for the mock LOWZ, CMASS1 and CMASS2 galaxies (Fig.~5 of \citet{2021arXiv210100113M}).}
\label{tab:validation}
\begin{center}
\begin{tabular}{l|llllll}\hline\hline
setup label & HOD & satellite gals. & $\dSigma$ & $\wproj$ & $\xi_{\pm}$ & description \\ \hline
3$\times$2pt & fid. & NFW & -- & -- & --\hspace{1em} &fiducial model \\
2$\times$2pt${}^\ast$& fid. & NFW & -- & -- & N/A & without $\xi_{\pm}$, using $\Delta z_{\rm ph}$ posterior from 3$\times$2pt analysis as a prior\\ 
cosmic shear${}^\ast$& N/A & N/A & N/A & N/A & -- & without $\dSigma$ and $\wproj$, using $\Delta z_{\rm ph}$ posterior from 3$\times$2pt analysis as a prior \\ \hline
nonfidNsat & fid. & NFW & $\checkmark$ & $\checkmark$ & --\hspace{1em}  & populate satellites irrespectively of centrals \\
sat-dm-dist& fid. & DM part. & $\checkmark$ & $\checkmark$ & --\hspace{1em} & populate satellites according to $N$-body particles  \\ 
sat-sub & fid. & subhalos & $\checkmark$ & $\checkmark$ & --\hspace{1em} & populate satellites according to subhalos \\ \hline
off-cent1 & fid. & NFW & $\checkmark$ & $\checkmark$ & --\hspace{1em} & all centrals off-centered, with Gaussian profile\\
off-cent2 & fid. & NFW & $\checkmark$ & $\checkmark$ & --\hspace{1em} & a fraction ($0.34$) of ``off-centered'' centrals, assuming 
Gaussian profile\\
off-cent3 & fid. & NFW & $\checkmark$ & $\checkmark$ & --\hspace{1em} & similar to ``off-cent1'', but with NFW profile\\
off-cent4 & fid. & NFW & $\checkmark$ & $\checkmark$ & --\hspace{1em} & similar to ``off-cent2'', but with NFW profile\\ \hline
baryon & fid. & NFW & $\checkmark$ & -- & --\hspace{1em} & mimic the baryonic effect of {\tt Illustris} on the 
halo mass profile\\ \hline
assembly-$b$-ext & fid. & NFW & $\checkmark$ & $\checkmark$ & --\hspace{1em} & populate galaxies according to 
concentrations of host halos\\ 
assembly-$b$ & fid. & NFW&  $\checkmark$ & $\checkmark$ & --\hspace{1em} & similar to ``assembly-$b$-ext'',
but introduce scatters\\ \hline
cent-incomp. & $\avrg{N_{\rm c}}$ mod. & NFW & $\checkmark$ & $\checkmark$ & --\hspace{1em} & include an ``incomplete'' selection of centrals \\ \hline
FoF-halo & mod. & FoF halos & -- & -- & $\checkmark$ &  use FoF halos to populate galaxies \\ \hline
HMCode v2015
& fid. & NFW & -- & -- & $\checkmark$ & $\xi_{\pm}$ is generated by HMCode v2015 with $A_{\rm bary}=1.6$ 
or $2.8$ \\ 
HMCode v2020
& fid. & NFW & -- & -- & $\checkmark$ & $\xi_{\pm}$ is generated by HMCode v2020 with $T_{\rm AGN}=7.3$
or $8.3$
 \\  \hline
$\Delta z_{\rm ph}^{\rm in}=-0.2$ & fid. & NFW & $\checkmark$ & -- & $\checkmark$\hspace{1em} & $\dSigma$ and $\xi_{\pm}$ with $\Delta z_{\rm ph}^{\rm in}=-0.2$, analyzed with a prior $\Pi(\Delta z_{\rm ph})={\cal U}(-1, 1)$  \\ 
$\Delta z_{\rm ph}^{\rm in}=-0.2$${}^{\dagger}$ & fid. & NFW & $\checkmark$ & -- & $\checkmark$\hspace{1em} &  $\dSigma$ and $\xi_{\pm}$ with $\Delta z_{\rm ph}^{\rm in}=-0.2$, analyzed with a prior $\Pi(\Delta z_{\rm ph})={\cal N}(0, 0.1)$ \\ \hline
4th-order PSF & fid. & NFW & -- & -- & $\checkmark$\hspace{1em} &  
Include the 4th-order moment PSF systematics into $\xi_{\pm}$
\\ 
\hline\hline
\end{tabular}
\end{center}
\end{table*}

\section{Details of Internal Consistency Tests}
\label{sec:consistency_tests}

\begin{figure}
    \includegraphics[width=\columnwidth]{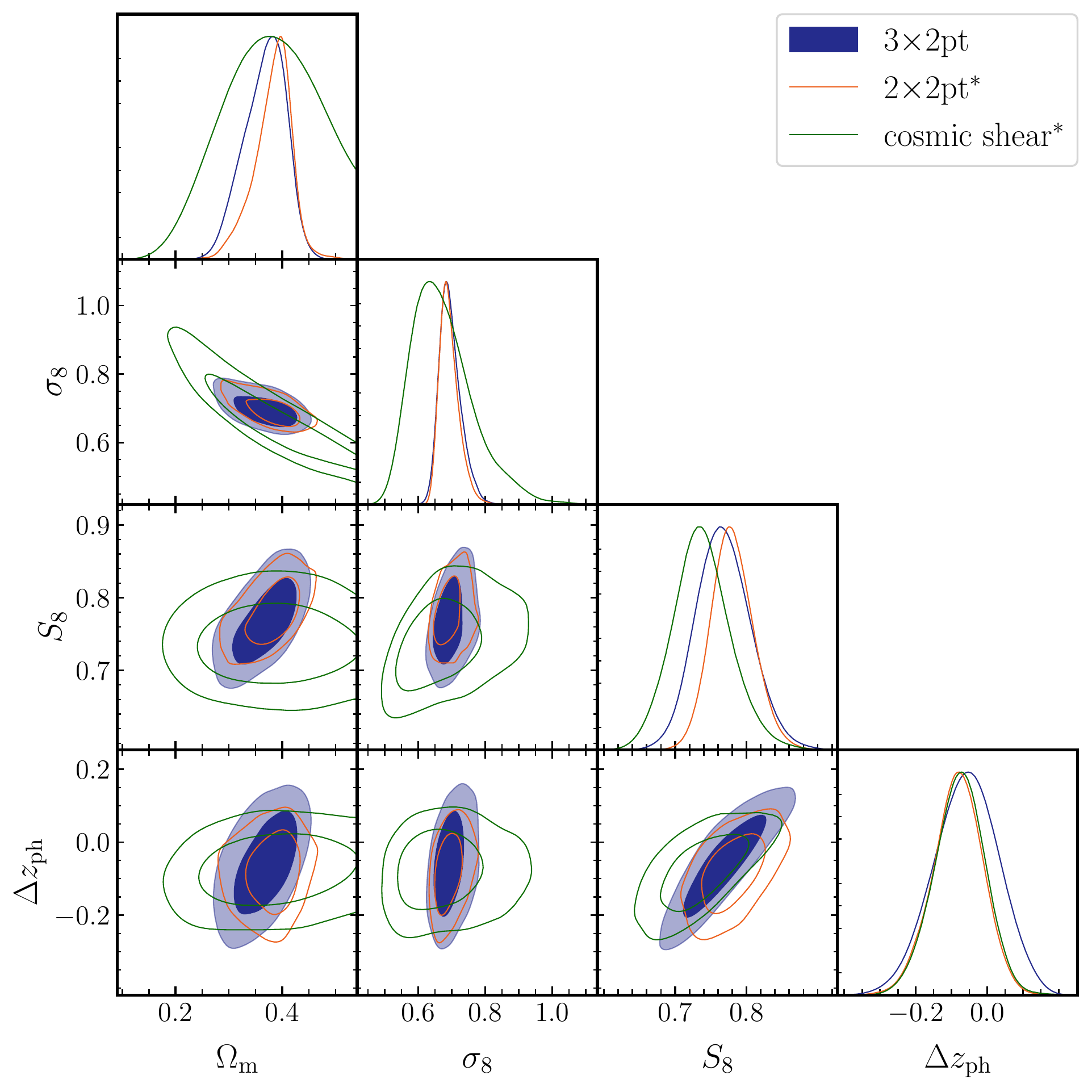}
    \caption{The posterior distributions for baseline ``3$\times$2pt,'' ``2$\times$2pt${}^{\ast}$ ,'' and ``cosmic shear${}^{\ast}$'' analysis setups in Table~\ref{tab:analysis_setups}. The contours show the 68\% and 96\% credible intervals. The constraints from ``2$\times$2pt${}^{\ast}$'' and ``cosmic shear${}^{\ast}$'' are consistent with the ``3$\times$2pt'' analysis. Note that since we use the prior on the residual photo-$z$ errors ($\Delta z_{\rm ph}$) for ``2$\times$2pt${}^{\ast}$'' and ``cosmic shear${}^{\ast}$'' derived from the ``3$\times$2pt'' analysis, the constraining power of ``2$\times$2pt${}^{\ast}$'' is similar to ``3$\times$2pt,'' but that of ``cosmic shear${}^{\ast}$'' is still weaker than ``3$\times$2pt.'' For the constraints from the 2$\times$2pt and cosmic shear analysis with the uniform prior on $\Delta z_{\rm ph}$, see the left panel in Fig.~\ref{fig:photoz_bias}.
    }    
    \label{fig:contour_3x2pt2x2ptcs_Ommsigma8S8dzp_hodmodesigma}
\end{figure}

\begin{figure}
    \includegraphics[width=\columnwidth]{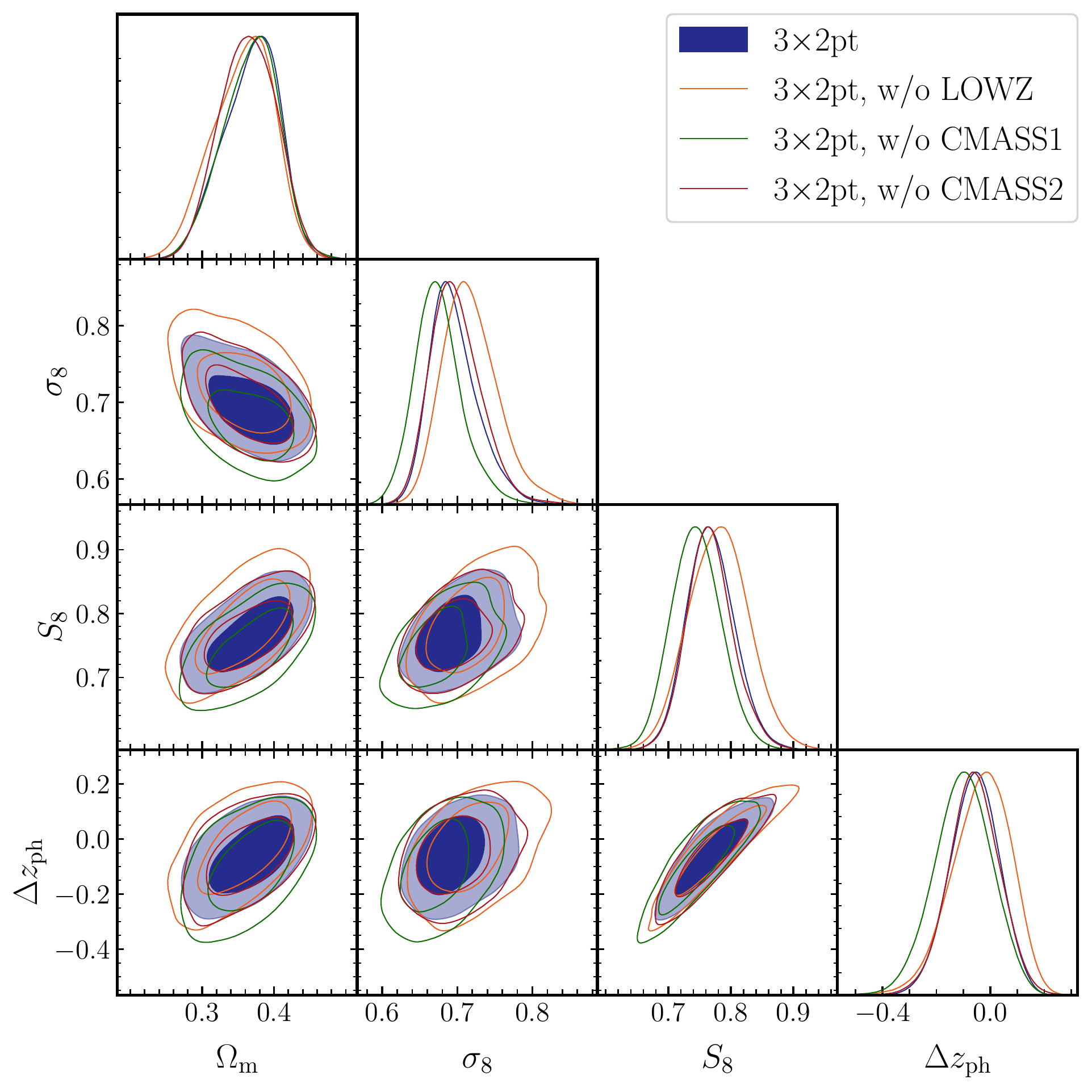}
    \caption{Similar to Fig.~\ref{fig:contour_3x2pt2x2ptcs_Ommsigma8S8dzp_hodmodesigma}, but the results from ``3$\times$2pt, w/o LOWZ,'' ``3$\times$2pt, w/o CMASS1,'' and ``3$\times$2pt, w/o CMASS2'' analysis setups in Table~\ref{tab:analysis_setups}. We do not see any significant changes in the cosmological constraints from the baseline ``3$\times$2pt'' analysis when excluding one of the lens samples in our analysis.
    }
    \label{fig:contour_wolowz_cmass1_or_cmass2}
\end{figure}

\begin{figure}
    \includegraphics[width=\columnwidth]{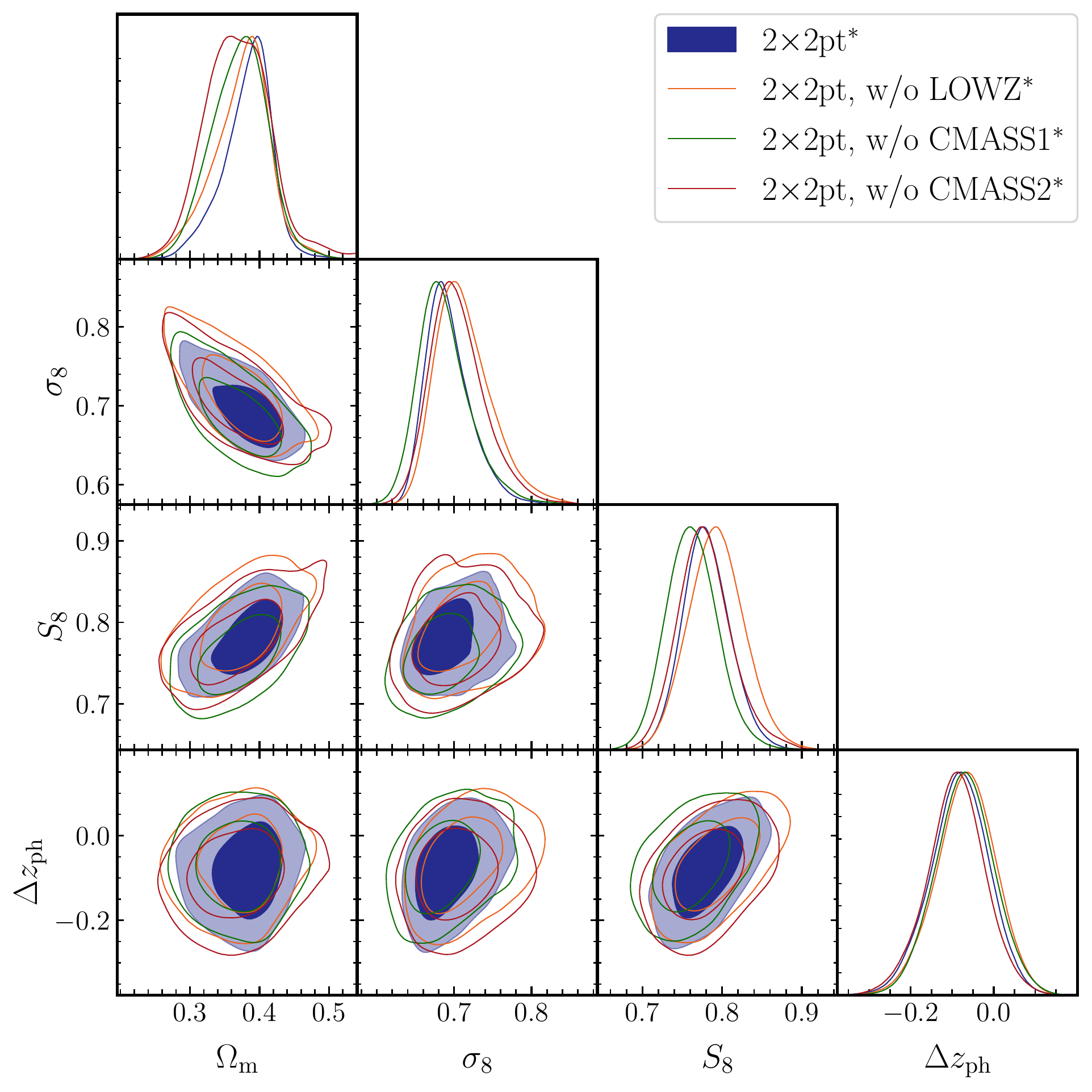}
    \caption{Similar to Fig.~\ref{fig:contour_3x2pt2x2ptcs_Ommsigma8S8dzp_hodmodesigma}, but the results from ``2$\times$2pt, w/o LOWZ${}^{\ast}$,'' ``2$\times$2pt, w/o CMASS1${}^{\ast}$,'' and ``2$\times$2pt, w/o CMASS2${}^{\ast}$'' analysis setups in Table~\ref{tab:analysis_setups}. We do not see any significant changes in the cosmological constraints from the baseline ``3$\times$2pt'' analysis when excluding one of the lens samples in our analysis.}
    \label{fig:contour_2x2pt_wolowz_cmass1_or_cmass2}
\end{figure}

\begin{figure}
    \includegraphics[width=\columnwidth]{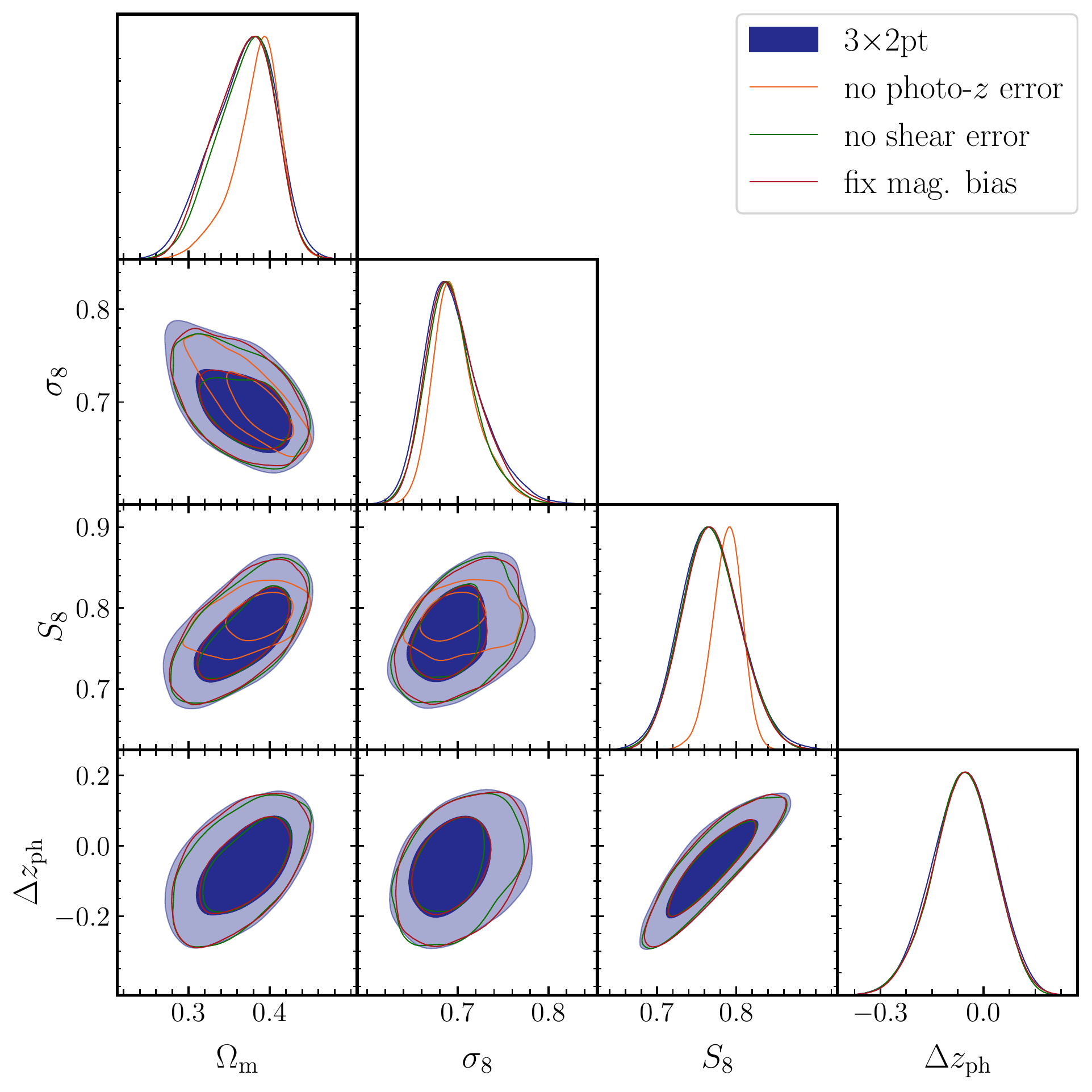}
    \caption{Similar to Fig.~\ref{fig:contour_3x2pt2x2ptcs_Ommsigma8S8dzp_hodmodesigma}, but the results from  ``no photo-$z$ error,'' ``no shear error,'' and ``fix mag. bias'' analysis setups in Table~\ref{tab:analysis_setups}. When the source redshift distribution is fixed, corresponding to the case ``no photo error'', the credible interval of the cosmological parameters are significantly tightened, but the central values are shifted. In particular, fixing the source redshift distribution as indicated from the photo-$z$ estimates rather than using the flat prior $\Pi(\deltapz)={\cal U}(-1,1)$ in our baseline analysis leads to a higher value of $S_8$ than that of the baseline analysis. We do not see any significant changes in the cosmological constraints for the other cases.}
    \label{fig:contour_fixing_each_of_obsnuisance}
\end{figure}

\begin{figure}
    \includegraphics[width=\columnwidth]{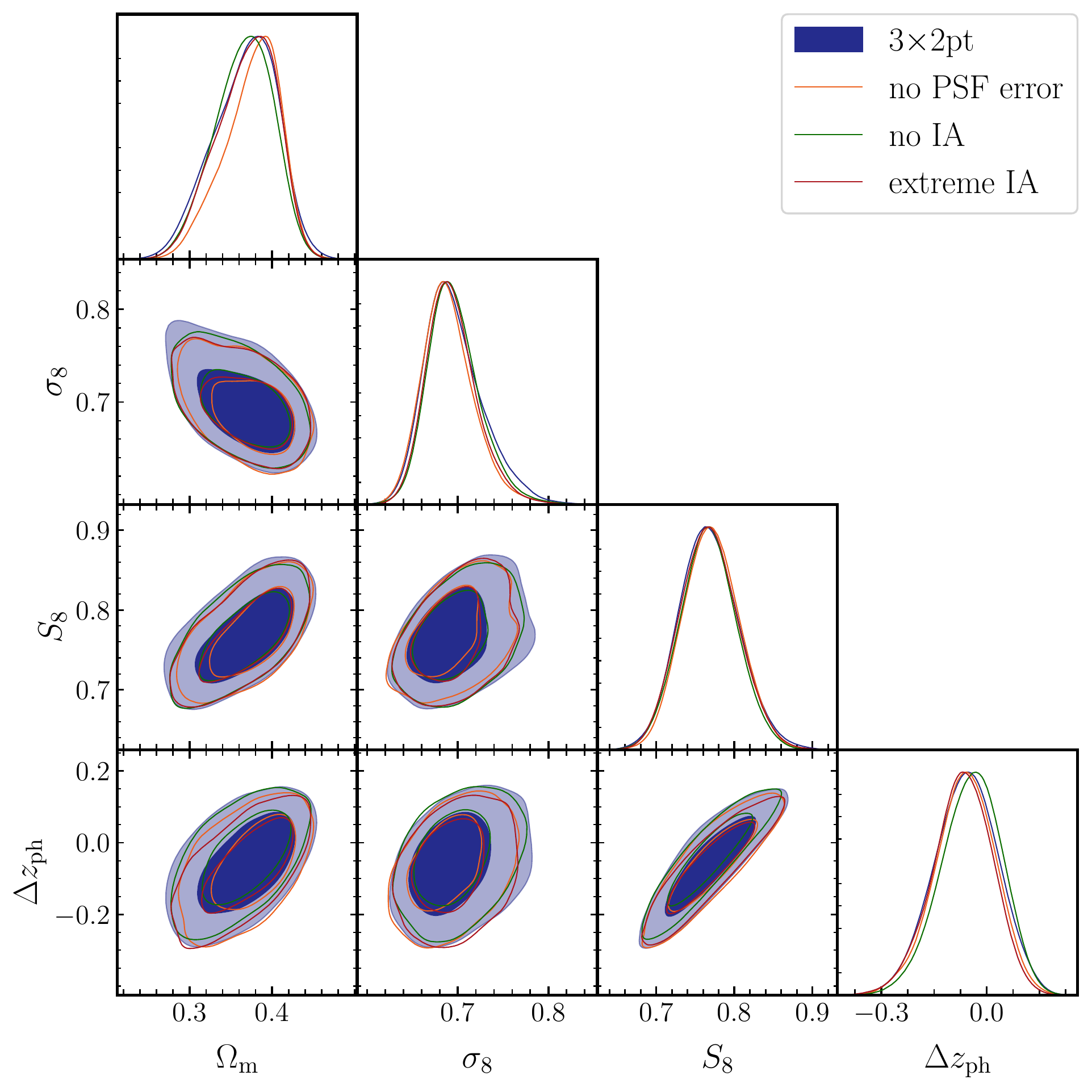}
    \caption{Similar to Fig.~\ref{fig:contour_3x2pt2x2ptcs_Ommsigma8S8dzp_hodmodesigma}, but the results from ``no PSF error,'' ``no IA error,'' and ``extreme IA'' analysis setups in Table~\ref{tab:analysis_setups}. For these setups, we do not see any significant changes in the cosmological constraints from the baseline analysis.}
    \label{fig:contour_noPSF_or_IA}
\end{figure}

\begin{figure}
    \includegraphics[width=\columnwidth]{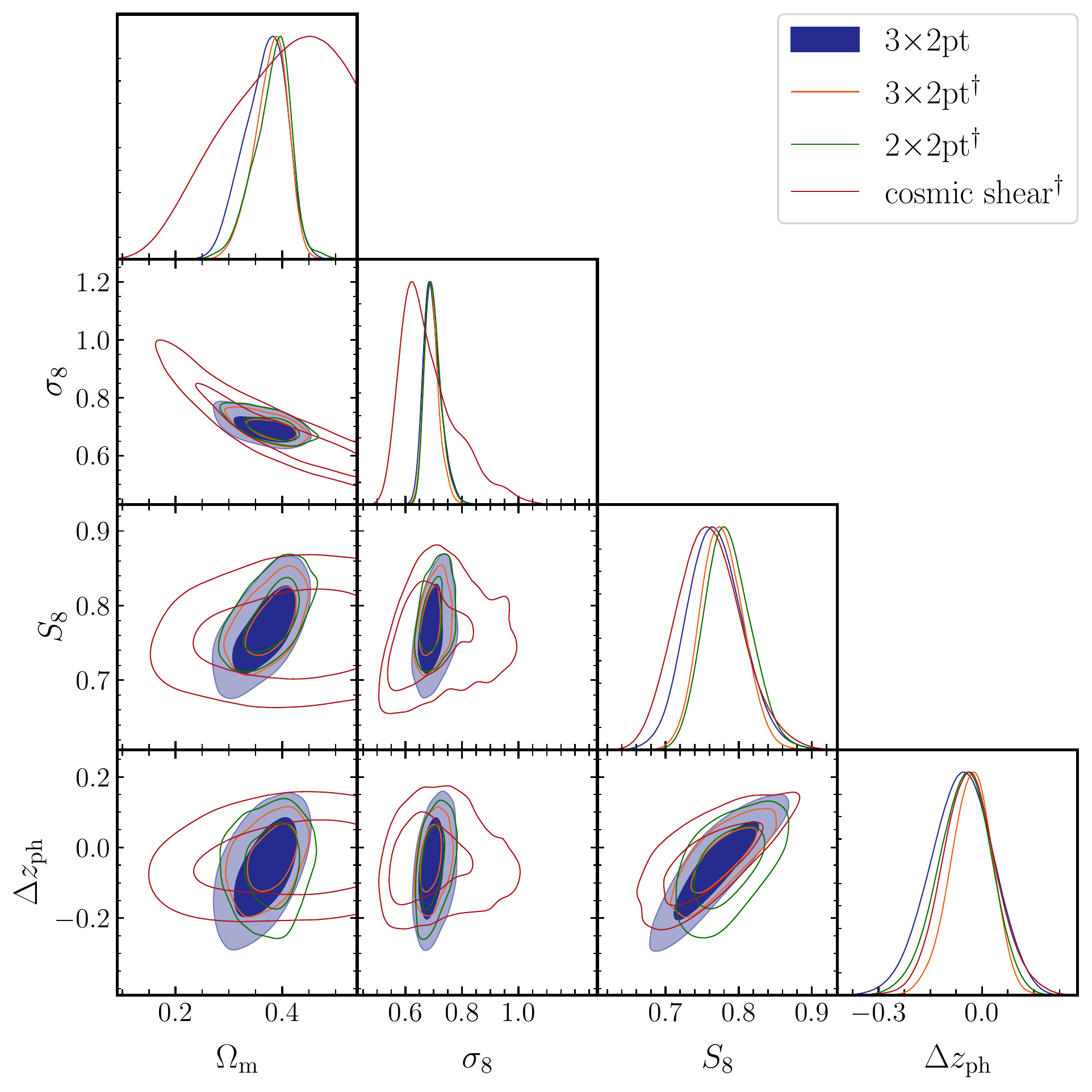}
    \caption{Similar to Fig.~\ref{fig:contour_3x2pt2x2ptcs_Ommsigma8S8dzp_hodmodesigma}, but the results from ``3$\times$2pt${}^{\dagger}$,'' ``2$\times$2pt${}^{\dagger}$,'' and ``cosmic shear${}^{\dagger}$'' analysis setups in Table~\ref{tab:analysis_setups}. Here the analyses with superscript ${}^\dagger$ uses the Gaussian prior on the 
    photo-$z$ error prior parameter,
    $\Pi(\deltapz)={\cal N}(0,0.1)$. The use of the Gaussian prior still affects the $S_8$ constraints for the ``3$\times$2pt${}^{\dagger}$,'' ``2$\times$2pt${}^{\dagger}$'' analyses.
    }
    \label{fig:contour_dpzN0prior}
\end{figure}

\begin{figure}
    \includegraphics[width=\columnwidth]{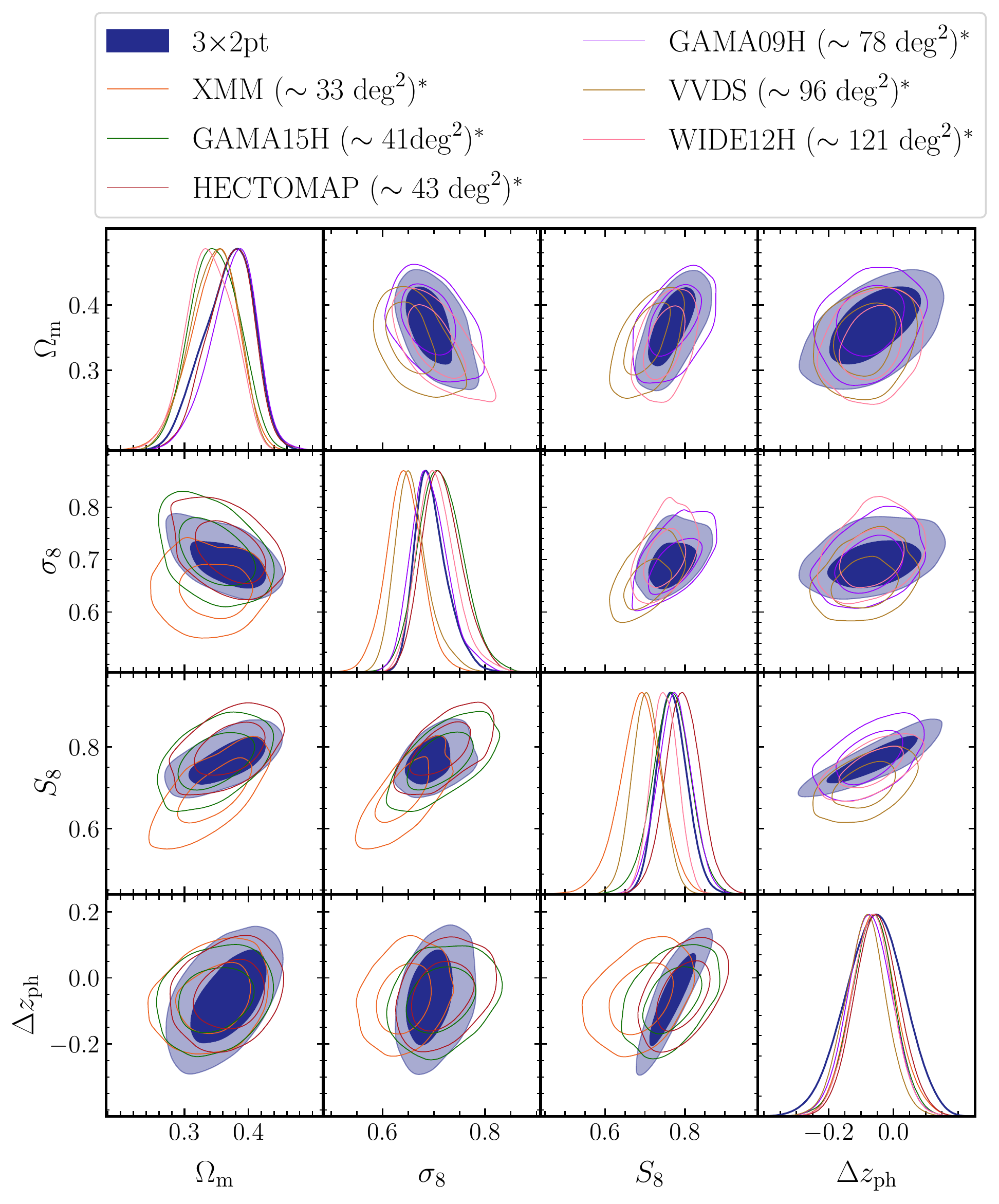}
    \caption{
    Similar to Fig.~\ref{fig:contour_3x2pt2x2ptcs_Ommsigma8S8dzp_hodmodesigma}, but the results from ``XMM $(\sim 33~\mathrm{deg}^2)$${}^{\ast}$,'' ``GAMA15H $(\sim 41\mathrm{deg}^2)$${}^{\ast}$,'' ``HECTOMAP $(\sim 43~\mathrm{deg}^2)$${}^{\ast}$,'' ``GAMA09H $(\sim 78~\mathrm{deg}^2)$${}^{\ast}$,'' ``VVDS $(\sim 96~\mathrm{deg}^2)$${}^{\ast}$,'' and ``WIDE12H $(\sim 121~\mathrm{deg}^2)$${}^{\ast}$'' analysis setups in Table~\ref{tab:analysis_setups}. Given that the analyses of each field are almost uncorrelated, the cosmological constraints are consistent with each other.}
    \label{fig:contour_eachfield}
\end{figure}

\begin{figure}
    \includegraphics[width=\columnwidth]{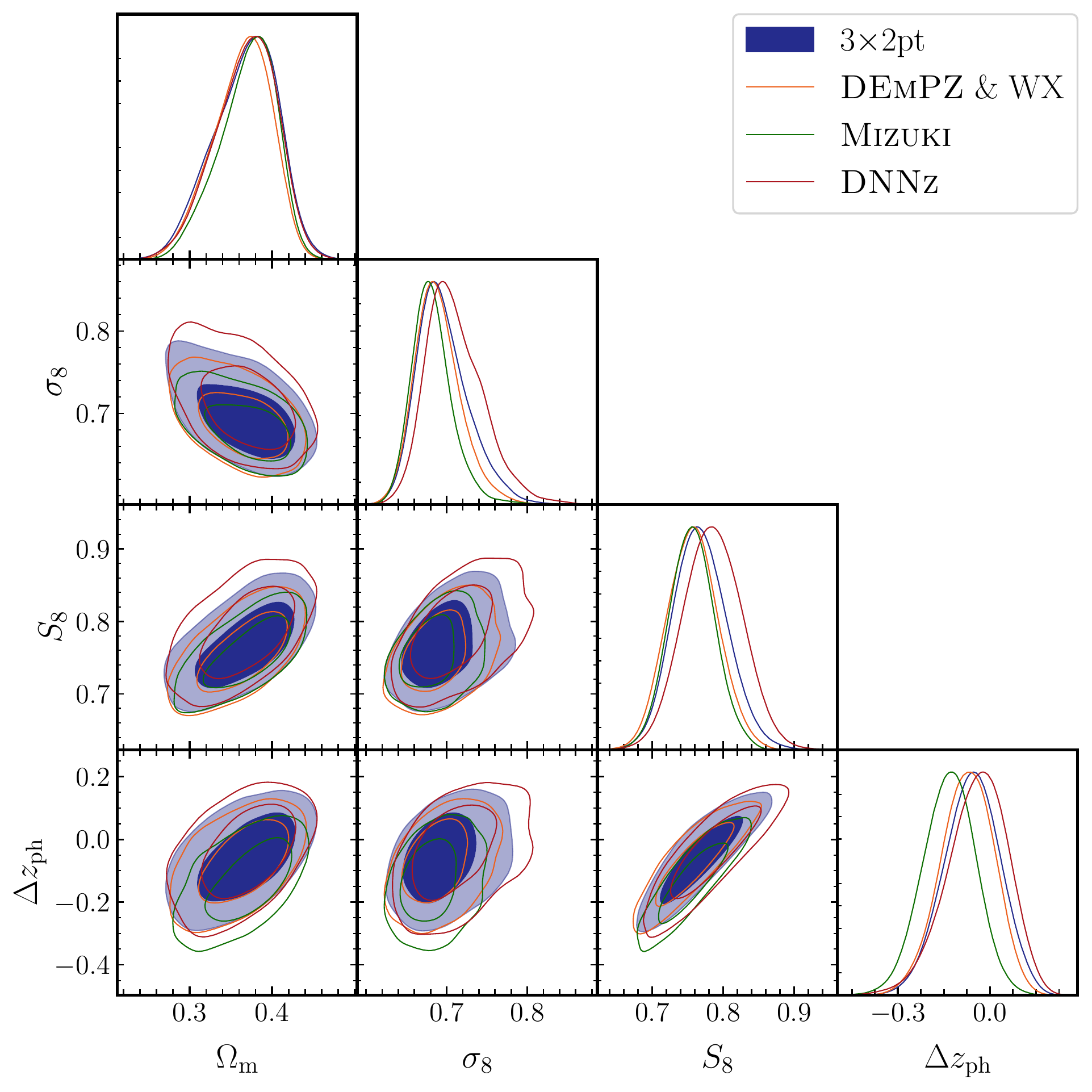}
    \caption{Similar to Fig.~\ref{fig:contour_3x2pt2x2ptcs_Ommsigma8S8dzp_hodmodesigma}, but the results from ``$\textsc{DEmPz}$ \& WX,'' ``$\textsc{Mizuki}$,'' and ``$\textsc{DNNz}$'' analysis setups in 
    Table~\ref{tab:analysis_setups}. ``The $\textsc{DNNz}$'' analysis exhibits the largest deviation in $S_8$, but it is still $\sim 0.5\sigma$.}
    \label{fig:contour_photoz}
\end{figure}

\begin{figure}
    \includegraphics[width=\columnwidth]{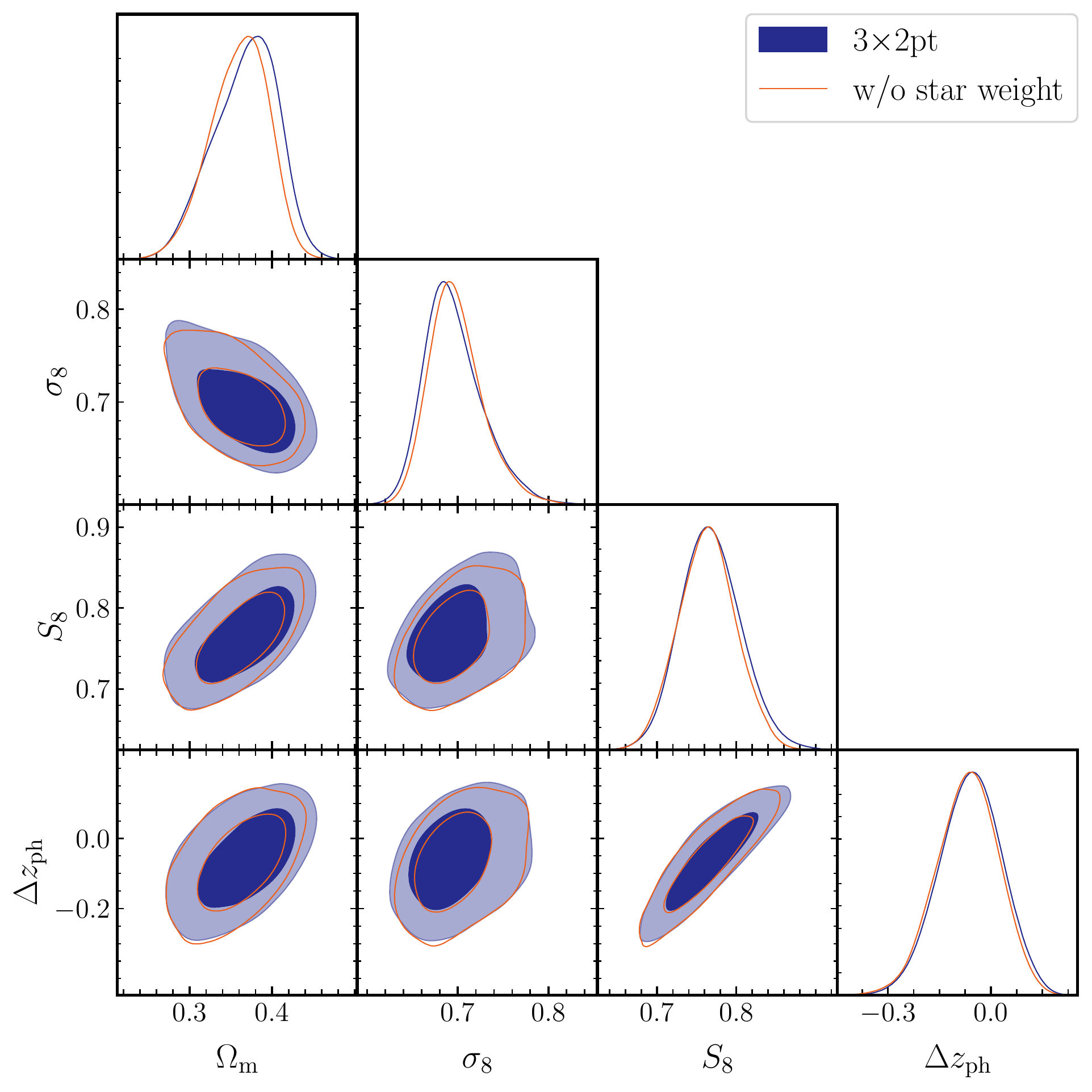}
    \caption{Similar to Fig.~\ref{fig:contour_3x2pt2x2ptcs_Ommsigma8S8dzp_hodmodesigma}, but the results from ``w/o star weight'' analysis setup in Table~\ref{tab:analysis_setups}. We do not see any significant changes in the cosmological constraints in the ``w/o star weight'' analysis.
    }
    \label{fig:contour_wo_star_weight}
\end{figure}

\begin{figure}
    \includegraphics[width=\columnwidth]{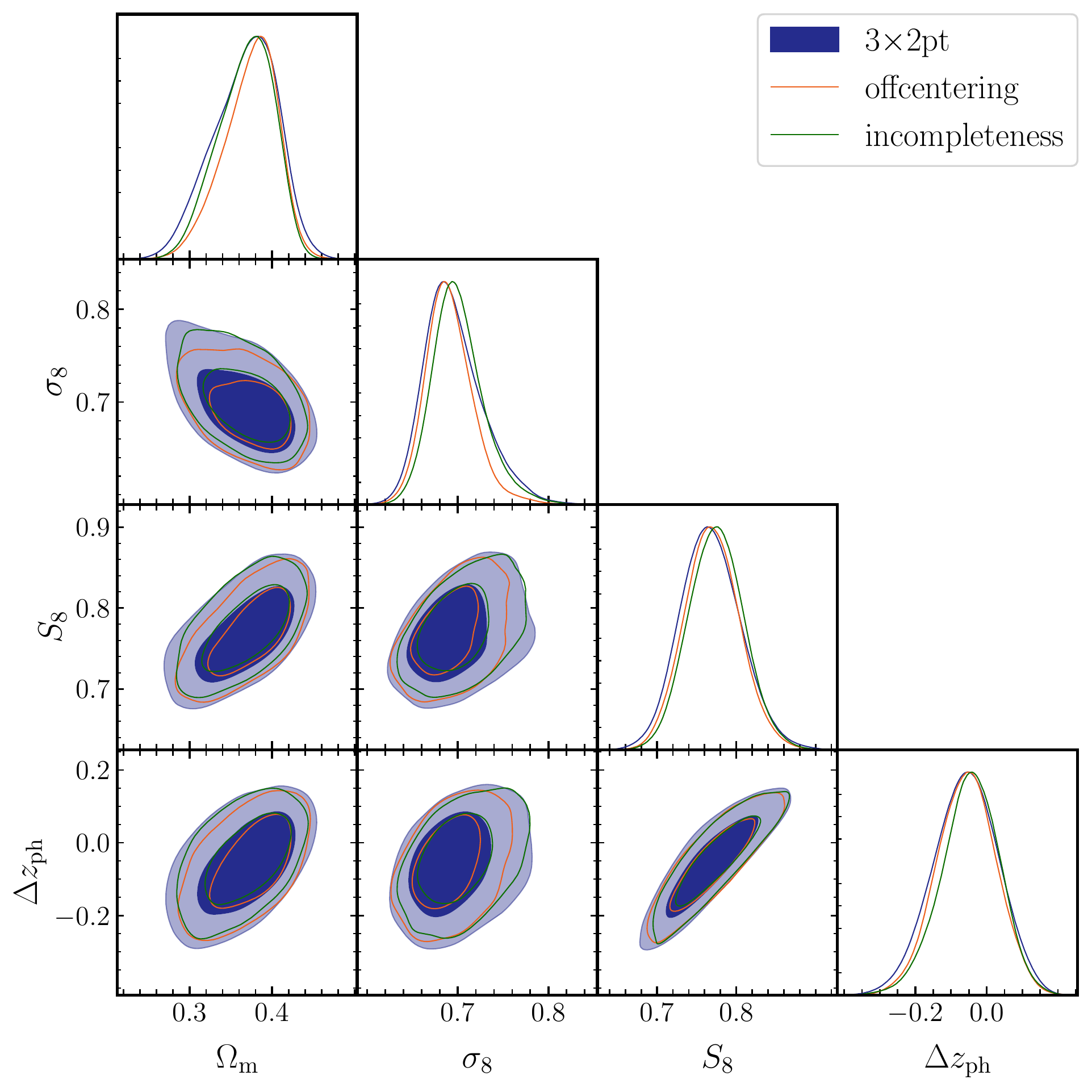}
    \caption{Similar to Fig.~\ref{fig:contour_3x2pt2x2ptcs_Ommsigma8S8dzp_hodmodesigma}, but the results from ``off-centering'' and ``incompleteness'' analysis setups in Table~\ref{tab:analysis_setups}. 
    We do not see any significant changes in the cosmological constraints in these analysis.}
    \label{fig:contour_offcen_incomp}
\end{figure}

In this section we show the results for various consistency tests for the different analysis setups 
and/or different subsets of the data vector, as listed in Table~\ref{tab:analysis_setups}. Table~\ref{tab:internal_consistency} shows the mean and 68\% upper and lower credible intervals with the MAP in parenthesis, i.e., the numbers plotted in Fig.~\ref{fig:summary-real-data}. Figs.~\ref{fig:contour_3x2pt2x2ptcs_Ommsigma8S8dzp_hodmodesigma}--\ref{fig:contour_offcen_incomp} show the 1- or 2-dimensional posterior distributions of the different analysis setups in which the same kind of consistency tests are grouped and compared with the fiducial 3$\times$2pt analysis.

\begin{table*}
    \centering
    \renewcommand{\arraystretch}{1.3}
    \caption{Summary of the main cosmological parameters constrained in this work, $\Omega_{\rm m}$, $\sigma_8$, and $S_8$. The estimates are presented in the format of $\text{mode}^{+68\%~\text{upper}}_{-68\%~\text{lower}}$ $(\text{MAP},~\text{mean})$. The analysis setup for each row is summarized in Table~\ref{tab:analysis_setups}.
    }
    \input{summary_Ommsigma8S8dpz_b2_dempz.tex}
    \label{tab:internal_consistency}
    \renewcommand{\arraystretch}{1}
\end{table*}


\begin{figure}
    \includegraphics[width=\columnwidth]{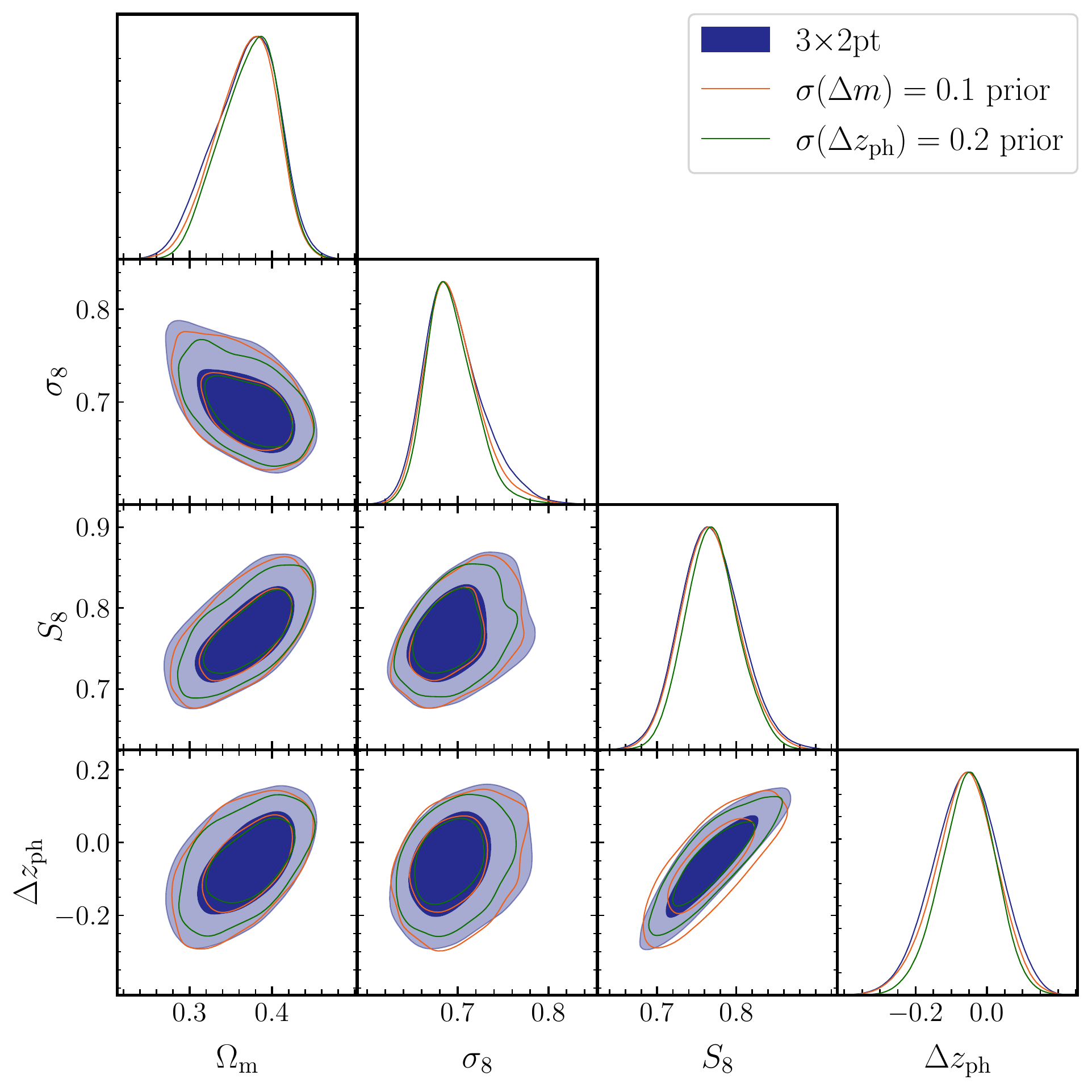}
    \caption{Similar to Fig.~\ref{fig:contour_3x2pt2x2ptcs_Ommsigma8S8dzp_hodmodesigma}, but the results from ``$\sigma(\Delta m)=0.1$ prior'' and ``$\sigma(\Delta z_{\rm ph})=0.2$ prior'' analysis setups in Table~\ref{tab:analysis_setups}. Note that these are the post-unblinding analyses. We do not see a significant change in the ``$\sigma(\Delta m)=0.1$ prior'' analysis, meaning that the degradation of constraining power is mostly due to the use of flat prior on the residual photo-$z$ errors. We see only a slight change in the ``$\sigma(\Delta z_{\rm ph})=0.2$'' analysis without improvement in the constraining power. This means that with this prior width, we are reaching to the limit of the flat prior.}
    \label{fig:contour_wide_prior}
\end{figure}
\begin{figure}
    \includegraphics[width=\columnwidth]{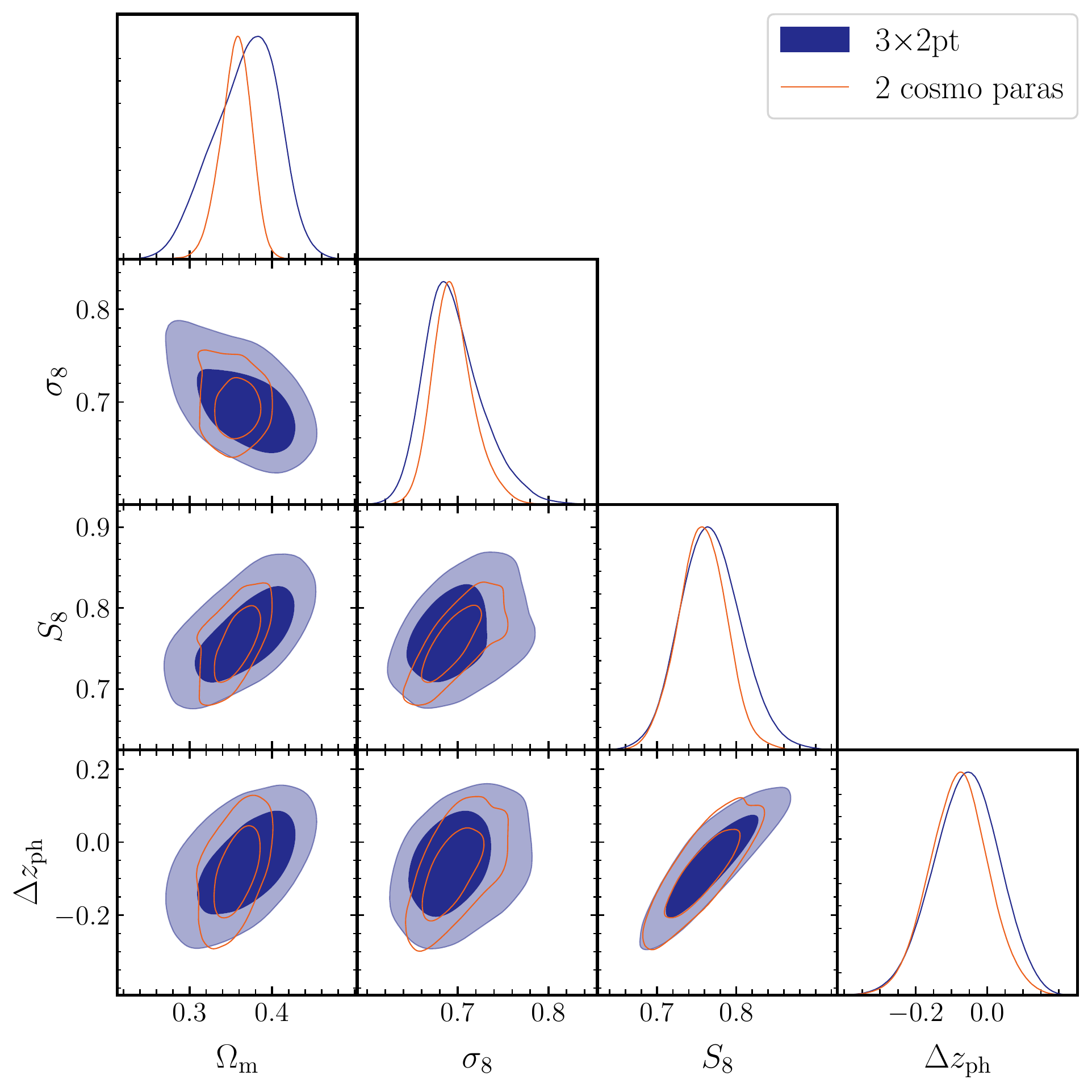}
    \caption{Similar to Fig.~\ref{fig:contour_3x2pt2x2ptcs_Ommsigma8S8dzp_hodmodesigma}, but the results from ``2 cosmo paras'' analysis setup in Table~\ref{tab:analysis_setups}. Note that this is the post-unblinding analyses. When fixing cosmological parameters other than $\Omega_{\rm de}$ and $\ln{(10^{10}A_{\rm s}})$ to the Planck {\it Planck} CMB constraints \citep{2020A&A...641A...6P}, we obtain beter cosmological constraints, but the shift in $\Delta z_{\rm ph}$ is unchanged.}
    \label{fig:contour_2cosmoparameters}
\end{figure}

\section{Posterior Distributions of All Parameters}
\begin{figure*}
    \includegraphics[width=2\columnwidth]{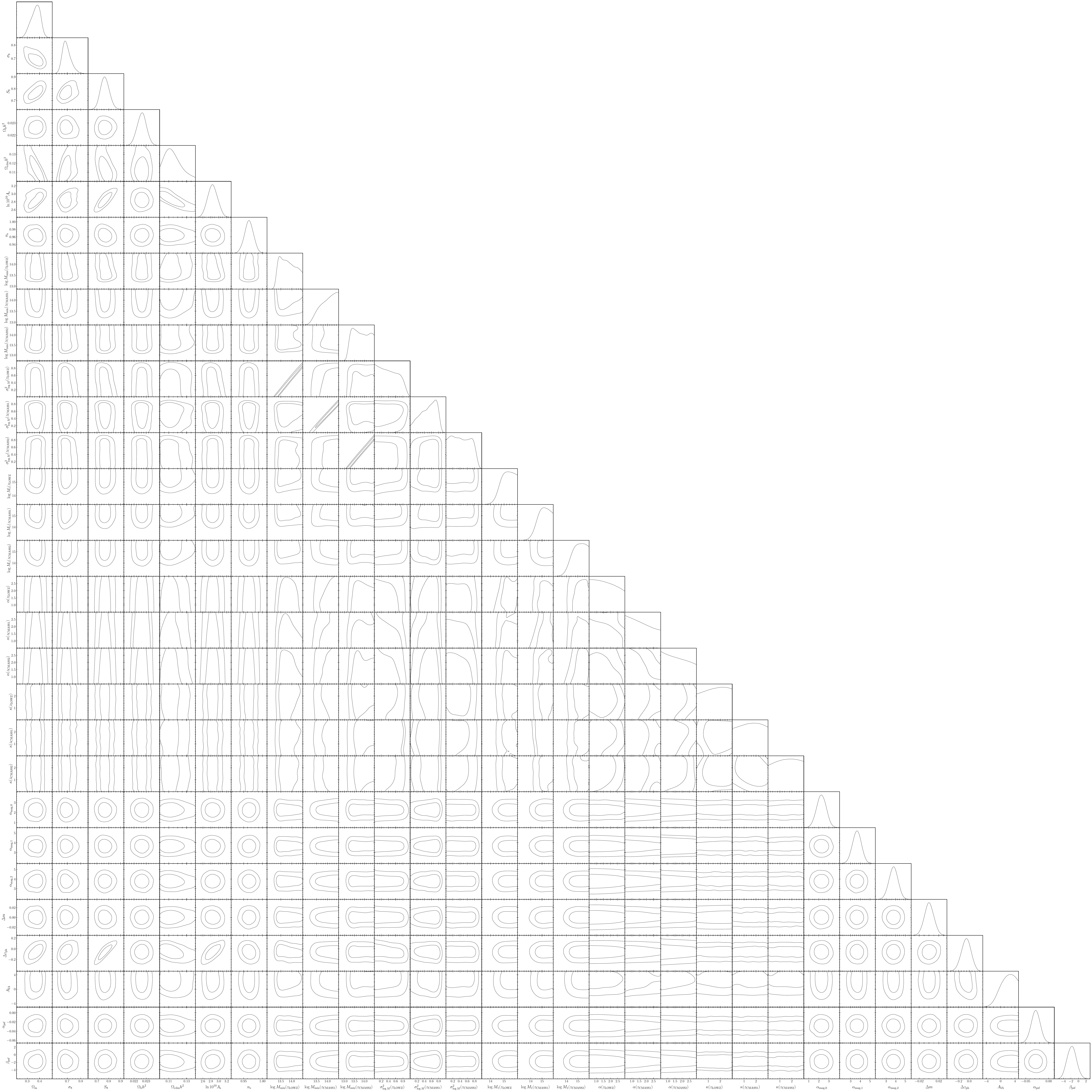}
    \caption{The posterior distributions of all parameters sampled in the baseline 3$\times$2pt analysis.}
    \label{fig:contour_full}
\end{figure*}

Fig.~\ref{fig:contour_full} shows the posterior distribution of all parameters sampled in our baseline analysis.

\section{Robustness of Parameter Sampling}
\label{sec:chain_convergence}
\subsection{Nestcheck}
\begin{figure}
    \includegraphics[width=\columnwidth]{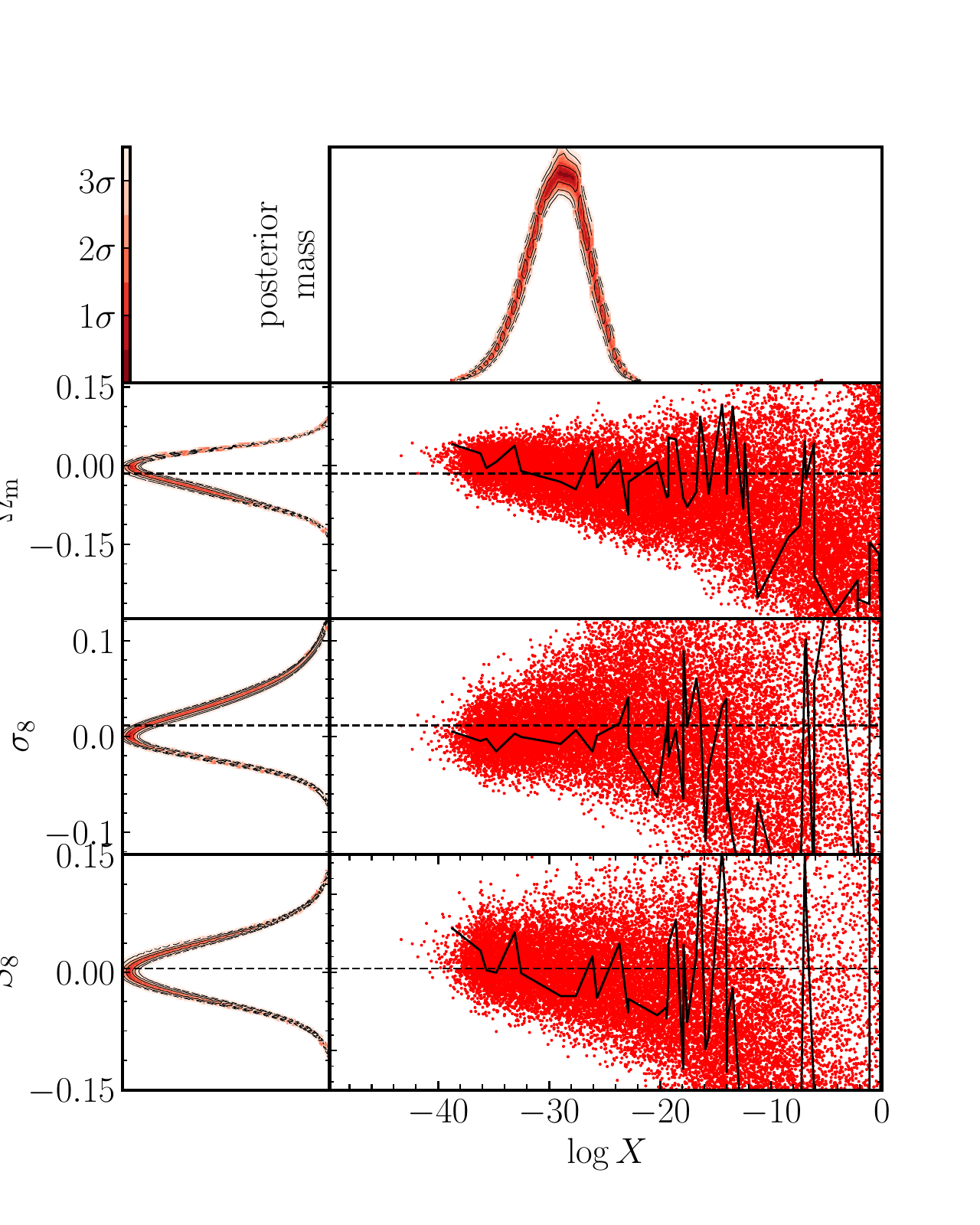}
    \caption{The result of {\tt nestcheck}\citep{higson2018nestcheck} for the baseline 3$\times$2pt analysis, sampled in the real data analysis. The top panel shows the posterior volume as a function of the prior volume, $X$. The left panels show the uncertainty of the posterior distribution from an input nested sampling chain, where the uncertainty is estimated by bootstrapping the chain.}
    \label{fig:nestcheck}
\end{figure}

In this section, we present the results of the convergence test of {\tt MultiNest} sampling for the baseline 3$\times$2pt analysis. We use the nestcheck diagnostic to test the convergence of the {\tt MultiNest} chain, implemented as {\tt nestcheck}\citep{higson2019diagnostic}. Fig.~\ref{fig:nestcheck} shows the result of the convergence test by nestcheck for the main cosmological parameters, $\Omega_{\rm m}, \sigma_8$, and $S_8$. In the top right panel, we can see that the chain covers sufficient posterior volume. The left panels show the uncertainty of the posterior distributions, estimated by bootstrapping the original {\tt MultiNest} chain, and indicating that our estimate of the posterior distributions is robust.

\subsection{Sampler difference}
\begin{figure}
    \includegraphics[width=\columnwidth]{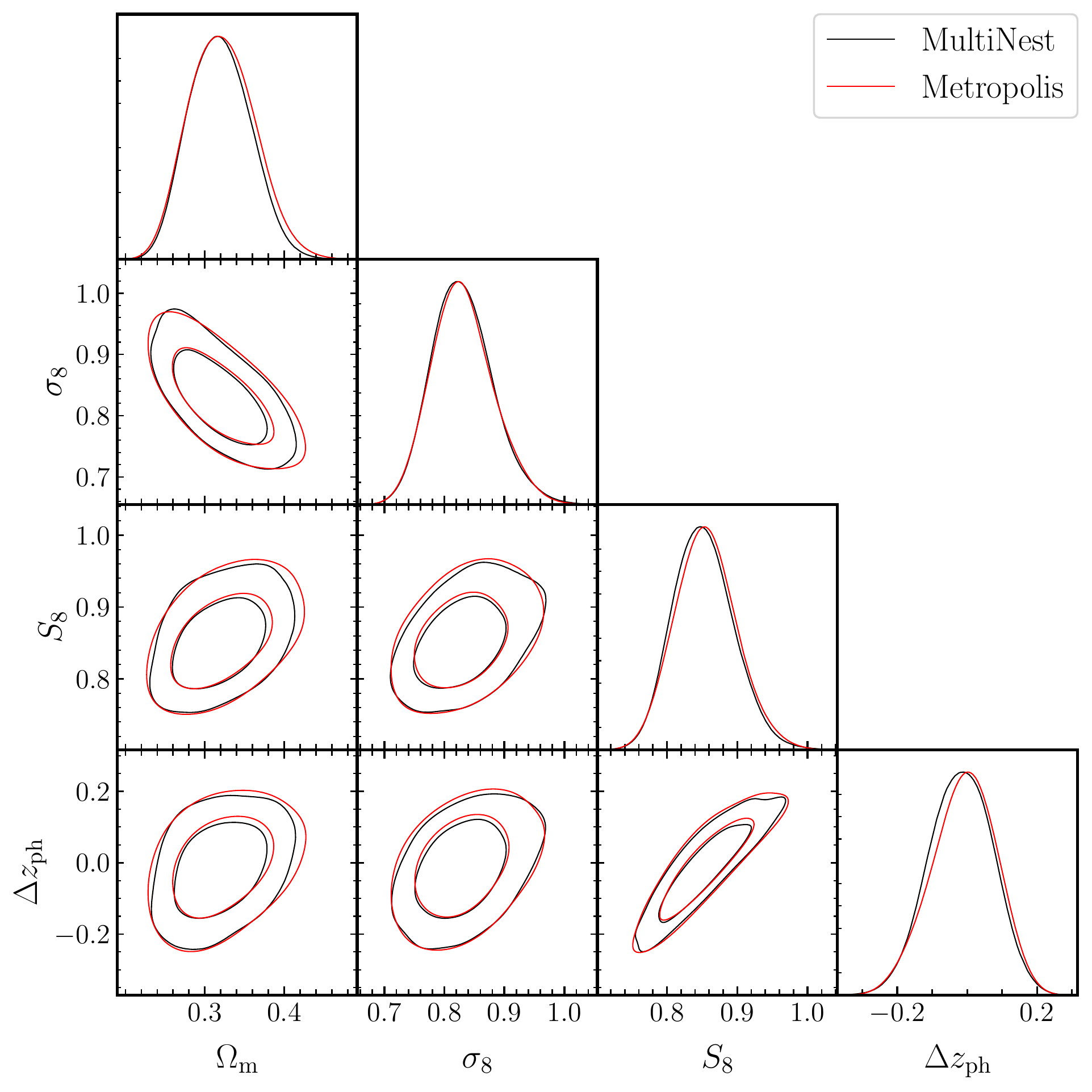}
    \caption{The comparison of posterior estimates of our baseline analysis from the nested sampling by {\tt MultiNest} and the Markov-Chain Monte Carlo sampling by the standard Metropolis algorithm. We do not see significant difference in the cosmological constraints from {\tt MultiNest} and the Metropolis algorithm, meaning {\tt MultiNest} provides sufficiently accurate estimate of credible levels.}
    \label{fig:mn-vs-mp}
\end{figure}
As an additional test of convergence of our parameter estimates, we compare the result of the nested sampling by {\tt MultiNest} to the result with the standard Metropolis algorithm in Fig.~\ref{fig:mn-vs-mp}. The difference between the posterior estimates is almost negligible (the difference in mode and 68\% credible internal is $\sim$2\% for $S_8$), and thus we conclude that our parameter inference by {\tt MultiNest} is robust.

\bibliography{refs-short}

\end{document}

%% file: param_prior_b2.tex
\begin{tabular}{ll}  \hline\hline
Parameter & Prior \\ \hline
\multicolumn{2}{l}{\hspace{-1em}\bf Cosmological parameters}\\
$\Omega_{\rm de}$           & ${\cal U}(0.4594, 0.9094)$\\
$\ln(10^{10}A_{\rm s})$     & ${\cal U}(1.0, 5.0)$\\
$\omega_{\rm c}$            & ${\cal U}(0.0998, 0.1398)$\\ 
$\omega_{\rm b}$            & ${\cal N}(0.02268, 0.00038)$\\
$n_{\rm s}$                 & ${\cal N}(0.9649, 3\times 0.0042)$\\ \hline
\multicolumn{2}{l}{\hspace{-1em}{\bf HOD parameters}} \\
$\log M_{\rm min}(z_i)$     & ${\cal U}(12.0,14.5)$ \\ 
$\sigma_{\log M}^2(z_i)$    & ${\cal U}(0.01,1.0)$ \\ 
$\log M_1(z_i)$             & ${\cal U}(12.0,16.0)$\\
$\kappa(z_i)$               & ${\cal U}(0.01,3.0)$\\ 
$\alpha(z_i)$               & ${\cal U}(0.5,3.0)$\\ \hline
\multicolumn{2}{l}{\hspace{-1em}\bf Magnification bias}\\
$\alpha_{\rm mag}$ (LOWZ)   & ${\cal N}(2.26,0.5)$ \\
$\alpha_{\rm mag}$ (CMASS1) & ${\cal N}(3.56,0.5)$ \\
$\alpha_{\rm mag}$ (CMASS2) & ${\cal N}(3.73,0.5)$ \\ \hline
\multicolumn{2}{l}{\hspace{-1em}\bf Residual photo-$z$/Shear errors}\\
$\Delta z_{\rm ph}$         & ${\cal U}(-1.0,1.0)$  
\\
$\Delta m$           & ${\cal N}(0.0,0.01)$ \\ \hline
\multicolumn{2}{l}{\hspace{-1em}\bf Residual PSF modeling errors}\\
$\alpha_{\rm psf}$          & ${\cal N}(-0.03, 0.01)$\\
$\beta_{\rm psf}$           & ${\cal N}(-1.66, 1.33)$\\
\hline
\multicolumn{2}{l}{\hspace{-1em}\bf IA contamination to cosmic shear}\\
$A_{\rm IA}$                & ${\cal U}(-5.0,5.0)$\\
\hline\hline
\multicolumn{2}{l}{\hspace{-1em}\bf Add. galaxy-halo connection paras} \\ 
\multicolumn{2}{l}{Off-centering parameters} \\
$p_{\rm off}(z_i)$          & ${\cal U}(0,1)$ \\
$R_{\rm off}(z_i)$          & ${\cal U}(0.01,1)$ \\
\multicolumn{2}{l}{\hspace{-1em} \bf Incompleteness parameters} \\
$\alpha_{\rm incomp}(z_i)$  & ${\cal U}(0,5)$\\ 
$\log M_{\rm incomp}(z_i)$  & ${\cal U}(12,15.3)$\\ \hline\hline
\end{tabular}

%% file: 3x2pt_result_b2_dempz.tex
\Omega_{\rm m}  &= 0.382^{+0.031}_{-0.047} (0.401) \, ,\nonumber \\
\sigma_8        &= 0.685^{+0.035}_{-0.026} (0.696) \, ,\nonumber \\
S_8             &= 0.763^{+0.040}_{-0.036} (0.805) \, ,\nonumber \\
\Delta z_{\rm ph}&= -0.05\pm 0.09

%% file: summary_Ommsigma8S8dpz_b2_dempz.tex
\begin{tabular}{llll}
\toprule
\hline\hline
{} &                          $\Omega_{\rm m}$ &                                $\sigma_8$ &                                     $S_8$ \\
\midrule
\hline
3$\times$2pt                                                   &  $0.382_{-0.047}^{+0.031} (0.401, 0.367)$ 
\hspace*{2em}&  $0.685_{-0.026}^{+0.035} (0.696, 0.696)$ \hspace*{2em}&  $0.763_{-0.036}^{+0.040} (0.805, 0.768)$ \\
2$\times$2pt ${}^{\ast}$                                       &  $0.397_{-0.040}^{+0.025} (0.413, 0.382)$ &  $0.683_{-0.024}^{+0.030} (0.678, 0.693)$ &  $0.776_{-0.027}^{+0.032} (0.796, 0.780)$ \\
cosmic shear ${}^{\ast}$                                       &  $0.380_{-0.089}^{+0.095} (0.454, 0.375)$ &  $0.632_{-0.066}^{+0.103} (0.623, 0.674)$ &  $0.735_{-0.040}^{+0.039} (0.767, 0.737)$ \\
\hline
3$\times$2pt, $R_{\rm min}=(4,6)~h^{-1}{\rm Mpc}$ ${}^{\ast}$  &  $0.312_{-0.040}^{+0.044} (0.347, 0.318)$ &  $0.759_{-0.050}^{+0.056} (0.753, 0.767)$ &  $0.785_{-0.028}^{+0.028} (0.809, 0.785)$ \\
3$\times$2pt, $R_{\rm min}=(8,12)~h^{-1}{\rm Mpc}$ ${}^{\ast}$ &  $0.355_{-0.053}^{+0.037} (0.405, 0.341)$ &  $0.690_{-0.035}^{+0.070} (0.660, 0.721)$ &  $0.760_{-0.029}^{+0.035} (0.767, 0.763)$ \\
\hline
3$\times$2pt, w/o LOWZ                                         &  $0.374_{-0.054}^{+0.031} (0.398, 0.357)$ &  $0.708_{-0.032}^{+0.040} (0.684, 0.718)$ &  $0.785_{-0.053}^{+0.043} (0.788, 0.781)$ \\
3$\times$2pt, w/o CMASS1                                       &  $0.380_{-0.048}^{+0.030} (0.415, 0.367)$ &  $0.670_{-0.029}^{+0.031} (0.658, 0.675)$ &  $0.742_{-0.038}^{+0.041} (0.774, 0.745)$ \\
3$\times$2pt, w/o CMASS2                                       &  $0.364_{-0.039}^{+0.041} (0.447, 0.364)$ &  $0.690_{-0.030}^{+0.036} (0.682, 0.699)$ &  $0.764_{-0.036}^{+0.036} (0.833, 0.767)$ \\
\hline
2$\times$2pt, w/o LOWZ ${}^{\ast}$                             &  $0.389_{-0.049}^{+0.028} (0.400, 0.374)$ &  $0.700_{-0.030}^{+0.039} (0.718, 0.713)$ &  $0.792_{-0.035}^{+0.034} (0.830, 0.793)$ \\
2$\times$2pt, w/o CMASS1 ${}^{\ast}$                           &  $0.381_{-0.047}^{+0.032} (0.402, 0.371)$ &  $0.677_{-0.028}^{+0.036} (0.653, 0.688)$ &  $0.760_{-0.031}^{+0.032} (0.756, 0.762)$ \\
2$\times$2pt, w/o CMASS2 ${}^{\ast}$                           &  $0.358_{-0.035}^{+0.055} (0.423, 0.369)$ &  $0.694_{-0.029}^{+0.039} (0.673, 0.705)$ &  $0.774_{-0.032}^{+0.033} (0.800, 0.779)$ \\
\hline
no photo-$z$ error                                             &  $0.394_{-0.031}^{+0.023} (0.394, 0.383)$ &  $0.690_{-0.021}^{+0.026} (0.693, 0.699)$ &  $0.791_{-0.020}^{+0.018} (0.794, 0.788)$ \\
no shear error                                                 &  $0.383_{-0.044}^{+0.027} (0.366, 0.369)$ &  $0.687_{-0.025}^{+0.028} (0.693, 0.694)$ &  $0.765_{-0.035}^{+0.038} (0.765, 0.768)$ \\
fix mag. bias                                                  &  $0.381_{-0.045}^{+0.029} (0.407, 0.367)$ &  $0.686_{-0.025}^{+0.032} (0.688, 0.696)$ &  $0.766_{-0.035}^{+0.037} (0.801, 0.769)$ \\
no PSF error                                                   &  $0.393_{-0.043}^{+0.024} (0.408, 0.375)$ &  $0.684_{-0.025}^{+0.029} (0.681, 0.690)$ &  $0.768_{-0.035}^{+0.037} (0.794, 0.770)$ \\
no IA                                                          &  $0.374_{-0.040}^{+0.032} (0.358, 0.366)$ &  $0.688_{-0.024}^{+0.031} (0.692, 0.696)$ &  $0.766_{-0.036}^{+0.035} (0.756, 0.767)$ \\
extreme IA                                                     &  $0.385_{-0.046}^{+0.027} (0.373, 0.369)$ &  $0.687_{-0.024}^{+0.029} (0.684, 0.694)$ &  $0.766_{-0.037}^{+0.037} (0.763, 0.768)$ \\
\hline
3$\times$2pt ${}^{\dagger}$                                    &  $0.389_{-0.036}^{+0.025} (0.394, 0.378)$ &  $0.689_{-0.025}^{+0.025} (0.692, 0.694)$ &  $0.773_{-0.027}^{+0.031} (0.794, 0.777)$ \\
2$\times$2pt ${}^{\dagger}$                                    &  $0.397_{-0.042}^{+0.024} (0.355, 0.381)$ &  $0.690_{-0.025}^{+0.030} (0.728, 0.699)$ &  $0.779_{-0.028}^{+0.035} (0.792, 0.786)$ \\
cosmic shear ${}^{\dagger}$                                    &  $0.451_{-0.108}^{+0.089} (0.509, 0.387)$ &  $0.624_{-0.063}^{+0.113} (0.611, 0.688)$ &  $0.756_{-0.043}^{+0.044} (0.795, 0.760)$ \\
\hline
2$\times$2pt                                                   &  $0.375_{-0.039}^{+0.029} (0.371, 0.364)$ &  $0.655_{-0.030}^{+0.028} (0.652, 0.661)$ &  $0.719_{-0.034}^{+0.039} (0.725, 0.727)$ \\
cosmic shear                                                   &  $0.228_{-0.070}^{+0.164} (0.280, 0.299)$ &  $0.655_{-0.109}^{+0.135} (0.710, 0.691)$ &  $0.624_{-0.070}^{+0.094} (0.686, 0.660)$ \\
\hline
XMM $(\sim 33~\mathrm{deg}^2)$ ${}^{\ast}$                     &  $0.356_{-0.044}^{+0.031} (0.361, 0.345)$ &  $0.641_{-0.033}^{+0.037} (0.647, 0.642)$ &  $0.693_{-0.055}^{+0.050} (0.710, 0.688)$ \\
GAMA15H $(\sim 41\mathrm{deg}^2)$ ${}^{\ast}$                  &  $0.344_{-0.035}^{+0.042} (0.368, 0.347)$ &  $0.708_{-0.040}^{+0.046} (0.707, 0.714)$ &  $0.768_{-0.047}^{+0.046} (0.782, 0.766)$ \\
HECTOMAP $(\sim 43~\mathrm{deg}^2)$ ${}^{\ast}$                &  $0.382_{-0.043}^{+0.029} (0.398, 0.370)$ &  $0.708_{-0.035}^{+0.043} (0.725, 0.717)$ &  $0.794_{-0.042}^{+0.045} (0.835, 0.795)$ \\
GAMA09H $(\sim 78~\mathrm{deg}^2)$ ${}^{\ast}$                 &  $0.387_{-0.042}^{+0.029} (0.411, 0.375)$ &  $0.681_{-0.029}^{+0.042} (0.663, 0.693)$ &  $0.773_{-0.041}^{+0.044} (0.776, 0.774)$ \\
VVDS $(\sim 96~\mathrm{deg}^2)$ ${}^{\ast}$                    &  $0.356_{-0.043}^{+0.030} (0.373, 0.346)$ &  $0.650_{-0.030}^{+0.034} (0.643, 0.659)$ &  $0.704_{-0.037}^{+0.038} (0.717, 0.705)$ \\
WIDE12H $(\sim 121~\mathrm{deg}^2)$ ${}^{\ast}$                &  $0.333_{-0.030}^{+0.045} (0.326, 0.340)$ &  $0.698_{-0.034}^{+0.039} (0.735, 0.706)$ &  $0.744_{-0.029}^{+0.038} (0.765, 0.749)$ \\
\hline
$\textsc{DEmPZ}$ \& WX                                         &  $0.374_{-0.040}^{+0.030} (0.403, 0.364)$ &  $0.682_{-0.024}^{+0.030} (0.673, 0.690)$ &  $0.758_{-0.038}^{+0.034} (0.780, 0.757)$ \\
$\textsc{Mizuki}$                                              &  $0.383_{-0.039}^{+0.028} (0.371, 0.370)$ &  $0.678_{-0.023}^{+0.024} (0.681, 0.682)$ &  $0.757_{-0.033}^{+0.031} (0.757, 0.756)$ \\
$\textsc{DNNz}$                                                &  $0.381_{-0.044}^{+0.031} (0.390, 0.368)$ &  $0.696_{-0.026}^{+0.043} (0.710, 0.711)$ &  $0.784_{-0.040}^{+0.042} (0.810, 0.786)$ \\
\hline
w/o star weight                                                &  $0.371_{-0.041}^{+0.029} (0.382, 0.360)$ &  $0.690_{-0.026}^{+0.031} (0.679, 0.698)$ &  $0.765_{-0.039}^{+0.033} (0.766, 0.763)$ \\
\hline
offcentering                                                   &  $0.386_{-0.039}^{+0.026} (0.378, 0.373)$ &  $0.685_{-0.022}^{+0.028} (0.682, 0.691)$ &  $0.767_{-0.032}^{+0.037} (0.766, 0.770)$ \\
incompleteness                                                 &  $0.381_{-0.041}^{+0.027} (0.406, 0.368)$ &  $0.694_{-0.024}^{+0.030} (0.679, 0.701)$ &  $0.776_{-0.037}^{+0.032} (0.790, 0.775)$ \\
\hline
$\sigma(\Delta m)=0.1$ prior                                   &  $0.381_{-0.044}^{+0.028} (0.401, 0.368)$ &  $0.685_{-0.024}^{+0.032} (0.701, 0.694)$ &  $0.765_{-0.037}^{+0.036} (0.811, 0.767)$ \\
$\sigma(\Delta z_{\rm ph})=0.2$ prior                          &  $0.387_{-0.045}^{+0.025} (0.395, 0.372)$ &  $0.683_{-0.020}^{+0.031} (0.697, 0.693)$ &  $0.768_{-0.034}^{+0.032} (0.800, 0.769)$ \\
2 cosmo paras                                                  &  $0.359_{-0.019}^{+0.018} (0.367, 0.356)$ &  $0.691_{-0.021}^{+0.023} (0.703, 0.695)$ &  $0.756_{-0.028}^{+0.031} (0.778, 0.757)$ \\
\bottomrule
\hline\hline
\end{tabular}